\documentclass[aps,prd,eqsecnum,showpacs,superscriptaddress,twocolumn,nofootinbib,
floatfix,preprintnumbers,altaffilletter]{revtex4}

\usepackage{graphicx,hyperref,amssymb,amsmath,longtable,rotating,color,epsfig,epsf,fancyhdr}

\begin{document}

\pagestyle{fancy}
\rhead[]{}
\lhead[]{}

\newcommand{\ee}[1]{\!\times\!10^{#1}}
\newcommand{\gws}{gravitational waves~}
\newcommand{\gw}{gravitational wave~}
\newcommand{\geo}{GEO\,600~}

\title{Upper limits on gravitational wave emission from 78 radio pulsars}

\input authorlist.tex

\email{matthew@astro.gla.ac.uk}

\date{\today}

\begin{abstract}
We present upper limits on the gravitational wave emission from 78
radio pulsars based on data from the third and fourth science runs
of the LIGO and GEO\,600 gravitational wave detectors. The data from
both runs have been combined coherently to maximise sensitivity. For
the first time pulsars within binary (or multiple) systems have been
included in the search by taking into account the signal modulation
due to their orbits. Our upper limits are therefore the first
measured for 56 of these pulsars. For the remaining 22, our results
improve on previous upper limits by up to a factor of 10. For
example, our tightest upper limit on the gravitational strain is 
$2.6\ee{-25}$ for PSR\,J1603$-$7202, and the equatorial ellipticity of
PSR\,J2124$-$3358 is less than $10^{-6}$. Furthermore, our strain
upper limit for the Crab pulsar is only 2.2 times greater than the
fiducial spin-down limit.
\end{abstract}

\preprint{LIGO-P060011-05-Z}

\maketitle


\section{Introduction}
This paper details the results of a search for gravitational wave
signals from known radio pulsars in data from the third and fourth
LIGO and GEO\,600 science runs (denoted S3 and S4). These runs were
carried out from $31^{\rm st}$ October 2003 to $9^{\rm th}$ January
2004 and $22^{\rm nd}$ February to $23^{\rm rd}$ March, 2005
respectively. We have applied, and extended, the search technique of
Dupuis and Woan \cite{dw} to generate upper limits on the \gw
amplitude from a selection of known radio pulsars, and infer upper
limits on their equatorial ellipticities. The work is a natural
extension our previous work given in Refs.~\cite{ab1, ab2}.

\subsection{Motivation}\label{sec:motivation}
To emit \gws a pulsar must have some mass (or mass-current)
asymmetry around its rotation axis. This can be achieved through
several mechanisms such as elastic deformations of the solid crust
or core or distortion of the entire star by an extremely strong
misaligned magnetic field (see Sec.~III of Ref.~\cite{Abbott:2006vg}
for a recent review). Such mechanisms generally result in a triaxial
neutron star which, in the quadrupole approximation and with
rotation and angular momentum axes aligned, would produce \gws at
twice the rotation frequency.  These waves would have a
characteristic strain amplitude at the Earth (assuming optimal
orientation of the rotation axis) of
\begin{equation}\label{Pulsarh0}
h_0 = \frac{16\pi^2G}{c^4}\frac{\varepsilon I_{zz}\nu^2}{r},
\end{equation}
where $\nu$ is the neutron star's spin frequency, $I_{zz}$ its
principal moment of inertia, $\varepsilon = (I_{xx} -
I_{yy})/I_{zz}$ its equatorial ellipticity, and $r$ its distance
from Earth \cite{jks}.

A rotating neutron star may also emit gravitational waves at
frequencies other than $2\nu$.  For instance, if the star is
undergoing free precession there will be gravitational wave emission
at (or close to) both $\nu$ and $2\nu$ \cite{Zimmermann:1979}.  In
general such a precession would modulate the time-of-arrival of the
radio pulses.  No strong evidence of such a modulation is seen in any of the
pulsars within our search band, although it might go unnoticed by radio
astronomers, either because the modulation is small (as would be the case if the
precession is occurring about an axis close to the pulsar beam axis) or because
the period of the modulation is very long.  However, this misalignment and
precession will be quickly damped unless sustained by some mechanism (e.g.
Ref.~\cite{JonesAndersson:2002}), and even with such a mechanism
calculations give strain amplitudes which would probably be too low compared to
LIGO sensitivities \cite{JonesAndersson:2002, Bonazzola:1996}.  For
these reasons, and for the reason discussed in \S\ref{sec:method}, we
restrict our search to twice the rotation frequency.  Of course, it
cannot be ruled out that there are in fact other gravitational wave
components, perhaps caused either by a stronger than expected
precession excitation mechanism or by an event in the pulsar's recent
past that has set it into a precessional motion which has not yet
decayed away.  A search for gravitational waves from the Crab pulsar
at frequencies other than twice the rotation frequency is currently
under way and will be presented elsewhere.

Known pulsars provide an enticing target for \gw searches as their
positions and frequencies are generally well-known through radio or
X-ray observations. As a result the signal search covers a much
smaller parameter space than is necessary when searching for signals
from unknown sources, giving a lower significance threshold.  In
addition, the deterministic nature of the waves allows a building
up of the signal-to-noise ratio by observing coherently for a
considerable time. The main drawback in a search for \gws from the
majority of known pulsars is that the level of emission is likely to
be lower than can be detected with current detector sensitivities.

Using existing radio measurements, and some reasonable assumptions,
it is possible to set an upper limit on the \gw amplitude from a
pulsar based purely on energy conservation arguments. If one assumes
that the pulsar is an isolated rigid body and that the observed
spin-down of the pulsar is due to the loss of rotational kinetic
energy as gravitational radiation (i.e.,
${\rm{d}}E_{\rm{rot}}/{\rm{d}}t = 4\pi^2I_{zz}\nu\dot{\nu}$) then the
gravitational wave amplitude at the Earth (assuming optimal orientation of
the rotation axis) would be
\begin{equation}\label{spindownUL}
h_\mathrm{sd} = \left(\frac{5}{2}\frac{GI_{zz}|\dot{\nu}|}{c^3r^2\nu}\right)^{1/2}.
\end{equation}
Of course these assumptions may not hold, but it would be surprising
if neutron stars radiated significantly more gravitational energy
than this. With these uncertainties in mind, searches such as the one 
described in this paper place {\em direct} upper limits on gravitational
wave emission from rotating neutron stars, and these limits are already
approaching the regime of astrophysical interest.

\subsection{Previous results}
Before the advent of large-scale interferometric detectors, there
was only a limited ability to search for \gws from known pulsars.
Resonant mass \gw detectors are only sensitive in a relatively narrow 
band around their resonant frequency and so cannot be used to target objects
radiating outside that band. A specific attempt to search for \gws
from the Crab pulsar at a frequency of $\sim 60$\,Hz was, however,
made with a specially designed aluminium quadrupole antenna
\cite{Hirakawa:1978, Suzuki:1995} giving a $1\sigma$ upper limit
of $h_0 \le 2\ee{-22}$. A search for \gws from what was then the
fastest millisecond pulsar, PSR\,J1939+2134, was conducted by Hough
{\it et al} \cite{Hough:1983} using a split bar detector, producing
an upper limit of $h_0 < 10^{-20}$.

The first pulsar search using interferometer data was carried out 
with the prototype 40\,m interferometer at Caltech by Hereld
\cite{Hereld:1983}. The search was again for \gws from
PSR\,J1939+2134, and produced upper limits of $h_0 < 3.1\ee{-17}$
and $h_0 < 1.5\ee{-17}$ for the first and second harmonics of the
pulsar's rotation frequency.

A much larger sample of pulsars is accessible to broadband 
interferometers. As of the beginning of 2005 the Australia Telescope 
National Facility (ATNF) online pulsar catalogue \cite{ATNF} 
listed\footnote{The catalogue is continually updated and as such 
now contains more objects.} 154 millisecond and young pulsars, all 
with rotation frequencies 
$> 25$\,Hz (\gw frequency $> 50$\,Hz) that fall within the design 
band of the LIGO and \geo interferometers, and the search for their 
\gws has developed rapidly since the start of data-taking runs in 
2002. Data from the first science run (S1) were used to perform a 
search for \gws at twice the rotation frequency from PSR\,J1939+2134 
\cite{ab1}. Two techniques were used in this search: one a frequency 
domain, frequentist search, and the other a time domain, Bayesian 
search which gave a 95\% credible amplitude upper limit of 
$1.4\ee{-22}$, and an ellipticity upper limit of $2.9\ee{-4}$ 
assuming $I_{zz} = 10^{38}\,{\rm kg}\,{\rm m}^2$.

Analysis of data from the LIGO S2 science run set upper limits on
the \gw amplitude from 28 radio pulsars \cite{ab2}. To do this, new
radio timing data were obtained to ensure the pulsars' rotational
phases could be predicted with the necessary accuracy and to check
that none of the pulsars had glitched. These data gave strain upper
limits as low as a few times $10^{-24}$, and several ellipticity
upper limits less than $10^{-5}$. The Crab pulsar was also studied
in this run, giving an upper limit a factor of $\sim 30$ greater
than the spin-down limit considered above. Prior to this article 
these were the most sensitive studies made. Preliminary
results for the same 28 pulsars using S3 data were given in Dupuis
(2004) \cite{Dupuis:2004}, and these are expanded below.

In addition to the above, data from the LIGO S2 run have been used
to perform an all-sky (i.e., non-targeted) search for continuous
wave signals from isolated sources, and a search for a signal from
the neutron star within the binary system Sco-X1
\cite{Abbott:2006vg}. An all-sky continuous wave search using the
distributed computing project
\href{http://einstein.phys.uwm.edu}{Einstein@home}\footnote{\url{http://einstein.phys.uwm.edu}}
has also been performed on S3 data \cite{einsteinathome}. These
searches use the same search algorithms, are fully coherent and are
ongoing using data from more recent (and therefore more sensitive)
runs. Additional continuous wave searches using incoherent
techniques are also being performed on LIGO data \cite{HoughPaper, PSH}.

Unfortunately the pulsar population is such that most have spin
frequencies that fall below the sensitivity band of current
detectors. In the future, the low-frequency sensitivity of VIRGO
\cite{virgo} and Advanced LIGO \cite{Creighton:2003} should allow
studies of a significantly larger sample of pulsars.

\subsection{The signal}
Following convention, we model the observed phase evolution of a
pulsar using a Taylor expansion about a fixed epoch time $t_0$:
\begin{eqnarray}\label{PhaseTaylorExp}
\phi(T) & = & \phi_0 + 2\pi{}\Big\{\nu_0(T-t_0) + \frac{1}{2}\dot{\nu}_0(T-t_0)^2 + \nonumber \\
& & \frac{1}{6}\ddot{\nu}_0(T-t_0)^3 + \ldots\Big\},
\end{eqnarray}
where $\phi_0$ is the initial (epoch) spin phase, $\nu_0$ and its time
derivatives are the pulsar spin frequency and spin-down coefficients at
$t_0$, and $T$ is the pulsar proper time.

The expected signal in an interferometer from a triaxial pulsar is
\begin{eqnarray}\label{PulsarSignal}
h(t) & = & \frac{1}{2}F_+(t;\psi)h_0(1+\cos^2\iota)\cos{2\phi(t)} + \nonumber \\
& & F_{\times}(t;\psi)h_0\cos{\iota}\sin{2\phi(t)},
\end{eqnarray}
where $\phi(t)$ is the phase evolution in the detector time $t$, $F_+$ and
$F_{\times}$ are the detector antenna patterns for the plus and
cross polarisations of gravitational waves, $\psi$ is the wave
polarisation angle, and $\iota$ is the angle between the rotation
axis of the pulsar and the line-of-sight. A gravitational wave
impinging on the interferometer will be modulated by Doppler, time
delay and relativistic effects caused by the motions of the Earth
and other bodies in the solar system. Therefore we need to transform
the `arrival time' of a wave-crest at the detector, $t$, to its
arrival time at the solar system barycentre (SSB) $t_{\rm b}$ via
\begin{equation}\label{TimeDelay}
t_{\rm b} = t + \delta{}t = t +
\frac{\mathbf{r}\cdot\hat{\mathbf{n}}}{c} + \Delta_{E_{\odot}} +
\Delta_{S_{\odot}},
\end{equation}
where $\mathbf{r}$ is the position of the detector with respect to
the SSB, $\hat{\mathbf{n}}$ is the unit vector pointing to the
pulsar, $\Delta_{E_{\odot}}$ is the special relativistic Einstein
delay, and $\Delta_{S_{\odot}}$ is the general relativistic Shapiro
delay \cite{TaylorWeisberg:1989}. Although pulsars can be assumed to
have a large velocity with respect to the SSB, it is conventional to
ignore this Doppler term and set $t_b = T$, as its proper motion is
generally negligible (see \S\ref{sec:upperlimits} for cases where
this assumption is not the case). For pulsars in binary systems,
there will be additional time delays due to the binary orbit,
discussed in \S\ref{binaries}.

\section{Instrumental performance in S3/S4}
The S3 and S4 runs used all three LIGO interferometers (H1 and H2 at
the Hanford Observatory in Washington, and L1 at the Livingston 
Observatory in Louisiana) in the USA and the GEO\,600 interferometer in Hannover, 
Germany. \geo did not run for all of S3, but had two main data taking periods
between which improvements were made to its sensitivity. All these
detectors had different duty factors and sensitivities.

\subsection{LIGO}
For S3 the H1 and H2 interferometers maintained relatively high
duty factors of $69.3\%$ and $63.4\%$ respectively. The L1
interferometer was badly affected by anthropogenic seismic noise 
sources during the day and thus had a duty factor of only $21.8\%$.

Between S3 and S4 the L1 interferometer was upgraded with better
seismic isolation. This greatly reduced the amount of time the
interferometer was thrown out of its operational state by 
anthropogenic noise, and allowed it to operate successfully during 
the day, with a duty factor of $74.5\%$ and a longest lock stretch 
of 18.7\,h. The H1 and H2 interferometers also both improved 
their duty factors to $80.5\%$ and $81.4\%$, with longest lock 
stretches of almost a day.
\begin{figure}[!htbp]
\includegraphics[width=0.45\textwidth]{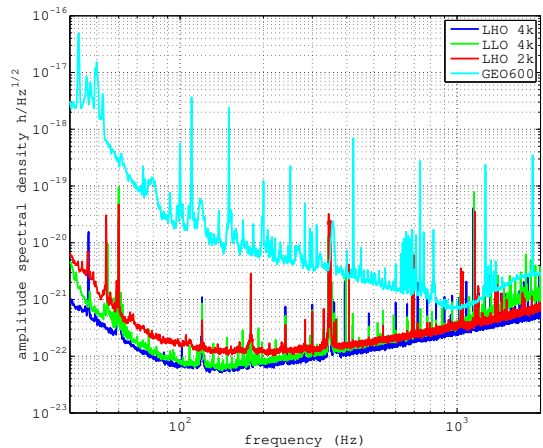}
\caption{Median strain amplitude spectral density curves for the LIGO and
GEO\,600 interferometers during the S4 run.}\label{fig:senscurvesS4}
\end{figure}
The typical strain sensitivities of all the interferometers during S4
can be seen in Fig.~\ref{fig:senscurvesS4}. This shows the LIGO detectors reach
their best sensitivities at about 150\,Hz, whilst GEO\,600 achieves its best sensitivity
at its tuned frequency of 1\,kHz.

\subsection{GEO\,600}
During S3 \geo was operated as a dual-recycled Michelson
interferometer tuned to have greater sensitivity to signals around 1\,kHz. The first
period of \geo participation in S3 was between $5^{\rm th}$ to
$11^{\rm th}$ November 2003, called S3\,I, during which the detector
operated with a 95.1\% duty factor. Afterwards, \geo was taken
offline to allow further commissioning work aimed at improving
sensitivity and stability. Then from $30^{\rm th}$ December 2003 to
$13^{\rm th}$ January 2004 \geo rejoined S3, called S3\,II,
 with an improved duty factor of 98.7\% and with more than one order of
magnitude improvement in peak sensitivity. During S3 there were
five locks of longer than 24 hours and one lock longer than 95 hours.
For more information about the performance of \geo during S3 see
Ref.~\cite{Smith:2004}.

\geo participated in S4 from $22^{\rm nd}$ February to $24^{\rm th}$
March 2005, with a duty factor of 96.6\%. It was operated in
essentially the same optical configuration as in S3. With respect to
S3, the sensitivity was improved more than an order of magnitude
over a wide frequency range, and close to two orders or magnitude
around 100\,Hz. For more information about \geo during S4 see
Ref.~\cite{Hild:2006}.

\subsection{Data quality}
When a detector is locked on resonance and all control loops are in their 
nominal running states and there are no
on-site work activities that are known to compromise the data, then the
data are said to be {\it science mode}. All science mode data are not of sufficient
quality to be analysed however, and may be flagged for exclusion.
Examples of such data quality flags are ones produced for epochs of
excess seismic noise, and the flagging of data corrupted by
overflows of photodiode ADCs.  For this analysis we use all science
mode data for which there is no corresponding data quality flag. For
S3 this gives observation times of 45.5\,days for H1; 42.1\,days for
H2; and 13.4\,days for L1. For S4 this gives observation times of
19.4\,days for H1; 22.5\,days for H2; and 17.1\,days for L1.

\section{The search method}\label{sec:method}

Our search method involves heterodyning the data using the phase model $\phi(t)$
to precisely unwind the phase evolution of the
expected signal, and has been discussed in detail in Ref.~\cite{dw}.
After heterodyning, the data are low-pass
filtered, using a ninth order Butterworth filter with a knee
frequency of 0.5\,Hz, and re-binned from the raw data sample rate of
16\,384\,Hz to 1/60\,Hz i.e., one sample per minute. The motion of
the detector within the solar system modulates the signal and this
is taken into account within the heterodyne by using a time delay
given in Eq.~(\ref{TimeDelay}), which transforms the signal to
the SSB. Signals from binary pulsar systems have an extra modulation
term in the signal, as discussed briefly below, and these we
targeted for the first time in S3/S4.

The search technique used here is currently only able to target
emission at twice the pulsar's rotation frequency. Emission near the
rotation frequency for a precessing star is likely to be offset from
the observed pulsation frequency by some small factor dependent on
unknown details of the stellar structure \cite{JonesAndersson:2002}.
As our search technique requires precise knowledge of the phase
evolution of the pulsar such an additional parameter cannot currently
be taken into account. For the emission at twice the rotation
frequency there is no extra parameter dependence on the frequency
and this is what our search was designed for.

We infer the pulsar signal parameters, denoted $\mathbf{a} = (h_0,
\phi_0, \cos{\iota}, \psi)$, from their (Bayesian) posterior
probability distribution (pdf) over this parameter space, assuming
Gaussian noise. The data are broken up into time segments over which
the noise can be assumed stationary and we analytically marginalise
over the unknown noise floor, giving a Student-t likelihood for the
parameters for each segment (see Ref.~\cite{dw} for the method).
Combining the segments gives an overall likelihood of
\begin{widetext}
\begin{equation}\label{equation:likelihood}
p(\{B_k\}|\mathbf{a}) \propto \prod_j^M \left(\sum_{k = 1 + \sum_{i=1}^{j-1} m_i}^{\sum_{i=1}^{j}
m_i} (\Re\{B_k\}-\Re\{y_k\})^2 + (\Im\{B_k\}-\Im\{y_k\})^2\right)^{-m_j},
\end{equation}
\end{widetext}
where each $B_k$ is a heterodyned sample with a sample rate of one
per minute, $M$ is the number of segments into which the whole data
set has been cut, $m_j$ is the number of data points in the $j$th
segment, and $y_{k}$, given by
\begin{equation}
y_k  = \frac{1}{4}F_+(t_k;\psi)h_0(1+\cos^2\iota)e^{i2\phi_0} -
\frac{i}{2}F_{\times}(t_k;\psi)h_0\cos{\iota}e^{i2\phi_0},
\end{equation}
is the \gw signal model evaluated at $t_{k}$, the time corresponding
to the $k$'th heterodyned sample. In Ref.~\cite{ab2} the value of
$m_j$ was fixed at 30 to give 30\,min data segments, and data that
was contiguous only on shorter timescales, and which could not be
fitted into one of these segments, was thrown out. In the analysis
presented here we have allowed segment lengths to vary from 5 to
30\,min, so we maximise the number of 30-minute segments whilst also
allowing shorter segments at the end of locked stretches to
contribute. The likelihood in Eq.~(\ref{equation:likelihood})
assumes that the data is stationary over each of these 30 minute (or
smaller) segments. This assumption holds well for our
data. Large outliers can also be identified and vetoed from the data,
for example those at the beginning of a data segment caused by
the impulsive ringing of the low-pass filter applied after the data 
is heterodyned.

The prior probabilities for each of the parameters are taken as
uniform over their respective ranges. Upper limits on $h_0$ are set
by marginalising the posterior over the nuisance parameters and then
calculating the $h_0^{95\%}$ value that bounds the cumulative probability
for the desired credible limit of 95\%:
\begin{equation}\label{eqn:upperlimit}
0.95 = \int_{0}^{h_0^{95\%}} p(h_0|\{B_k\}) {\rm d}h_0.
\end{equation}

\subsection{Combining data}
In the search of Ref.~\cite{ab2} the combined
data from the three LIGO interferometers were used to improve the
sensitivity of the search. This was done by forming the joint
likelihood from the three {\it independent} data sets:
\begin{equation}\label{eqn:jointlikelihood}
p({B_k}|\mathbf{a})_{\rm Joint} = p({B_k}|\mathbf{a})_{\rm H1} \,.\,
p({B_k}|\mathbf{a})_{\rm H2} \,.\, p({B_k}|\mathbf{a})_{\rm L1}.
\end{equation}
This is valid provided the data acquisition is coherent between
detectors and supporting evidence for this is presented in
\S\ref{injections}. It is of course a simple matter to extend
Eq.~(\ref{eqn:jointlikelihood}) to include additional likelihood
terms from other detectors, such as GEO\,600.

In this analysis we also combine data sets from two different
science runs. This is appropriate because S3 and S4 had comparable
sensitivities over a large portion of the spectrum. Provided the
data sets maintain phase coherence between runs, this combination can
simply be achieved by concatenating the data sets from the two runs
together for each detector.

\begin{figure}[!htbp]
\includegraphics[width=0.45\textwidth]{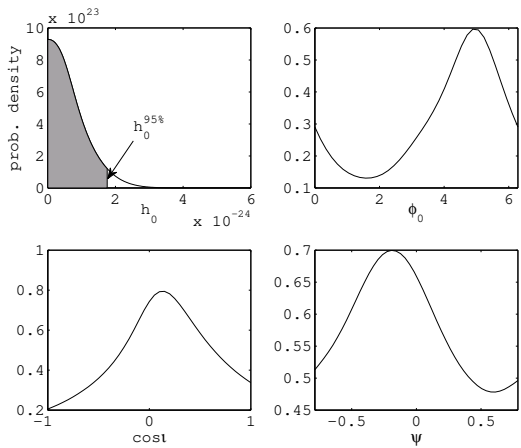}
\caption{The marginalised posterior pdfs for the four unknown pulsar
parameters $h_0$, $\phi_0$, $\cos{\iota}$ and $\psi$, for
PSR\,J0024$-$7204C using the joint data from the three LIGO
detectors over S3 and S4.}\label{fig:examplePDF}
\end{figure}

An example of the posterior pdfs for the four unknown pulsar
parameters of PSR\,J0024$-$7204C (each marginalised over the three
other parameters) is shown in Fig.~\ref{fig:examplePDF}.
The pdfs in Fig.~\ref{fig:examplePDF} are from the joint analysis
of the three LIGO detectors using the S3 and S4 data, 
all combined coherently. The shaded area in the $h_0$ posterior 
shows the area containing 95\% of the
probability as given by Eq.~(\ref{eqn:upperlimit}). In this
example the posterior on $h_0$ is peaked at $h_0=0$, though any
distribution that is credibly close to zero is consistent with
$h_0=0$.  Indeed an upper limit can formally be set even when the
bulk of the probability is well away from zero (see the discussion
of hardware injections in \S\ref{injections}).

\subsection{Binary models}\label{binaries}
Our previous known pulsar searches \cite{ab1, ab2} have excluded
pulsars within binary systems, despite the majority of pulsars within our detector
band being in such systems.  To address this, we have included an
additional time delay to transform from the binary system barycentre
(BSB) to pulsar proper time, which is a stationary reference frame
with respect to the pulsar. The code for this is based on the widely
used radio pulsar timing software TEMPO \cite{TEMPO}.  The algorithm
and its testing are discussed more thoroughly in
Ref.~\cite{Pitkin:2006}.

There are five principal parameters describing a Keplerian orbit:
the time of periastron, $T_0$, the longitude of periastron,
$\omega_0$, the eccentricity, $e$, the period, $P_b$, and the
projected semi-major axis, $x=a\sin{i}$. These describe the majority
of orbits very well, although to fully describe the orbit of some
pulsars requires additional relativistic parameters. The basic
transformation and binary models below are summarised by Taylor and
Weisberg \cite{TaylorWeisberg:1989} and Lange {\it et al.}\
\cite{Lange:2001}, and are those used in TEMPO. The transformation
from SSB time $t_b$ to pulsar proper time $T$ follows the form of
Eq.~(\ref{TimeDelay}) and is
\begin{equation}\label{SSBtoPPT}
t_b  = T + \Delta_{\rm{R}} + \Delta_{\rm{E}} + \Delta_{\rm{S}},
\end{equation}
where $\Delta_{\rm{R}}$ is the Roemer time delay giving the
propagation time across the binary orbit, $\Delta_{\rm{E}}$ is the
Einstein delay and gives gravitational redshift and time dilation
corrections, and $\Delta_{\rm{S}}$ is the Shapiro delay and gives
the general relativistic correction (see
Ref.~\cite{TaylorWeisberg:1989} for definitions of these delays).

The majority of binary pulsars can be described by three orbital
models: the Blandford-Teukolsky (BT) model, the low eccentricity
(ELL1) model, and the Damour-Deruelle (DD) model (see
Refs.~\cite{TaylorWeisberg:1989, Lange:2001, TEMPO} for further
details of these models). These different models make different
assumptions about the system and/or are specialised to account for
certain system features. For example, the ELL1 model is used in
cases where the eccentricity is very small, and therefore periastron
is very hard to define, in which case the time and longitude of
periastron will be highly correlated and have to be reparameterised
to the Laplace-Lagrange parameters \cite{Lange:2001}. When a binary
pulsar's parameters are estimated from radio observations using
TEMPO the different models are used accordingly. These models can be
used within our search to calculate all the associated time delays
and therefore correct the signal to the pulsar proper time, provided
we have accurate model parameters for the pulsar.

\section{Pulsar selection}
The noise floor of the LIGO detectors increases rapidly below about
50\,Hz, so pulsar targets were primarily selected on their
frequency. The choice of a 50\,Hz \gw frequency cut-off (pulsar spin
frequency of 25\,Hz) is somewhat arbitrary, but it also loosely
reflects the split between the population of fast
(millisecond/recycled and young) pulsars and slow pulsars.

All 154 pulsars with spin frequencies $> 25$\,Hz were taken from the
ATNF online pulsar catalogue \cite{ATNF} (described in
Ref.~\cite{Manchester:2005}). The accuracy of these parameters
varies for each pulsar and is dependent on the time span, density of
observations and the noise level of the timing observations. Clearly
it is important to ensure that parameter uncertainties do not lead
to unacceptable phase errors in the heterodyne. Pulsars are not
perfect clocks, so the epoch of the parameters is also important as
more recent measurements will better reflect the current state of
the pulsar. Importantly, there is near-continuous monitoring of the
Crab pulsar at Jodrell Bank Observatory, and as such its parameters
are continuously updated \cite{CrabEphemeris}.

Precise knowledge of the phase evolution of each target pulsar is
vital for our analysis, and possible effects that may lead to a
departure from the simple second-order Taylor expansion are
discussed below.

\subsection{Pulsar timing}
Using TEMPO, we obtained the parameters of 75 pulsars from the 
regular observation programs carried out at Jodrell Bank Observatory 
and the Parkes Telescope (see Ref.~\cite{Hobbs:2004} for details of 
the techniques used for this). For 37 of these the timings spanned 
the period of S3. These same model parameters were used to extrapolate 
the pulsar phases to the period of S4. The effect of parameter
uncertainties on this extrapolation is discussed in
\S\ref{sec:errors}, but is only important in its effect on the
extrapolated phase. For those pulsars observed during S3 the
interpolation is taken to be free from significant error.

The parameters for 16 additional pulsars (for which new timings were
not available) were taken directly from the ATNF catalogue, selected
using criteria described in the following section. The parameters of
the X-ray pulsar PSR\,J0537$-$6910 were taken from
Ref.~\cite{Marshall:2004} and those for the Crab pulsar from the
Jodrell Bank monthly ephemeris \cite{CrabEphemeris}. The remaining
61 pulsars (from the original list of 154) were not timed with
sufficient confidence and were excluded from the search. This
included many of the newly discovered pulsars (for example the 21
millisecond pulsars in the Terzan 5 globular cluster
\cite{Ransom:2005}) for which accurate timing solutions have yet to
be published. We therefore had a catalogue of 93 timed pulsars for
our \gw search.

\subsection{Error propagation in source parameters}\label{sec:errors}
The impact of parameter uncertainties on the search was assessed
for both the S3 and S4 runs. At some level there are positional,
frequency and frequency derivative uncertainties for all the target
pulsars, and for pulsars in binary system there are also
uncertainties associated with all the binary orbital parameters.
Some of these uncertainties are correlated, for example the error on
frequency could affect the accuracy of the first frequency
derivative, and the binary time of periastron and longitude of
periastron are also highly correlated.

We took a `worst-case scenario' approach by adding and subtracting
the quoted uncertainties from the best-fit values of all the
parameters to determine the combination which gave a maximum phase
deviation, when propagated over the period of the run (either S3 or
S4), from the best fit phase value calculated over the same time
period. For example if we assume $\phi(t_{\rm S3})$  given by
Eq.~(\ref{PhaseTaylorExp}) (ignoring for simplicity the $\phi_0$
and $\ddot{\nu}$ terms) is the best fit phase over the time span of
S3, $t_{\rm S3}$, the maximum phase uncertainty is
\begin{eqnarray}
\Delta\phi_{\rm err} & = & \rm{max}\Big[\Big|\phi(t_{\rm S3}) \pm
2\pi\Big\{(\nu\pm\sigma_{\nu})(t_{\rm
S3}\pm\sigma_{t_{\rm S3}} \nonumber \\
& & + \frac{1}{2}(\dot{\nu}\pm\sigma_{\dot{\nu}})(t_{\rm S3}\pm\sigma_{t_{\rm S3}})^2 +
\ldots\Big\}\Big|\Big],
\end{eqnarray}
where the $\sigma$s are the uncertainties on the individual parameters.
Correlations between the parameters mean that this represents an
upper limit to the maximum phase uncertainty, sometimes greatly
overestimating its true value.

There are 12 pulsars with overall phase uncertainty $>30^{\circ}$ in S3,
which we take as the threshold of acceptability. A $30^{\circ}$ phase 
drift could possibly give a factor of $\sim 1-\cos{30^{\circ}} = 0.13$ 
in loss of sensitivity for a signal. Nine of these are
in binary systems (PSRs\,J0024$-$7204H, J0407+1607, J0437$-$4715,
J1420$-$5625, J1518+0205B, J1709+2313, J1732$-$5049, J1740$-$5340 and
J1918$-$0642) and in five of these $T_0$ and $\omega_0$ contribute
most to the phase uncertainty. For the three isolated pulsars
(PSRs\,J0030+0451, J0537$-$6910, and J1721$-$2457) the phase error 
is dominated by uncertainties in frequency and/or position.

Applying the same criterion to the time-span of S4 we find that
PSR\,J1730$-$2304 rises above the limit. For this pulsar its
parameter uncertainties do not affect it for the S3 analysis as it
was timed over this period, however when extrapolating over the 
time of the S4 run the uncertainties become non-negligible.

In total there are 13 pulsars rejected over the combined run. This
highly conservative parameter check reduces our 93 candidate pulsars to 80.

\subsection{Timing noise}
Pulsars are generally very stable rotators, but there are phenomena
which can cause deviations in this stability, generically known as
timing noise. The existence of timing noise has been clear since the
early days of pulsar astronomy and appears as a random walk in
phase, frequency or frequency derivative of the pulsar about the
regular spin-down model given in Eq.~(\ref{PhaseTaylorExp})
\cite{CordesHelfand:1980}. The strength of this effect was
quantified in Ref.~\cite{CordesHelfand:1980} as an \emph{activity
parameter} $A$, referenced to that of the Crab pulsar, and in
Ref.~\cite{Arzoumanian:1994} as a \emph{stability parameter}
$\Delta_8$. $A$ is based on the logarithm of the ratio of the rms
residual phase of the pulsar, after removal of the timing model, to
that of the Crab pulsar over an approximately three-year period.
$\Delta_8$ is not based on the stochastic nature of the Crab
pulsar's timing noise and and is defined for a fixed time
($10^8$\,s) as
\begin{equation}\label{delta8}
\Delta_8 = \log{\left(\frac{1}{6\nu}|\ddot{\nu}|\times (10^8\,{\rm s})^3\right)}.
\end{equation}
This assumes that the measured value of $\ddot{\nu}$ is dominated by
the timing noise rather than the pulsar's intrinsic second spin-down
derivative. Although generally true, this assumption is not valid
for the Crab pulsar and PSR\,J0537$-$6910, where a non-timing noise
dominated $\ddot{\nu}$ can be measured between
glitches\footnote{These two pulsars are among the most prolific
glitchers, and in any global fit to their parameters the value of
$\ddot{\nu}$ would most likely be swamped by the glitch events.}.
This quantity relates to the pulsar clock error caused by timing
noise. The value of $\ddot{\nu}$ is so small as to be unmeasurable
for most pulsars, although an upper limit can often be defined.
Arzoumanian {\it et al.} \cite{Arzoumanian:1994} deduce, by eye, a
linear relationship between $\Delta_8$ and $\log{\dot{P}}$ of
\begin{equation}\label{delta8slope}
\Delta_8 = 6.6 + 0.6\log{\dot{P}},
\end{equation}
where $\dot{P} = -\dot{\nu}/\nu^2$ is the period derivative.

As defined, $\Delta_8$ is a somewhat imprecise indicator of the
timing noise, not least because the time span of $10^8$ seconds
chosen by Arzoumanian {\it et al.} was simply the length of their
data set. A preferred measure may simply be the magnitude and sign
of $\ddot{P}$, but we shall continue to use the $\Delta_8$ parameter
as our timing noise magnitude estimate for the current analysis. A
thorough study of timing noise, comparing and contrasting the
various measures used, will be given in Ref.~\cite{Hobbs:2006} (also
see Refs.~\cite{Hobbs:2004, Hobbs:2005}).

There is a definite correlation between the $\Delta_8$ parameters,
spin-down rate and age. Young pulsars, like the Crab pulsar,
generally show the most timing noise. The categorisation of the type
of timing noise (i.e., phase, frequency or frequency derivative) in
Ref.~\cite{CordesHelfand:1980} allowed them to ascribe different
processes for each. The majority of pulsars studied showed
frequency-type noise, possibly a result of random fluctuations in
the star's moment of inertia. The actual mechanism behind the process
is still unknown, with Cordes and Greenstein
\cite{CordesGreenstein:1981} positing and then ruling out several
mechanisms inconsistent with observations.

Timing noise intrinsically linked to motions of the electromagnetic
emission source or fluctuations in the magnetosphere, rather than
the rotation of the pulsar, is important in the search for
gravitational waves as it may allow the relative phase of the
electromagnetic and gravitational signals to drift. The implications
of timing noise in this context are discussed by Jones
\cite{Jones:2004}. He gives three categories of timing noise, not
necessarily related to the three types of timing noise given by
Cordes and Helfand \cite{CordesHelfand:1980}, having different
effects on any search. If all parts of the neutron star are strongly
coupled on short timescales, there should be no difference between
the electromagnetic phase and \gw phase. If the timing noise were
purely a magnetospheric fluctuation then phase wandering caused by
timing noise would not be seen in the \gw emission. The third
possibility, whereby the electromagnetic emission source wanders
with respect to the mass quadrupole, could result from a weak
exchange of angular momentum between the parts of the star
responsible for electromagnetic and \gw emission.  Jones describes
the ratio of the electromagnetic and gravitational timing noise
phase residuals ($\Delta\Phi$) by a parameter $\alpha =
\Delta\Phi_{\rm{gw}}/\Delta\Phi_{\rm{em}}$, with the three types of
timing noise described above corresponding to $\alpha = 1$, 0 and
$-I_{\rm em}/I_{\rm gw}$ respectively, where the $I$s represent the
moments of inertia of the electromagnetic and gravitational wave
producing components. In principle this factor could be included as
another search parameter. However, given the cost of including an
extra parameter in this search, and given that it is plausible that
that all parts of a neutron star are tightly coupled on the
timescales of interest here, we will assume rigid coupling between
the two components, i.e set $\alpha=1$, corresponding to the
gravitational and electromagnetic signals remaining perfectly in
phase.

\begin{figure}[!htbp]
\includegraphics[width=0.45\textwidth]{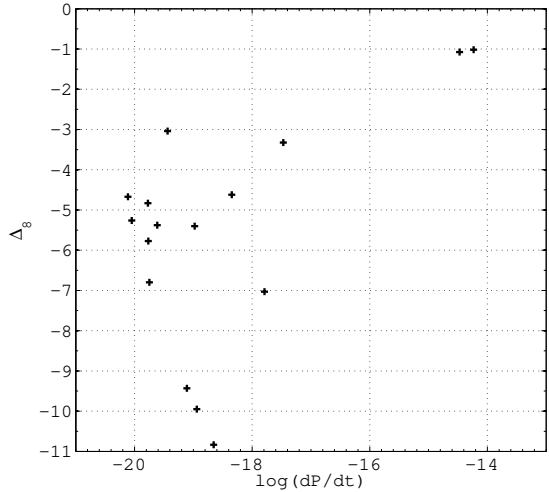}
\caption{The values of $\Delta_8$ for our selection of pulsars with measured
$\ddot{\nu}$.}\label{delta8Figure}
\end{figure}

The Crab pulsar is regularly monitored \cite{CrabEphemeris} on
timescales that are sufficiently short to allow its timing noise to
be effectively removed using a second heterodyne procedure
\cite{PitkinWoan:2004}. Like the Crab pulsar,
PSR\,J0537$-$6910 is young, has a high glitch rate and also shows
high levels of timing noise \cite{Marshall:2004}. Unfortunately,
unlike the Crab pulsar, we have no regular ephemeris for it that
covers our data set, and timing irregularities are likely to be too
great for historical data to be of use. We therefore have  excluded
PSR\,J0537$-$6910 from the analysis. For less noisy pulsars we still
need a method of estimating the effect of timing noise on phase
evolution that does not rely on continuous observation. One such
estimate is the $\Delta_8$ parameter given by Eq.~(\ref{delta8}),
which can provide a measure of the cumulative phase error. For those
pulsars with a measured $\ddot{\nu}$ we use this estimate to obtain 
a corresponding value of $\Delta_8$ as shown in Fig.~\ref{delta8Figure}.
This should provide a reasonable estimate of the timing noise over the 
timespan of the pulsar observation. Again we apply our criterion that 
cumulative phase errors of $> 30^{\circ}$ are unacceptable. In 
Fig.~\ref{delta8Figure} there are four pulsars (those with the four 
largest $\Delta_8$ values), with 
measured $\ddot\nu$, for which this is the case, and therefore timing 
noise could be a problem (having already noted the Crab
pulsar and PSR\,J0537$-$6910 as exceptions): PSRs\,J1748$-$2446A,
J1823$-$3021A, J1913+1011, and J1952+3252.  For pulsars with no
measured $\ddot{\nu}$ we use the approximate linear relation between
the period derivative $\dot{P}$ and $\Delta_8$ given in
Eq.~(\ref{delta8slope}). The low $\dot{P}$ values for these
pulsars implies that timing noise will be negligible.

In addition to the above, there are some pulsars in globular clusters 
for which there is no $\ddot{\nu}$ and for which $\dot{P}$ is negative
($\dot{\nu}$ is positive), so no value of $\Delta_8$ can be assigned
either through Eqs.~(\ref{delta8}) or (\ref{delta8slope}). For these
pulsars the value of $\dot{\nu}$ (and therefore $\ddot{\nu}$) must
be rather small to have been affected by motions within the cluster
(discussed more in \S\ref{results}), so timing noise should again be
negligible.

For pulsars which were re-timed over the period of S3 timing noise
will be negligible (for the S3 analysis at least) as any timing
noise, which usually has variations on time-scales of several months
to years, will have been absorbed in the parameter estimation. PSRs 
J1748$-$2446A and J1823$-$3021A were re-timed over S3 meaning that 
their S3 results will stand, although the other two will not. 
However, being conservative we will remove all four pulsars with 
large values of $\Delta_8$, and PSR\,J0537$-$6910, in which timing 
noise could be problematic, from the S4 and joint analysis. Note that 
PSR\,J0537$-$6910 is vetoed by both the parameter error criterion and 
our timing noise criterion.

This reduces our final number of well parameterised pulsar targets to 
78 for the S3 analysis and 76 for the S4 and joint analyses. 
The 76 pulsars include 21 of the 28 from the previous study of 
Abbott {\it et al.} \cite{ab2}, and so through our selection criterion we lose the
following 7 previously analysed pulsars: PSRs
J0030$+$0451, J1721$-$2457, J1730$-$2304, J1823$-$3021A, J1910$-$5959B,
J1913$+$1011, and J1952$+$3252. The same selection rules were not applied 
over S2, especially of note was that no timing noise criterion was considered, 
which accounts for three of the pulsars we lose between the two analyses. Also 
our $30^{\circ}$ rule was strictly applied, which the other four pulsars 
just exceeded.

The analysis was actually performed on all 93 timed pulsars mentioned above, 
however the various parameter uncertainties preclude us setting upper limits 
on a total of 15 of these.

\section{Hardware injections}\label{injections}
For analysis validation purposes simulated gravitational wave
signals for a variety of sources (bursts, pulsars, inspirals and
stochastic) have been mechanically injected into the LIGO
interferometers during science runs. During S2 two pulsar signals
were injected \cite{ab2}. This was increased to 10 injections in the
LIGO instruments for S3 and 12 for S4 covering a wider range of
signal parameters. Extracting and understanding these injections has
been invaluable in validating the analysis.

The hardware injection signals are produced using software (under
LALApps \cite{LALapps}), which was largely developed independently
of the extraction code. However, the codes do share the same solar
system barycentring and detector antenna response function
routines, both of which have been extensively checked against other
sources (e.g. checks against TEMPO in Refs.~\cite{dw, Pitkin:2006}).

The signals were added into each of the three LIGO detectors via the
position control signal going to the end test mass in one arm. Control signals in the digital
servos that maintain optical cavities on resonance were summed with
fake pulsar waveforms, modulating mirror positions to mimic the
effect of a real spinning compact object (i.e. differential length
motions with frequency and amplitude modulations appropriate for a
given sky position, frequency and spin-down). Furthermore, as the
digital fake waveforms have to be converted to analog coil currents
of suspended optics, the injected waveforms have to be divided by
the transfer function of the output chain (predominantly the
pendulum), in order to produce the desired differential length
response of the cavity.

The extraction of these injections is described in detail in
Appendix~\ref{app:injections}. They show the relative phase
consistency between the detectors over the course of a run. This
means that a joint analysis combining the data from all detectors is
valid. The injection plots (see Figs.~\ref{S3PulsarInj} and
\ref{S4PulsarInj}) show what we would expect our posterior plots
to look like given a detection i.e. strongly peaked pdfs with
very small probability at $h_0 = 0$, as compared to those in
Fig.~\ref{fig:examplePDF} where $h_0$ peaks at zero.

\section{Results}\label{results}

\subsection{Upper limits}\label{sec:upperlimits}
Here we present 95\% degree-of-belief upper limits on the amplitude
of gravitational waves ($h_0$) from the 78 pulsars identified above.
The value of $h_0$ is independent of any assumptions about the
neutron star other than it is emitting \gws at twice its rotation
frequency. The results will also be presented in terms of the pulsars'
equatorial ellipticity $\varepsilon$, which under the assumption of
triaxiality is related to $h_0$ via Eq.~(\ref{Pulsarh0}) by
\begin{equation}\label{h0epsilon}
\varepsilon = 0.237\left(\frac{h_0}{10^{-24}}\right)\left(\frac{r}{1\,{\rm
kpc}}\right)\left(\frac{1\,{\rm Hz}}{\nu}\right)^2\left(\frac{10^{38}\,{\rm kg}\,{\rm
m}^2}{I_{zz}}\right).
\end{equation}
To obtain an upper limit on $\varepsilon$ from that for $h_0$ we assume
a fiducial moment of inertia value of $I_{zz} = 10^{38}$\,kg\,${\rm
m}^2$. We discuss below in Sec.~\ref{izz-dependence} the effect of relaxing this
assumption. Pulsar distances are taken from the ATNF catalogue \cite{ATNF} and
are generally derived from the radio dispersion measures, with errors
estimated to be of order 20\%, although in some cases even this can be an 
underestimate. A critical review of pulsar distance measurements can be 
found in Ref.~\cite{Frail:1990}.

\begin{figure*}[!htbp]
\includegraphics[width=0.95\textwidth]{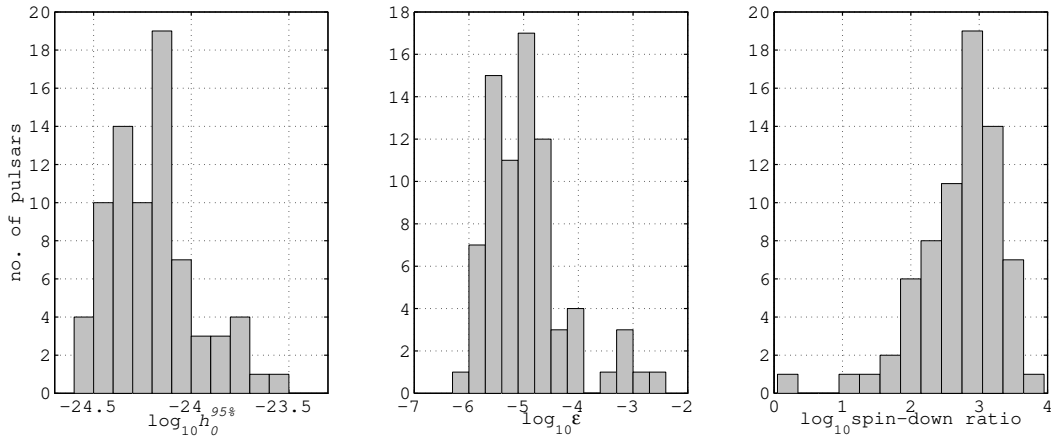}
\caption{Histograms of the log of amplitude, ellipticity, and ratio
of spin-down to \gw upper limits for the combined LIGO S3 and S4
run.}\label{ResultsHistogram}
\end{figure*}

\begin{figure}[!htbp]
\includegraphics[width=0.5\textwidth]{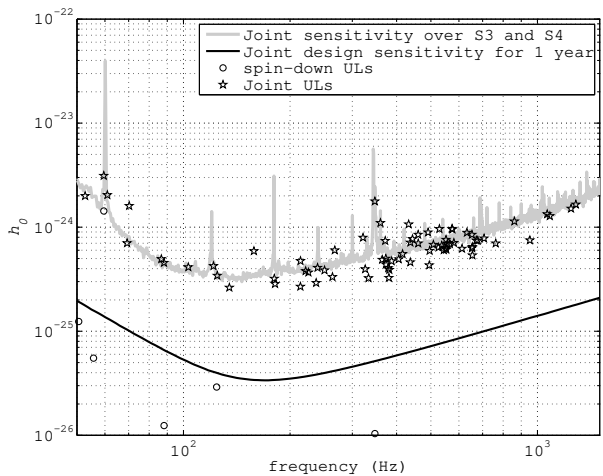}
\caption{The combined S3 and S4 upper limit results on the amplitude
of \gws for 76 pulsars using LIGO data compared to the joint
sensitivity curve.}\label{ResultsSensitivity}
\end{figure}

All upper limit results from the individual S3 and S4 runs along with
results from the combined run, with and without GEO\,600 included,
are given in Appendix~\ref{resultstables} in
Tables~\ref{table:allresults} and \ref{table:allresultsGEO}. The
GEO\,600 data only provides comparable sensitivities to LIGO at
frequencies greater than 1000\,Hz, and are therefore only used in
the search for PSR\,J1939+2134 (at the time the fastest known millisecond
pulsar) in S3, and additionally  PSR\,J1843$-$1113 in S4 and the
combined run. Inclusion of GEO 600 does not significantly
change the joint upper limits for these pulsars. For the majority 
of pulsars the lowest upper limits come from the combined S3/S4 data 
set, although for 14 pulsars
(PSRs\,J0024$-$7204I, J0024$-$7204S, J0024$-$7204U, J0621+1002, J1045$-$4509, J1757$-$5322
J1802$-$2124, J1804$-$2717, J1857+0943, J1910$-$5959D, J1910$-$5959E, J1911+0101B, J2129$-$5721 and
J2317+1439) the S4 results alone provide a lower limit. The combined S3 and S4 run results are
presented in histogram form in Fig.~\ref{ResultsHistogram}.
Fig.~\ref{ResultsSensitivity} shows the results compared to a joint LIGO S4 upper limit
estimate curve, taken as the best sensitivity during S4.
The joint upper limit sensitivity curve for the three detectors can
be estimated by combining the detector one-sided power spectral
densities (PSDs) via
\begin{eqnarray}
S(f) & = & \left(\frac{T_{\rm obs~H1}}{S_h(f)_{\rm H1}} + \frac{T_{\rm obs~H2}}{S_h(f)_{\rm
H2}} + \frac{T_{\rm obs~L1}}{S_h(f)_{\rm L1}} \right)^{-1}, \\ \nonumber
h_0^{95\%} & = & 10.8\sqrt{S(f)},
\end{eqnarray}
where $S_h(f)$ is the PSD and $T_{\rm obs}$ is each detector's live
time (using the associated duty factor of each interferometer during
the run). The factor of 10.8 is given in Ref.~\cite{dw} and was calculated 
through simulations with Gaussian noise\footnote{In Ref.~\cite{ab2} a similar plot to
Fig.~\ref{ResultsSensitivity} is shown for the S2 data using a factor 
of 11.4 in the relation between the upper limit and PSD. This 
definition comes from using the $\mathcal{F}$-statistic search method 
and setting a 1\% false alarm rate and 10\% false dismissal rate for 
signals given the underlying detector PSD \cite{ab1}}.

The results are also compared to the upper limit deduced from the
observed spin-down via Eq.~(\ref{spindownUL}) making the assumption
that all rotational energy is lost through \gw emission. The
spin-down limit is seen as a natural crossing point after which
gravitational-wave data, including upper limits, have a likely
bearing on the nature of the neutron star. The spin-down upper limit
will obviously depend on $\dot{\nu}$. This value however, can be
masked by radial and transverse motions of the object (see
Ref.~\cite{PulsarAstronomy} for discussion of these effects). The
Shklovskii effect \cite{Shklovskii:1970}, in which the pulsar has a
large transverse velocity $v$, will cause an apparent rate of change
in the pulsar's period of
\begin{equation}\label{ShklovskiiEffect}
\dot{P}_{\rm S} = \frac{v^2}{rc}P.
\end{equation}
Its $1/r$ dependence makes this effect more prominent for nearby
pulsars. In the ATNF catalogue \cite{ATNF} values of the intrinsic
period derivative $\dot{P}_{\rm int} = \dot{P}-\dot{P}_{\rm S}$ can
be obtained where this effect has been corrected for. This provides
a measure of intrinsic (rather than apparent)
spin-down\footnote{Note that the heterodyne procedure still needs to
make use of the measured spin-down rather than the intrinsic
spin-down as these Doppler effects will have the same effect on the
gravitational waves.} and, when available, is used in the spin-down
ratio results.

The observed value of $\dot{P}_{\rm obs}$ will also differ from its intrinsic
value, $\dot{P}_{\rm int}$, if the pulsar is accelerating -- a
likely scenario in the gravitational field of a globular cluster
\cite{PulsarAstronomy}. Any line-of-sight component to the acceleration, 
$a_{||}$, will give an observed value of 
\begin{equation}\label{eq:GCspindown}
\dot{P}_{\rm obs} = \dot{P}_{\rm int} + \frac{a_{||}}{c}P
\end{equation}
where $P$ is the spin period \cite{Phinney:1993}. These effects can 
cause pulsars to have apparent spin-ups (seen in quite a large number 
of globular cluster pulsars) although are only strong enough to greatly 
affect pulsars with intrinsically small period derivatives. There are 
still many globular clusters for which the radial accelerations have 
not been measured, therefore no firm spin-down upper limit can be set, 
making the direct \gw results a unique limit.

Highlights of the combined S3/S4 results include the tightest strain
upper limit set on a pulsar of $h_0^{95\%} = 2.6\ee{-25}$ for
PSR\,J1603$-$7207, the smallest ellipticity at $\varepsilon =
7.1\ee{-7}$ for PSR\,J2124$-$3358, and the closest upper limit to
the spin-down limit at a ratio of 2.2 for the Crab pulsar
(PSR\,J0534+2200).

\subsection{Dependence on the moment of inertia}
\label{izz-dependence}
The pulsar ellipticity results detailed above assume a moment of
inertia of $10^{38}\,{\rm kg}\,{\rm m}^2$, which is the standard
fiducial number used in the literature. However, modern
theoretically computed equations of state (EOS) generally predict
somewhat larger moments of inertia for stars more massive than
1\,$M_\odot$, a group which includes all neutron stars with measured
masses (see Ref.~\cite{Thorsett:1999}). Therefore the dependence on
moment of inertia should be considered.

Bejger, Bulik, and Haensel \cite{Bejger:2005} give an overview of
the theoretical expectations for the moment of inertia. Their Fig.~2
plots the moment of inertia vs.\ mass for several theoretically
predicted types of EOS. The maximum moment of inertia they find
(after varying the mass of the star) is 2.3 times the fiducial
value, with stars of 1.4\,$M_\odot$ having moments of inertia
1.2--2.0 except for one outlying type of EOS. Typically the maximum
moment of inertia occurs for a neutron star mass of 1.7\,$M_\odot$
or more. Recently masses greater than 1.6--1.7\,$M_\odot$ with 95\%
confidence have been measured \cite{Nice:2005, Ransom:2005} for some
systems, making this reasonable to consider. More recently Lackey
\cite{Lackey:thesis} found the highest moment of inertia to be
$3.3\ee{38}$\,kg\,m$^2$ for EOS G4 of Lackey, Nayyar, and Owen
\cite{Lackey:2005tk}. This is a relativistic mean-field EOS similar
to the GNH family considered by Bejger, Bulik, and Haensel
\cite{Bejger:2005} but contains no exotic phases of matter such as
hyperons or quarks. Consequently, we consider the range of
theoretically predicted moments of inertia to be approximately
1--3$\times10^{38}$\,kg\,m$^2$.

There have been recent attempts to infer neutron star moments of
inertia from observations. Bejger and Haensel \cite{Bejger:2002,
Bejger:2003} derived a value for the Crab pulsar's moment of inertia
by equating the spin-down power to the observed electromagnetic
luminosity and inferred acceleration of the nebula. However, this
(extremely high) value is dominated by the assumptions about the
highly uncertain mass and mass distribution of the nebula as well as
the relativistic wind from the pulsar, and thus cannot yet be
considered to give a reliable value. The double pulsar system
J0737-3039 shows great promise for tighter measurements of the
moment of inertia (and constraints on the EOS) in the near future
\cite{Morrison:2004df, Lattimer:2004nj, Bejger:2005, Kramer:2006}.
However, for the moment we are left with the theoretical range
quoted above.

As suggested in Ref.~\cite{Pitkin:2005}, instead of using
Eq.~(\ref{h0epsilon}) to set a limit on $\varepsilon$ assuming a
value of $I_{zz}$, one can use it to set a limit on the neutron star
quadrupole moment $\approx I_{zz}\varepsilon$ without relying on any
assumption about $I_{zz}$. The limit on the quadrupole moment can
then be used to help define an exclusion region in the
$I\textrm{-}\varepsilon$ plane. This exclusion region allows one to
read off an upper limit on $\varepsilon$ as a function of the
EOS-dependent moment of inertia. The spin-down can also be used to
provide exclusion regions via the relation
\begin{equation}\label{eq:spindownIe}
I_{zz} = \frac{5}{512\pi^4}\frac{|\dot{\nu}|c^5}{G\nu^5}\frac{1}{\varepsilon^2}.
\end{equation}
Theoretical contributions to the exclusion regions come from predictions of
the maximum moment of inertia and ellipticity.
In terms of the exclusion region, our observational upper limits on $h_0$
are far from contributing except for the Crab pulsar, to which we now turn.

\subsection{The Crab pulsar - PSR\,J0534+2200}\label{CrabPulsarResults}
Of the known radio pulsars, the Crab pulsar has often been
considered one of the most promising sources of gravitational waves.
This is due to its youth and large spin-down rate, leading to a
relatively large spin-down upper limit several orders of magnitude
higher than for most other pulsars. The high rate of glitching in
the pulsar also provides possible evidence of asymmetry. One glitch
model favoured for the Crab pulsar involves a change in the pulsar
ellipticity, and breaking of the crust, as the star settles to its
new equilibrium state as it spins down \cite{PulsarAstronomy}. In
the 1970s, estimates of \gw strains were spurred on by the
experimenters producing novel technologies which allowed the
possibility of probing these low strains, with Zimmermann
\cite{Zimmermann:1978} producing estimates of \gw strains from the
Crab pulsar ranging from $h_0 \approx 2\ee{-25}-10^{-29}$.

The first searches for \gws from the Crab pulsar were carried out
using specially designed resonant bar detectors, with frequencies of
around 60\,Hz \cite{Hirakawa:1978}. The most recent result using
such a bar was from 1993 and gave a $1\sigma$ upper limit of $h_0
\le 2\ee{-22}$ \cite{Suzuki:1995}. This upper limit was passed in
the LIGO S2 run, which gave $h_0^{95\%} = 4.1\ee{-23}$ \cite{ab2}.
Using Eq.~(\ref{spindownUL}), and taking
$I_{zz}=10^{38}$\,kg\,${\rm m}^2$ and $r = 2$\,kpc, gives a
spin-down upper limit for the Crab pulsar of $h_0 < 1.4\ee{-24}$,
about a factor of 30 below the S2 observational upper limit.
However, the S2 limit on the Crab was at the time the closest approach to
the spin-down limit obtained for any pulsar.

\begin{figure}[!htbp]
\includegraphics[width=0.45\textwidth]{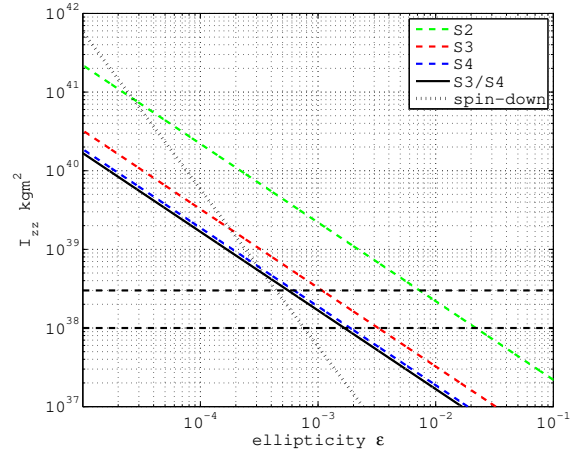}\caption{The moment of
inertia--ellipticity plane for the Crab pulsar over the S2, S3 and S4 runs.
The areas to the right of the diagonal lines are the experimentally excluded
regions. The horizontal lines represent theoretical upper and lower limits
on the moment of inertia as mentioned in \S\ref{izz-dependence}.
Theoretical upper limits on the ellipticity are much more uncertain, the
highest being a few times $10^{-4}$.
\label{CrabIeplane}}
\end{figure}

Our new results for the Crab pulsar (and the other 77 targets) are
shown in Table~\ref{table:allresults}. The
results improve by up to an order of magnitude over those from the
S2 run, and the majority of this improvement was between
the S2 and S3 runs. The results for the Crab pulsar over the S2, S3
and S4 runs are plotted on the $I$--$\varepsilon$ plane in
Fig.~\ref{CrabIeplane}.
The diagonal lines in Fig.~\ref{CrabIeplane} mark the lower boundaries
of exclusion regions on this plane using our upper limits obtained
for the different runs. The dashed black diagonal line marks the lower
boundary of the upper limit from spin-down as given in
Eq.~(\ref{eq:spindownIe}). The dashed horizontal black lines give
lower and upper bounds on the moment of inertia of
1--3$\times10^{38}\,{\rm kg}\,{\rm m}^2$, as given by our arguments
in \S\ref{izz-dependence}. It can be seen that our experimental
results currently only beat the spin-down limit for moments of
inertia at values greater than almost double the maximum of our theoretical
range. However, over this range the ratio of the gravitational to
spin-down upper limit ranges from 2.2 at the lowest value to only
1.3 at the largest value.

The spin-down limit in fact over-estimates the strongest possible signal
because we know that much of the spin-down energy of the Crab goes into
powering the nebula through electromagnetic radiation and relativistic
particle winds. Thus it is interesting to ask how far we would need to beat the spin-down
limit to have a chance of detecting a signal allowing for what is known
about the non \gw spin-down. Palomba \cite{Palomba:2000} uses the observed braking index 2.51 of the
Crab pulsar with a simple model of spin-down through gravitational
radiation (braking index 5) combined with some other mechanism (braking
index a free parameter) to place an upper limit of about $\varepsilon \le
3\times10^{-4}$. This is about 2.5 times lower than the spin-down limit and 5.5 times
lower than our result (for $I_{zz} = 10^{38}\,{\rm kg}\,{\rm m}^2$).

The Crab pulsar experienced two glitches between S3 and S4, a large
glitch on $6^{\rm th}$ September 2004 and a smaller glitch on
$22^{\rm nd}$ November 2004 \cite{CrabEphemeris}. The effect of
glitches on the relative phase between the electromagnetic pulse and
any possible \gw signal is unknown, so there is uncertainty whether
the (phase-coherent) combined S3/S4 result is valid. The combined
result stands, but the reader should be aware that it includes the
assumption of trans-glitch phase coherence.

\section{Astrophysical interpretation}
We have produced new, tight, upper limits on \gw signal strength
from a large selection of known pulsars and for the Crab pulsar we
are very near the fiducial limit set by spin-down arguments.

It can be seen from Table~\ref{table:allresults} and
Fig.~\ref{ResultsHistogram} that, for the majority of pulsars, the
gravitational wave detector upper limits  are at least one hundred
times above those from the spin-down argument, so is there anything
that we can take from the results in terms of astrophysics?

First we should note that spin-down limits on gravitational wave
luminosity are plausible, but model dependent. They assume a model
for the structure of the neutron star (for instance, that it is not
accreting and is rigidly rotating, in addition to assumptions about
its equation of state) and they take dispersion measure distance as
a consistently good measure of true distance. There is some
considerable uncertainty associated with all of these assumptions.
In contrast our observations set direct limits on a source's
gravitational wave strain.

Secondly, for globular cluster pulsars the spin-down measured from radio
timing observations is a combination of the spin-down intrinsic to the
pulsar and acceleration along the line-of-sight $a_{||}$ in the cluster's 
gravitational potential (see Eq.~(\ref{eq:GCspindown})). In general,
the magnitude and sign of the acceleration is unknown but the
intrinsic $\dot{P}_{\rm int}>0$ of millisecond pulsars is usually small
and often smaller than the extrinsic contribution. Only if
$\dot{P}_{\rm obs} <0$ one can be sure that $a_{||}<0$. Therefore,
the limits derived from our gravitational wave observations provide 
the only direct limits on $\dot{P}_{\rm int}$ which are independent
from biasing kinematic effects. These can be combined with the observed
spin-down to provide a limit on the acceleration in the cluster, 
i.e.~$a_{||} \ge c ( \dot{P}_{\rm obs} - \dot{P}_{\rm gw}^{\rm limit})/P$.

Finally, it is interesting to note that our ellipticity limits are
well into the range permitted by some models of strange quark stars
or hybrid stars ($\varepsilon \sim$ a few times $10^{-4} - 10^{-5}$)
and are reaching into the range permitted by more conventional
neutron star EOSs ($\varepsilon \sim$ a few times $10^{-7}$)
\cite{Owen:2005}.

Currently the fifth LSC science run (S5) is underway, and this
promises to beat the Crab pulsar spin-down limit within a few months
of its start.
For many other pulsars we should be able to reach amplitude upper
limits of $< 1\ee{-25}$ and ellipticities of $\sim 1\ee{-7}$.

\begin{acknowledgements}
The authors gratefully acknowledge the support of the United States
National Science Foundation for the construction and operation of
the LIGO Laboratory and the Particle Physics and Astronomy Research
Council of the United Kingdom, the Max-Planck-Society and the State
of Niedersachsen/Germany for support of the construction and
operation of the GEO600 detector. The authors also gratefully
acknowledge the support of the research by these agencies and by the
Australian Research Council, the Natural Sciences and Engineering
Research Council of Canada, the Council of Scientific and Industrial
Research of India, the Department of Science and Technology of
India, the Spanish Ministerio de Educacion y Ciencia, The National
Aeronautics and Space Administration, the John Simon Guggenheim
Foundation, the Alexander von Humboldt Foundation, the Leverhulme
Trust, the David and Lucile Packard Foundation, the Research
Corporation, and the Alfred P. Sloan Foundation.
\end{acknowledgements}

\newpage
\appendix
\section{Injections}\label{app:injections}
\subsection{S3 injections}
An initial analysis of the S3 pulsar injections is given in
Ref.~\cite{Dupuis:2004}. The data have since been re-analysed with
more recent versions of the detector calibrations, the results of
which are presented here. For S3 initially 10 pulsars signals were
injected, with a further one added at the end of the run to be in
coincidence with a single injection into GEO\,600
\cite{Weiland:2004}. The majority of injection parameters were
decided upon randomly, although pulsar frequencies were chosen to
avoid major instrumental or calibration lines, and amplitudes were
dependent on the frequency. The injections were split into two
groups of five, where values of $h_0$ were calculated to give two
each with signal-to-noise ratios of approximately 3, 9, 27, 81 and
243. The parameter values are shown in Table~\ref{InjectionParams}.
\begingroup
\squeezetable
\begin{table*}[!htbp]
\caption{\label{InjectionParams} The parameter values for the pulsar hardware
injections in S3 and S4.}
\begin{tabular}{c | c c c c c c c c c}
\hline \hline
P\textsc{ulsar} & $\alpha$ (rads) & $\delta$ (rads) & $\nu_{\rm gw}$ (Hz) & $\dot{\nu}_{\rm gw}$
(Hz/s) & $h_0$ (S3) & $h_0$ (S4) &
$\phi_0$ (rads) & $\iota$ (rads) & $\psi$ (rads) \\
\hline
0 & 1.25 & -0.98 & 265.5 & $-4.15\ee{-12}$ & $9.38\ee{-25}$ & $4.93\ee{-25}$ & 2.66 & 0.65 & 0.77 \\
1 & 0.65 & -0.51 & 849.1 & $-3.00\ee{-10}$ & $8.49\ee{-24}$ & $4.24\ee{-24}$ & 1.28 & 1.09 & 0.36 \\
2 & 3.76 & 0.06 & 575.2 & $-1.37\ee{-13}$ & $1.56\ee{-23}$ & $8.04\ee{-24}$ & 4.03 & 2.76 & -0.22 \\
3 & 3.11 & -0.58 & 108.9 & $-1.46\ee{-17}$ & $6.16\ee{-23}$ & $3.26\ee{-23}$ & 5.53 & 1.65 & 0.44 \\
4 & 4.89 & -0.21 & 1430.2 & $-2.54\ee{-8}$ & $1.01\ee{-21}$ & $4.56\ee{-22}$ & 4.83 & 1.29 & -0.65
\\
5 & 5.28 & -1.46 & 52.8 & $-4.03\ee{-18}$ & $1.83\ee{-23}$ & $9.70\ee{-24}$ & 2.23 & 1.09 & -0.36 \\
6 & 6.26 & -1.14 & 148.7 & $-6.73\ee{-9}$ & $5.24\ee{-24}$ & $2.77\ee{-24}$ & 0.97 & 1.73 & 0.47 \\
7 & 3.90 & -0.36 & 1221.0 & $-1.12\ee{-9}$ & $2.81\ee{-23}$ & $1.32\ee{-23}$ & 5.24 & 0.71 & 0.51 \\
8 & 6.13 & -0.58 & 194.3 & $-8.65\ee{-9}$ & $6.02\ee{-23}$ & $3.18\ee{-23}$ & 5.89 & 1.50 & 0.17 \\
9 & 3.47 & 1.32 & 763.8 & $-1.45\ee{-17}$ & $1.61\ee{-22}$ & $8.13\ee{-24}$ & 1.01 & 2.23 & -0.01 \\
G\textsc{eo} & 0.78 & -0.62 & 1125.6 & $-2.87\ee{-11}$ & $7.5\ee{-22}$ & * & 1.99 & 0.84 & 0.37 \\
\hline \hline
\end{tabular}
\end{table*}
\endgroup
The 10 initial signals were injected into the LIGO detectors for
approximately the first half of the run, then turned off for two
weeks, to ensure data were present that was not artificially
contaminated, and then turned back on with the two loudest signals
removed. The simultaneous injection with \geo was switched on near
the end of the run.

\begin{figure*}[h]
\begin{tabular}{l l l}
\includegraphics[width=0.3\textwidth]{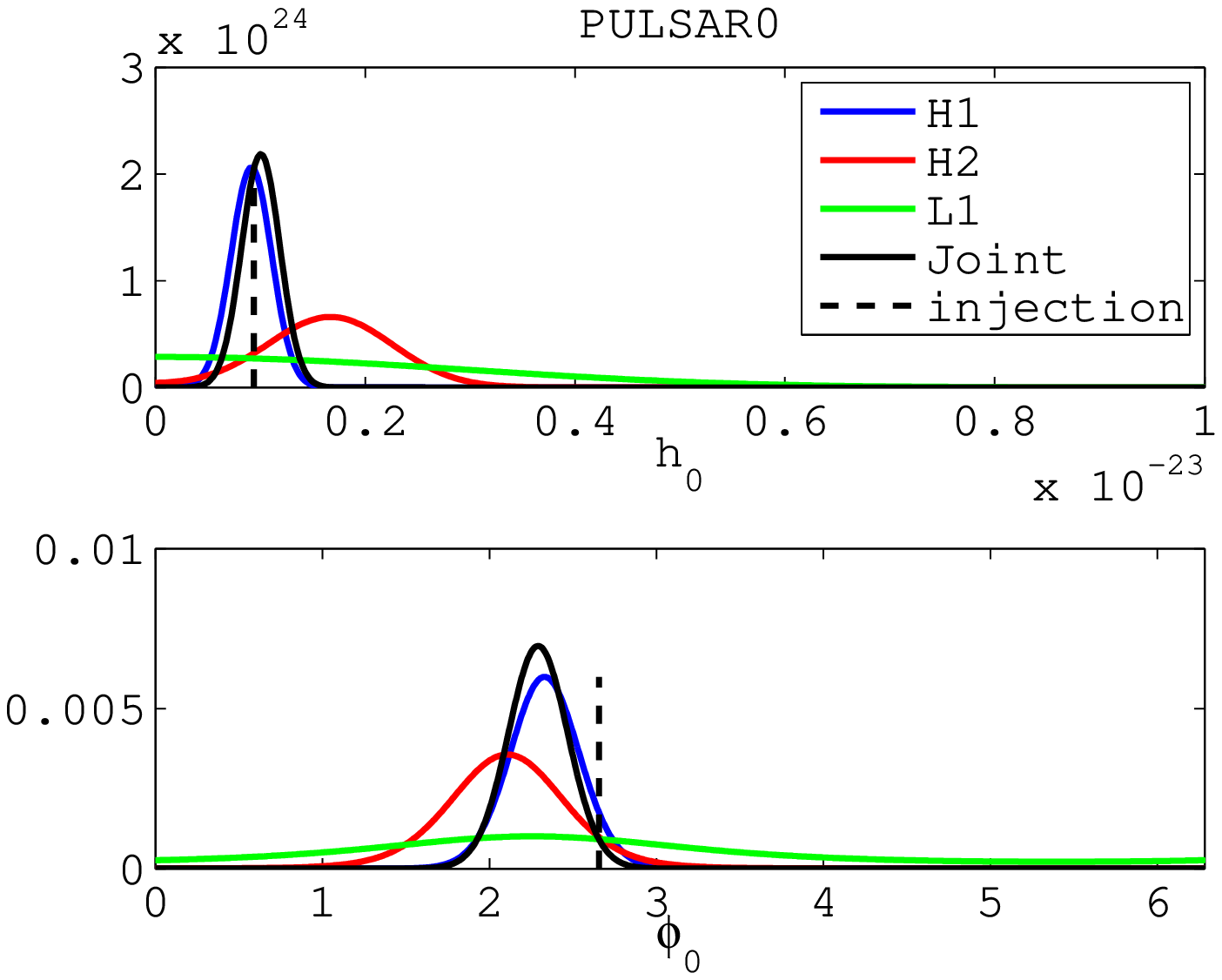} &
\includegraphics[width=0.3\textwidth]{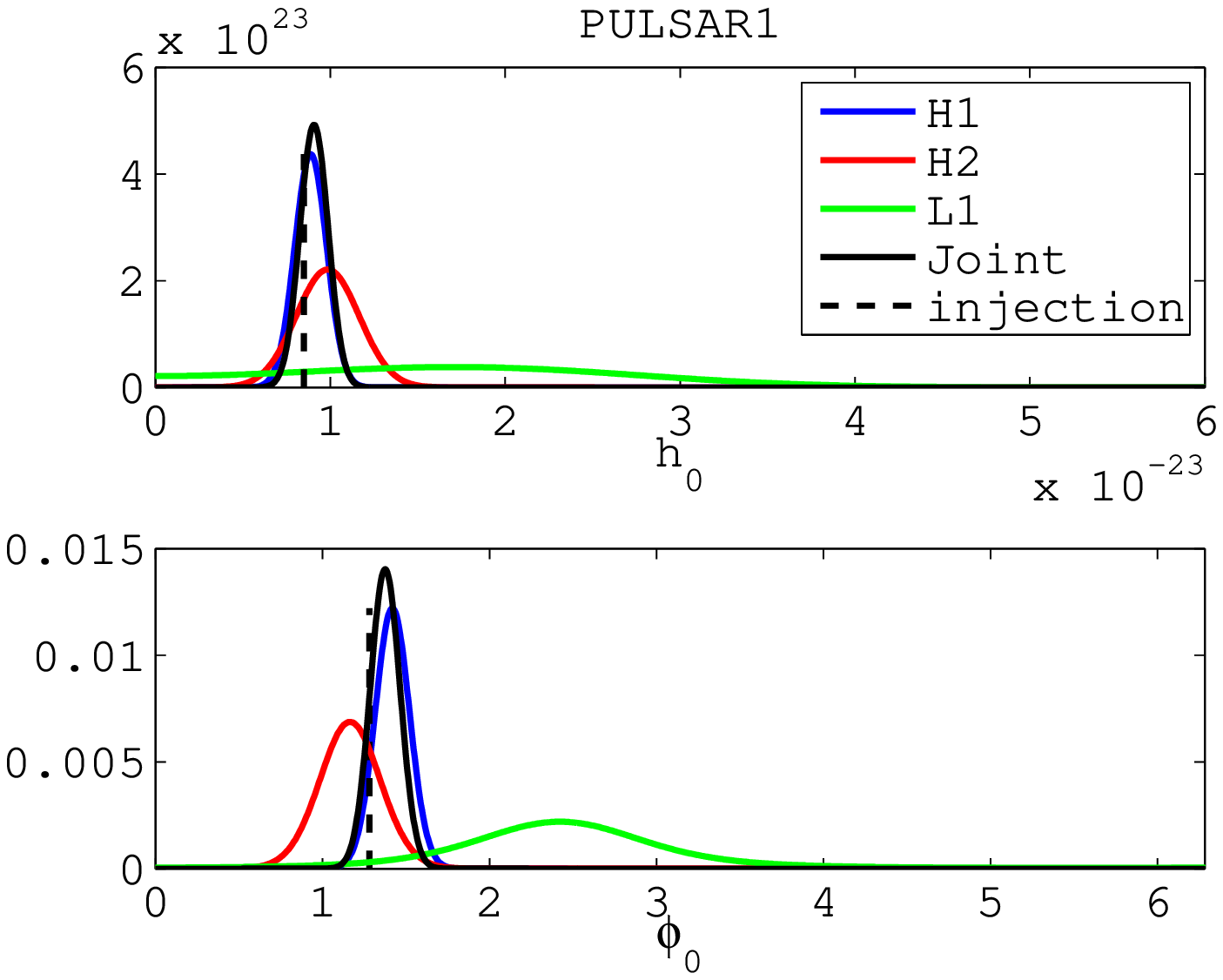} &
\includegraphics[width=0.3\textwidth]{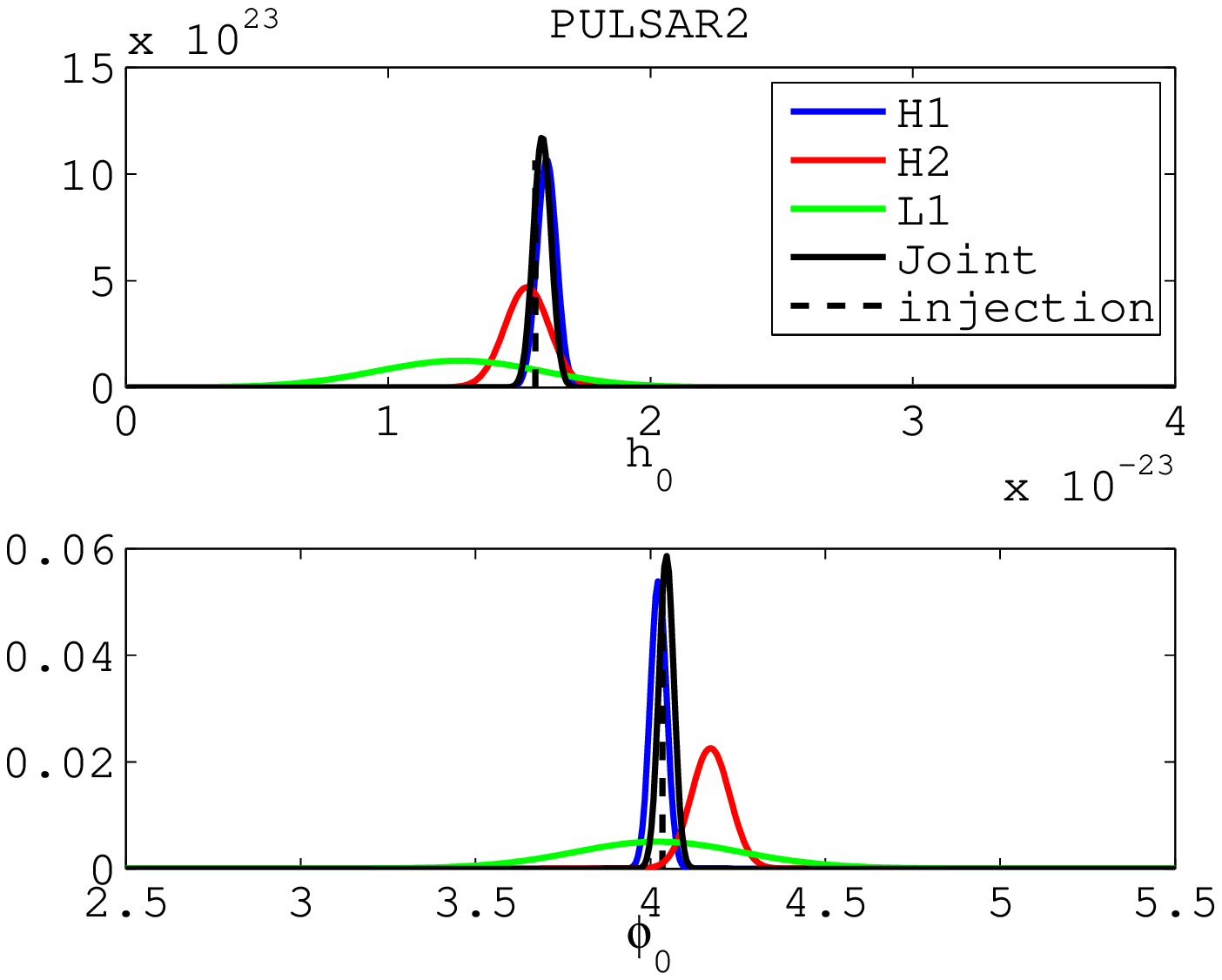} \\
\includegraphics[width=0.3\textwidth]{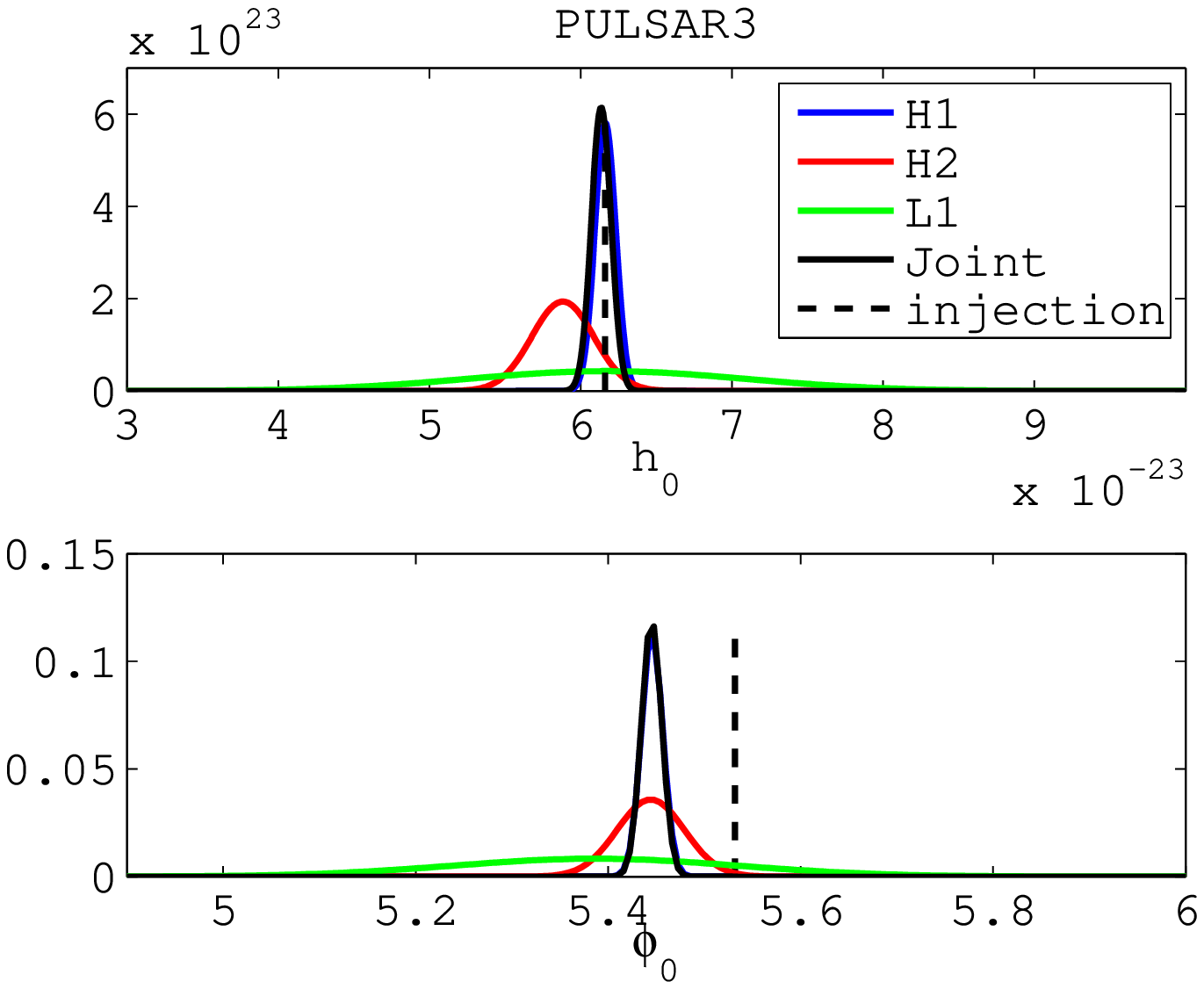} &
\includegraphics[width=0.3\textwidth]{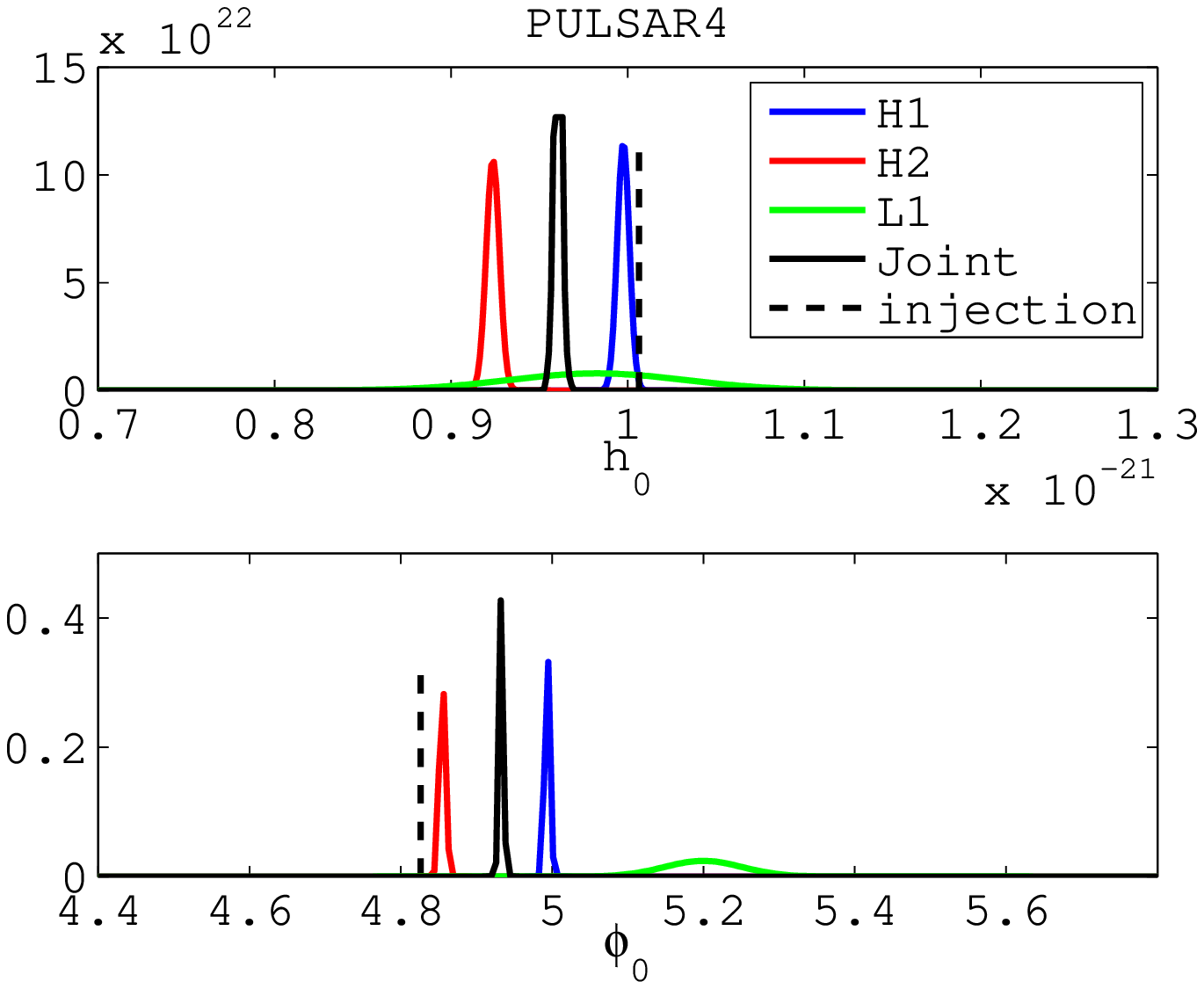} &
\includegraphics[width=0.3\textwidth]{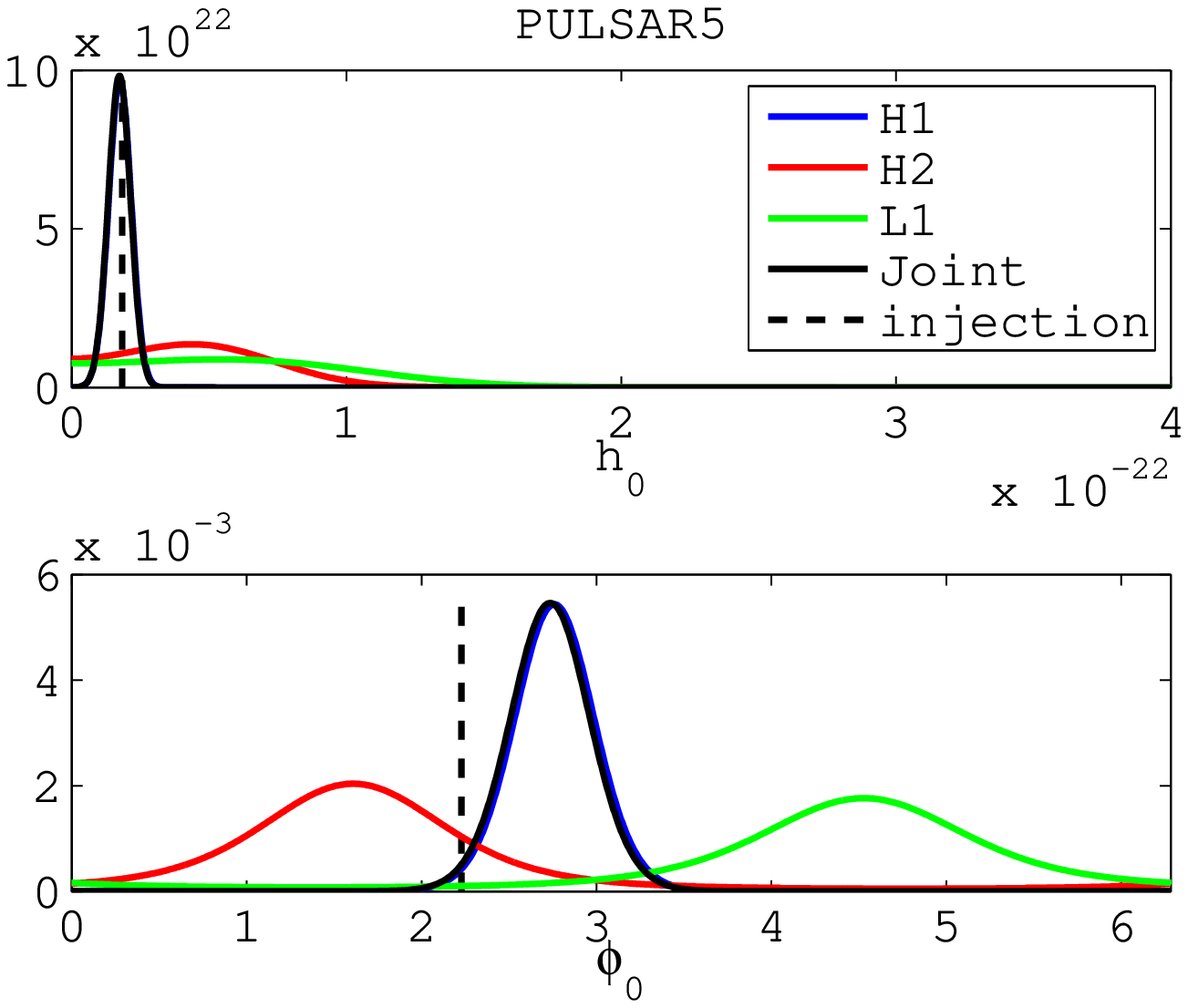} \\
\includegraphics[width=0.3\textwidth]{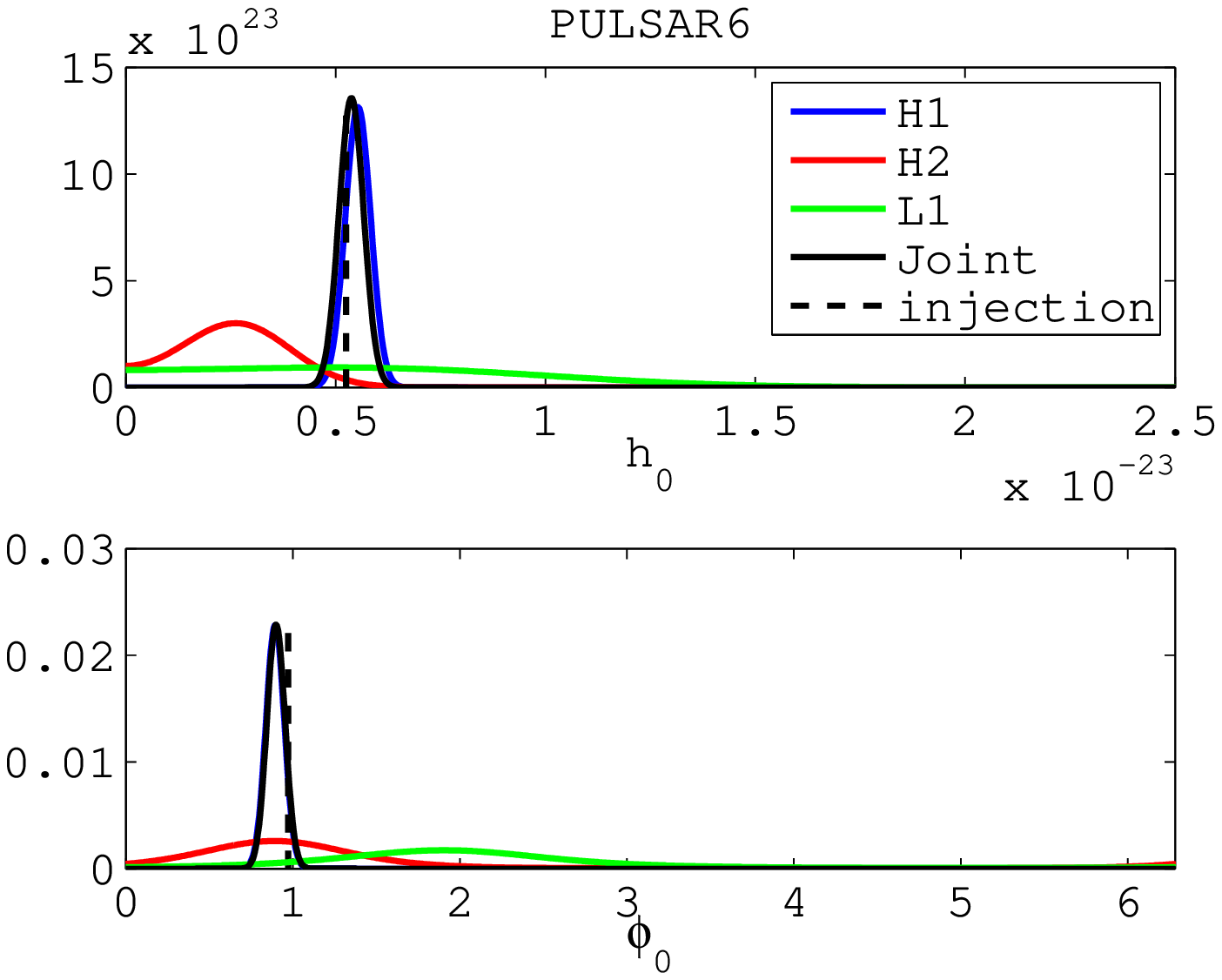} &
\includegraphics[width=0.3\textwidth]{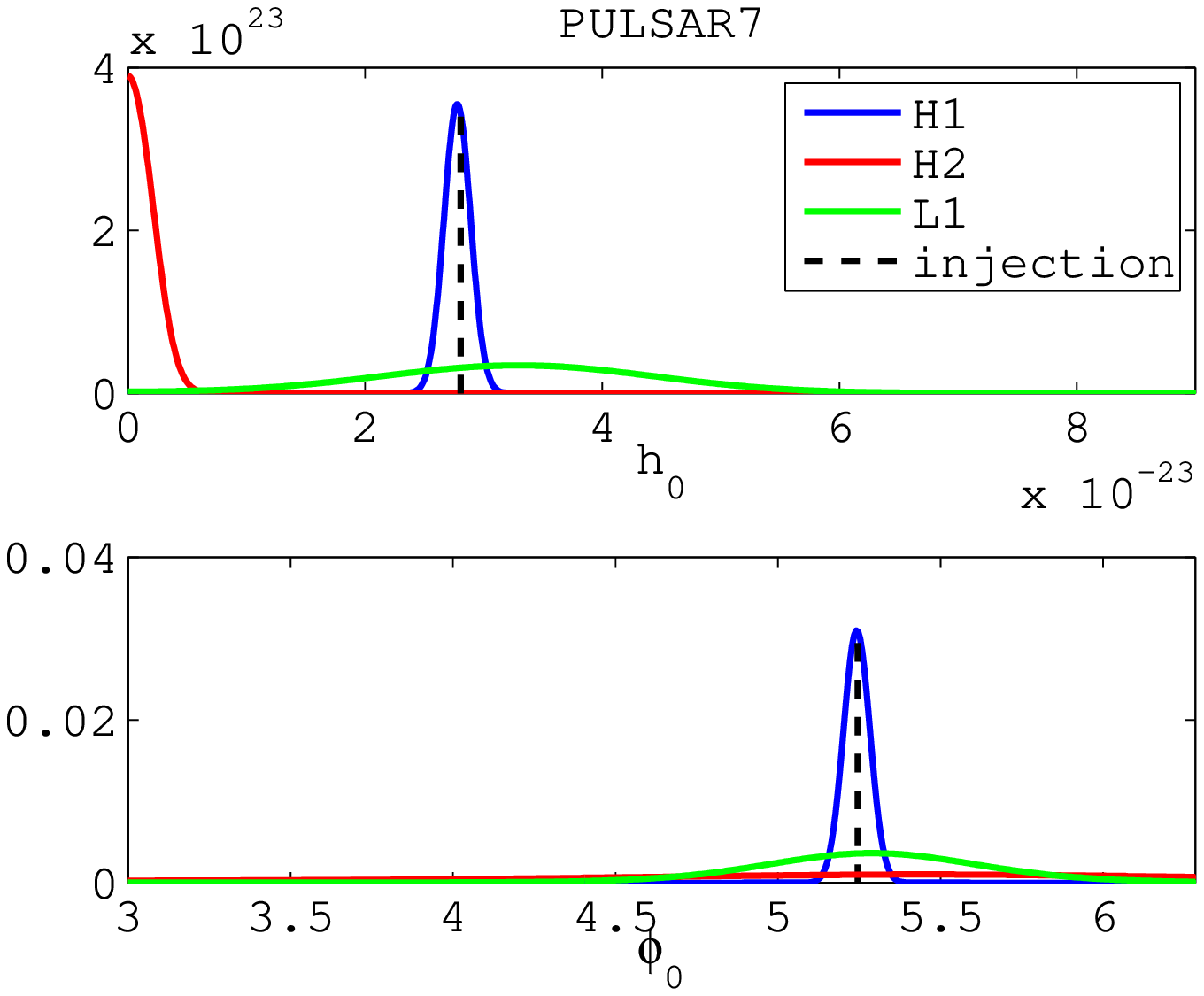} &
\includegraphics[width=0.3\textwidth]{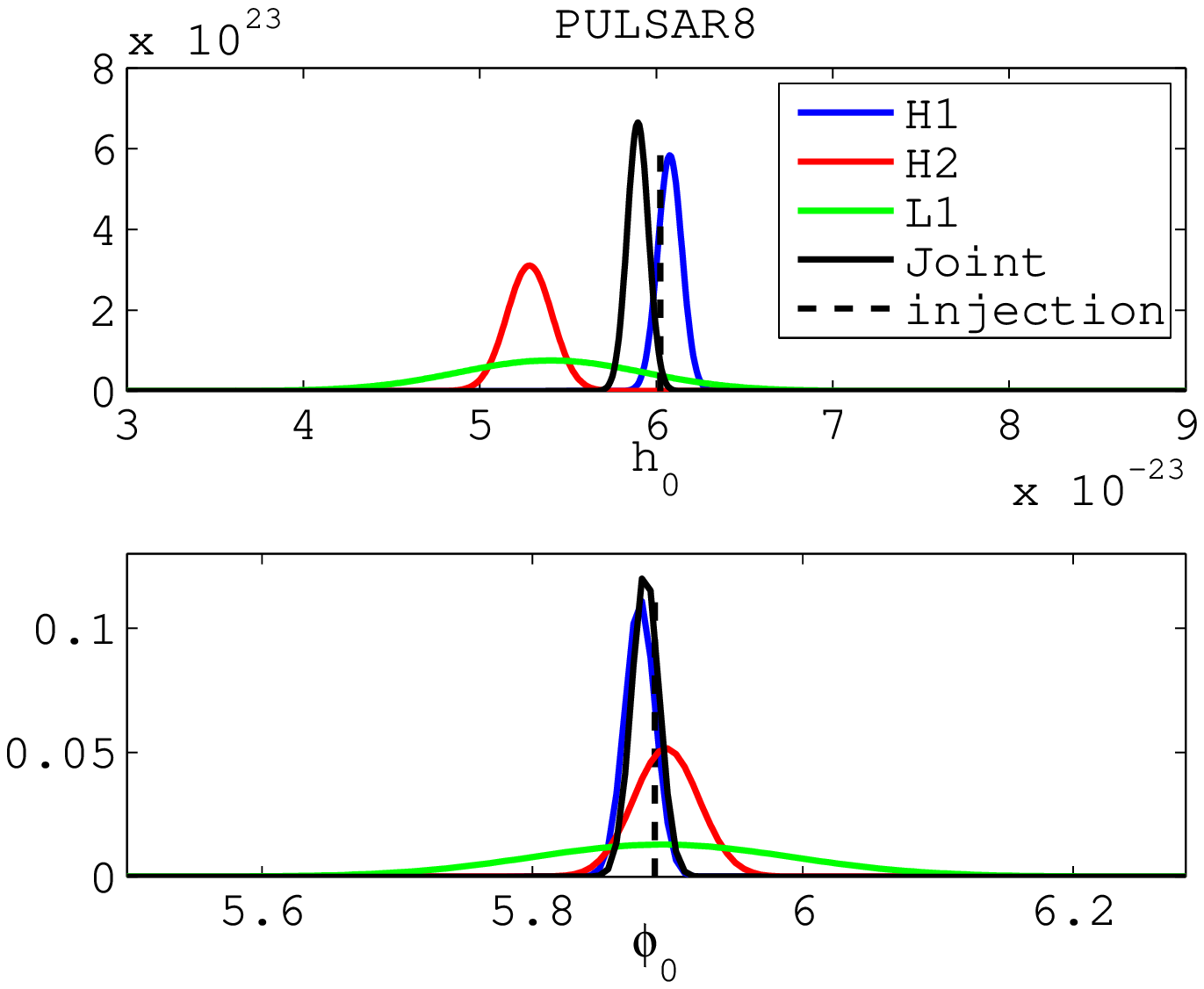} \\
 & \includegraphics[width=0.3\textwidth]{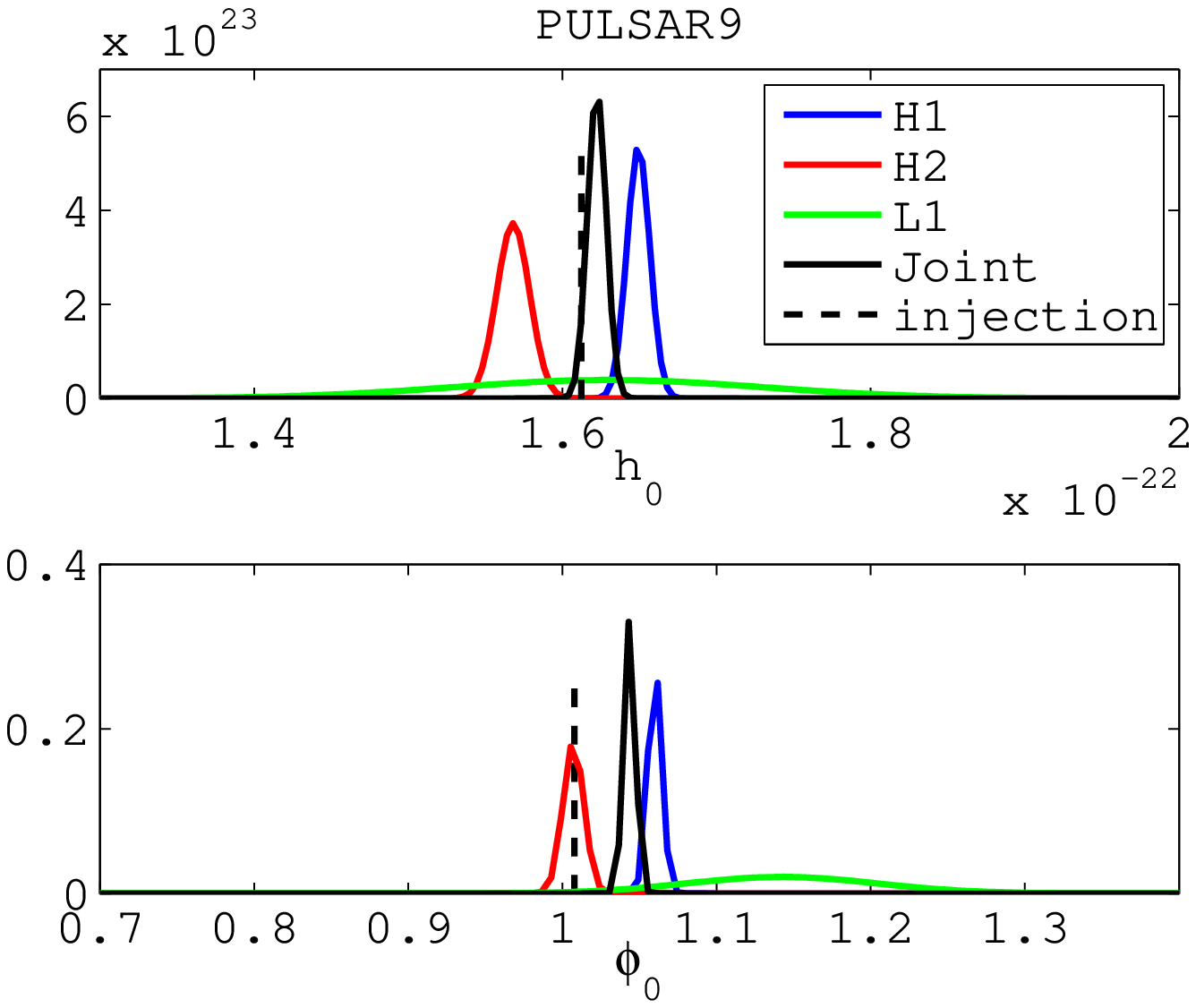} & \\
\end{tabular}
\caption{The pdfs of $h_0$ and $\phi_0$ for 10 isolated pulsar
injections into the LIGO detectors during S3. The anomaly seen in
P\textsc{ulsar}7 is discussed in the text.}\label{S3PulsarInj}
\end{figure*}

These signals were extracted from the data using the analysis
techniques described in \S\ref{sec:method} and Ref.~\cite{dw}. The
two most important parameters for checking that the calibration of
the instruments was correct were the amplitude and initial phase, so
in the Bayesian parameter estimation procedure the $\iota$ and
$\psi$ parameters were held fixed at their known values. This was
done because the correlations between $h_0$ and $\cos{\iota}$ and
$\phi_0$ and $\psi$, respectively could lead to the marginalised
posterior probability distributions (pdfs) for each parameter being
distorted or spread out (see Ref.~\cite{Dupuis:2004} for examples of
this). The extracted pdfs of $h_0$ and $\phi_0$ for each of the
injections, after corrections described below in
\S\ref{sec:calissues}, can be seen in Fig.~\ref{S3PulsarInj}.

For the vast majority of signals the extracted pdfs overlap with the
injected value. For the strongest injections with the largest
signal-to-noise ratios the pdfs are rather narrow, and any
uncertainties in the calibration become evident, with a maximum
offset of the order of 10--15\%. The far wider pdfs associated with
the L1 signal injections reflect the lower L1 sensitivity
and lower duty factor compared with the H1 and H2 detectors. It can
be seen that the injected phases for each detector agree with each
other to within a few degrees and are within the uncertainty of the
method. This provides some evidence that there is phase coherence
between the detectors and that a joint analysis, combining the data
from all the detectors, is possible.

Two main discrepancies have been identified as operational mistakes
made during the injection procedure: P\textsc{ulsar}7 was injected
into H2 with a much lower amplitude than intended, and remained
undetected, therefore no joint analysis was performed, and
P\textsc{ulsar}0 was injected into H1 with an amplitude 1.6 times
larger than intended.

The injection of the signal into GEO\,600 is described in
Ref.~\cite{Weiland:2004}, and its analysis is described in
Ref.~\cite{Dupuis:2004}. It was found that the injection performed
during S3 was badly contaminated and could not be used. However, a
subsequent injection performed shortly after S3 has verified that
the signal parameters were correctly injected and extracted,
validating the injection hardware and analysis software.

\subsection{S4 injections}\label{S4injections}
The 10 injections used in S3 were used again for S4 to create
artificial signals in the LIGO interferometers. However, their
amplitudes were adjusted to give approximately the same
signal-to-noise ratios as seen in S3, taking into account the better
sensitivity during the S4 run. For all except P\textsc{ulsar}9 the
$h_0$ values were reduced by a $\sim$ half, with P\textsc{ulsar}9
being so strong that its amplitude was reduced by a factor of $\sim$
20. These signals were injected for the second half of the run from
$8^{\rm th}$ March 2005 onwards. The updated $h_0$ values are shown
in Table~\ref{InjectionParams}. There were also an additional two
signals (P\textsc{ulsar}10 and 11), simulated to be from pulsars in
binary systems, injected for the last day of the run. The binary
pulsar injections allowed the testing of the binary timing code
described in \S\ref{binaries} as the injection code and extraction
code were written independently.  The binary injection signal
parameters for P\textsc{ulsar}10 and P\textsc{ulsar}11 were taken
from P\textsc{ulsar}3 and 8 respectively, with the frequencies
changed, and amplitudes increased to make sure they were visible
over the short injection time. The frequency, amplitude and binary
system parameters are shown in Table~\ref{S4BinInjectionParams}. The
binary system parameters were chosen to have one in a relatively
eccentric orbit and one in a circular orbit. We chose fairly short
periods, so that they would have completed or nearly completed at
least one full orbit during the injection. The $T_0$ values are
given in the pulsar rest frame.
\begingroup
\squeezetable
\begin{table*}[!htbp]
\caption{\label{S4BinInjectionParams} The parameter values for the S4 binary pulsar
hardware injections.}
\begin{tabular}{c | c c c c c c c}
\hline \hline
P\textsc{ulsar} & $\nu_{\rm gw}$ (Hz) & $h_0$ & $T_0$ (MJD) & $P_b$ (days) & $e$ & $\omega_0$ (degs) & $a\sin{i}$ (secs) \\
\hline
10 & 250.6 & $1.30\ee{-22}$ & 51749.71156482 & 1.35405939 & 0.0 & 0.0 & 1.65284 \\
11 & 188.0 & $5.21\ee{-22}$ & 52812.92041176 & 0.31963390 & 0.180567 & 322.571 & 2.7564 \\
\hline\hline
\end{tabular}
\end{table*}
\endgroup
For the recovery of the binary system injections the BT model was used,
although as no relativistic parameters were included any of the models
could have been used.

\begin{figure*}[h]
\begin{tabular}{l l l}
\includegraphics[width=0.3\textwidth]{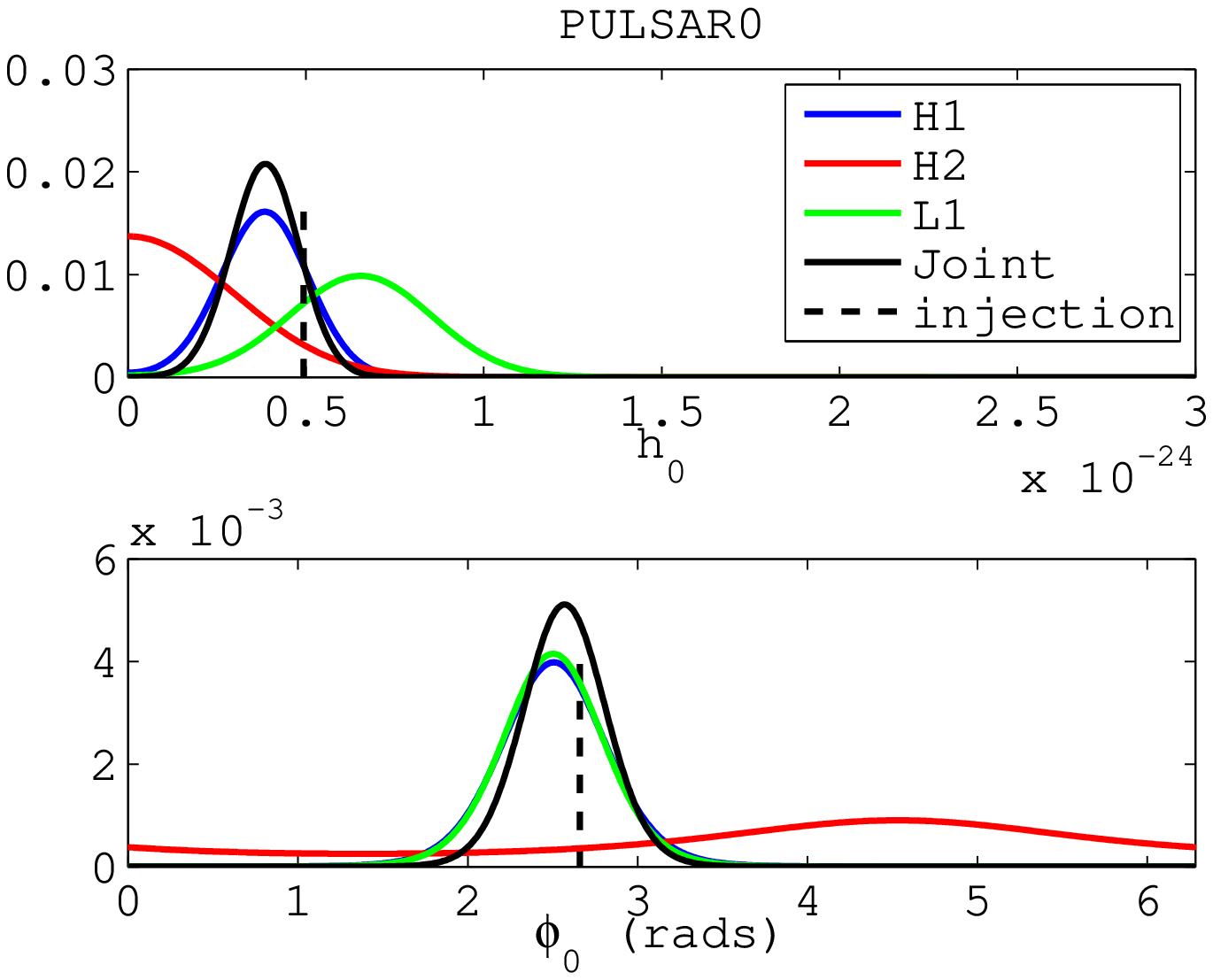} &
\includegraphics[width=0.3\textwidth]{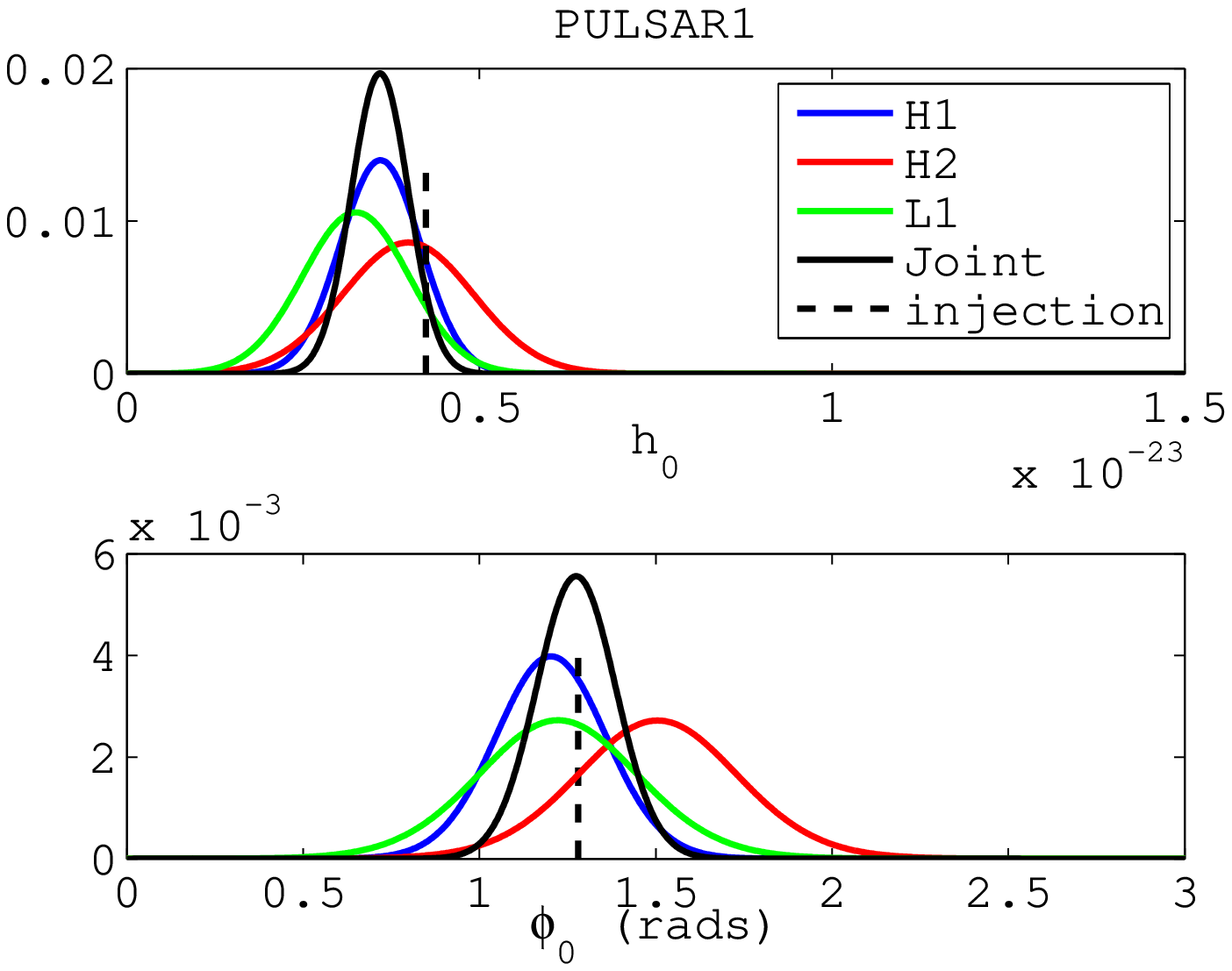} &
\includegraphics[width=0.3\textwidth]{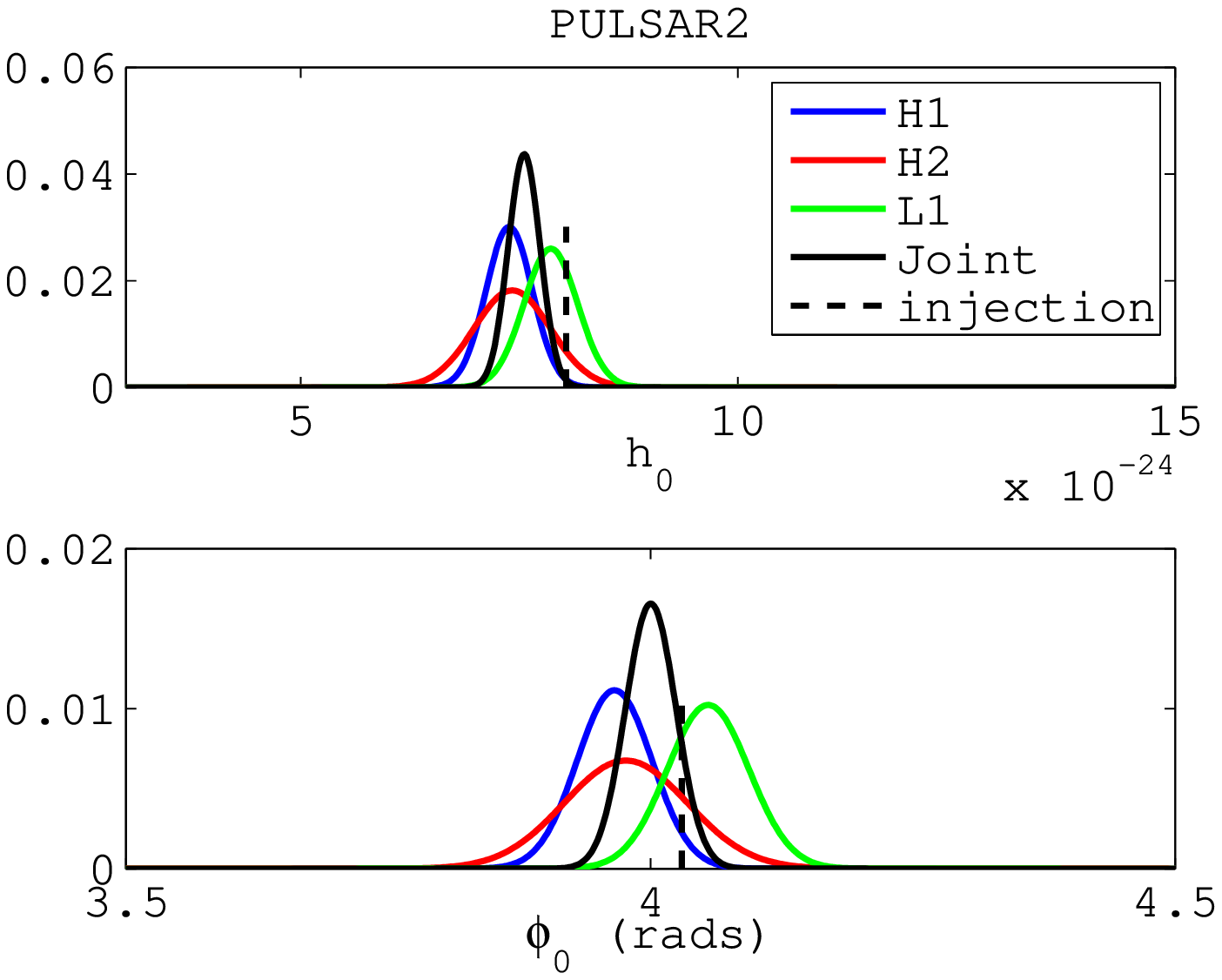} \\
\includegraphics[width=0.3\textwidth]{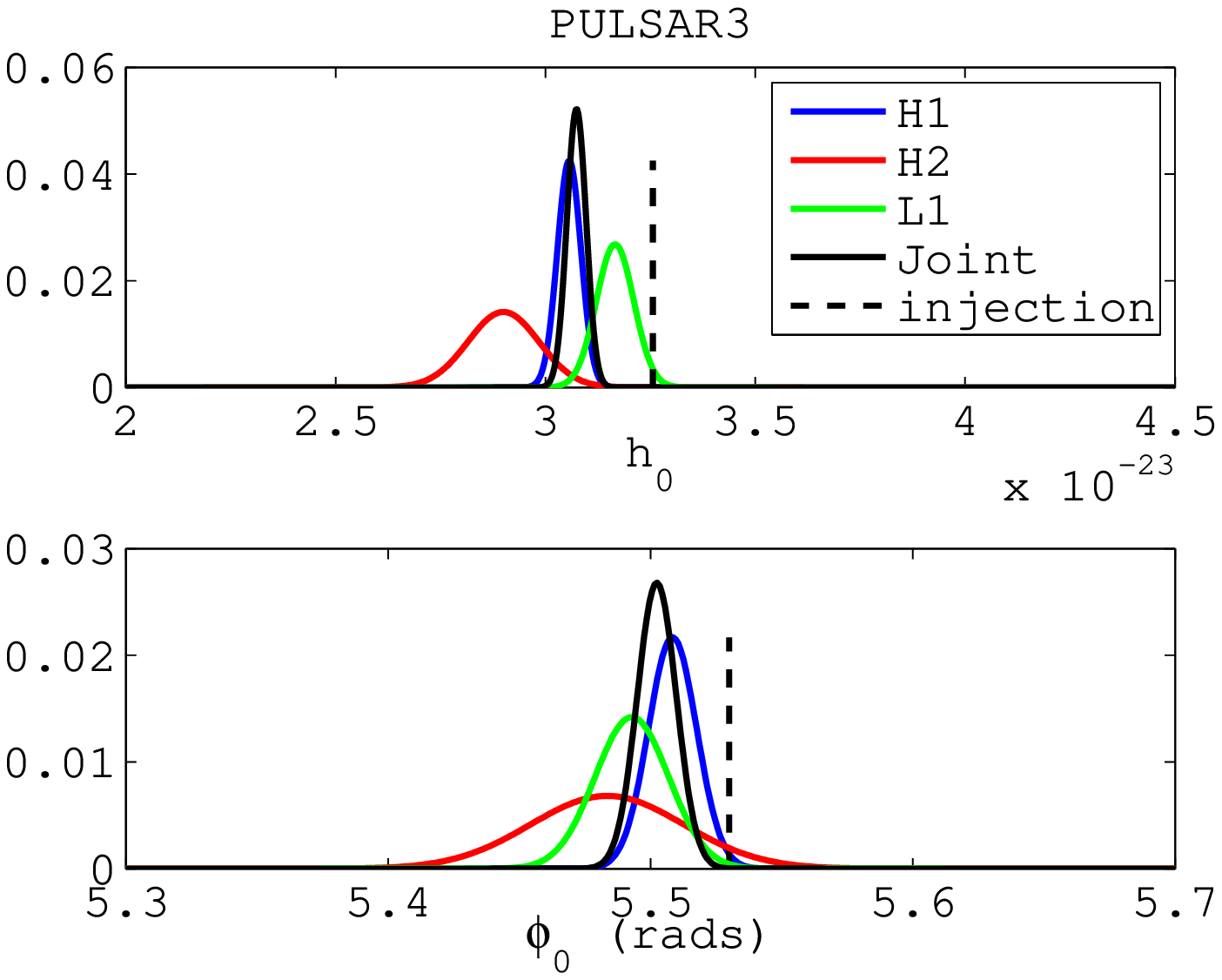} &
\includegraphics[width=0.3\textwidth]{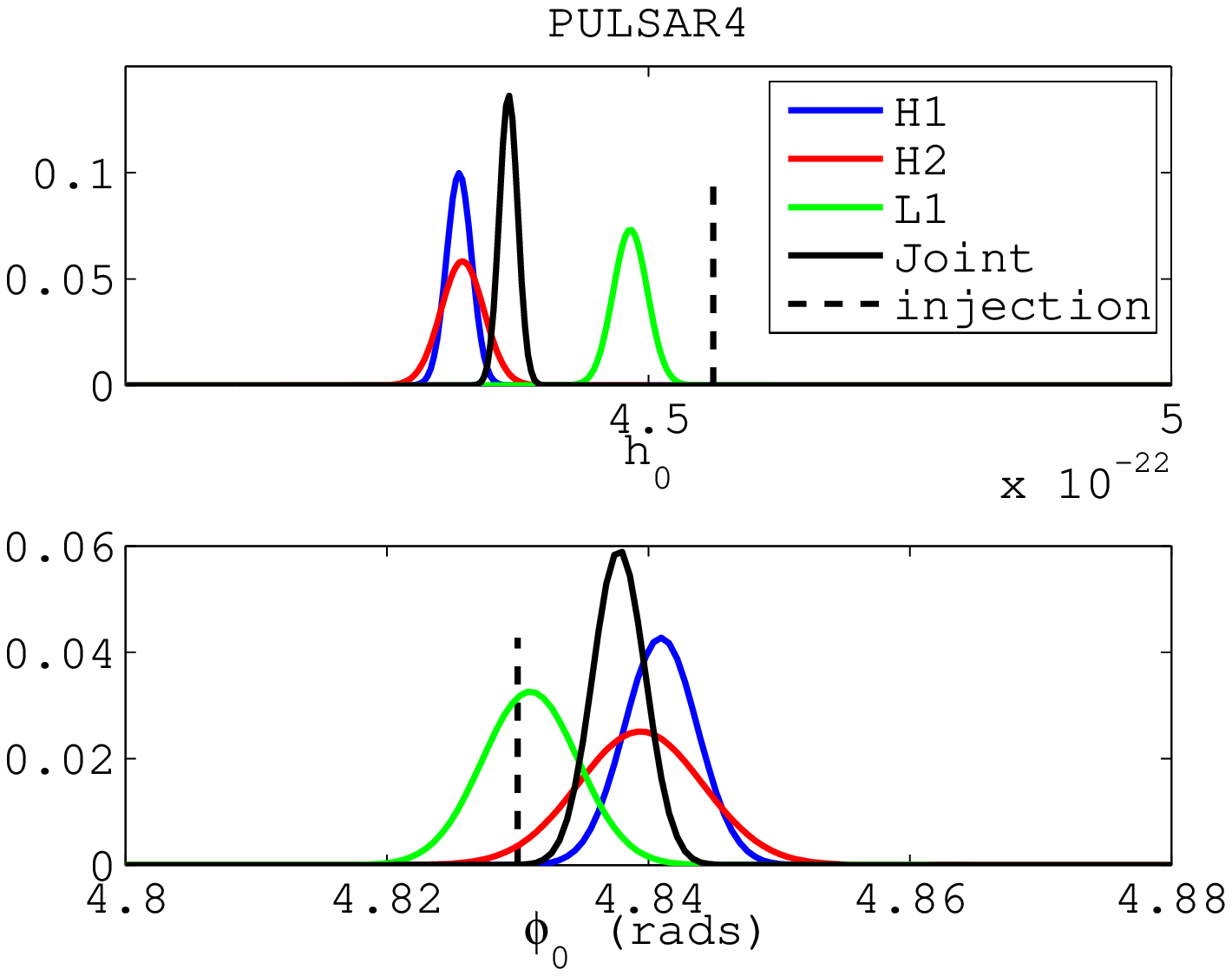} &
\includegraphics[width=0.3\textwidth]{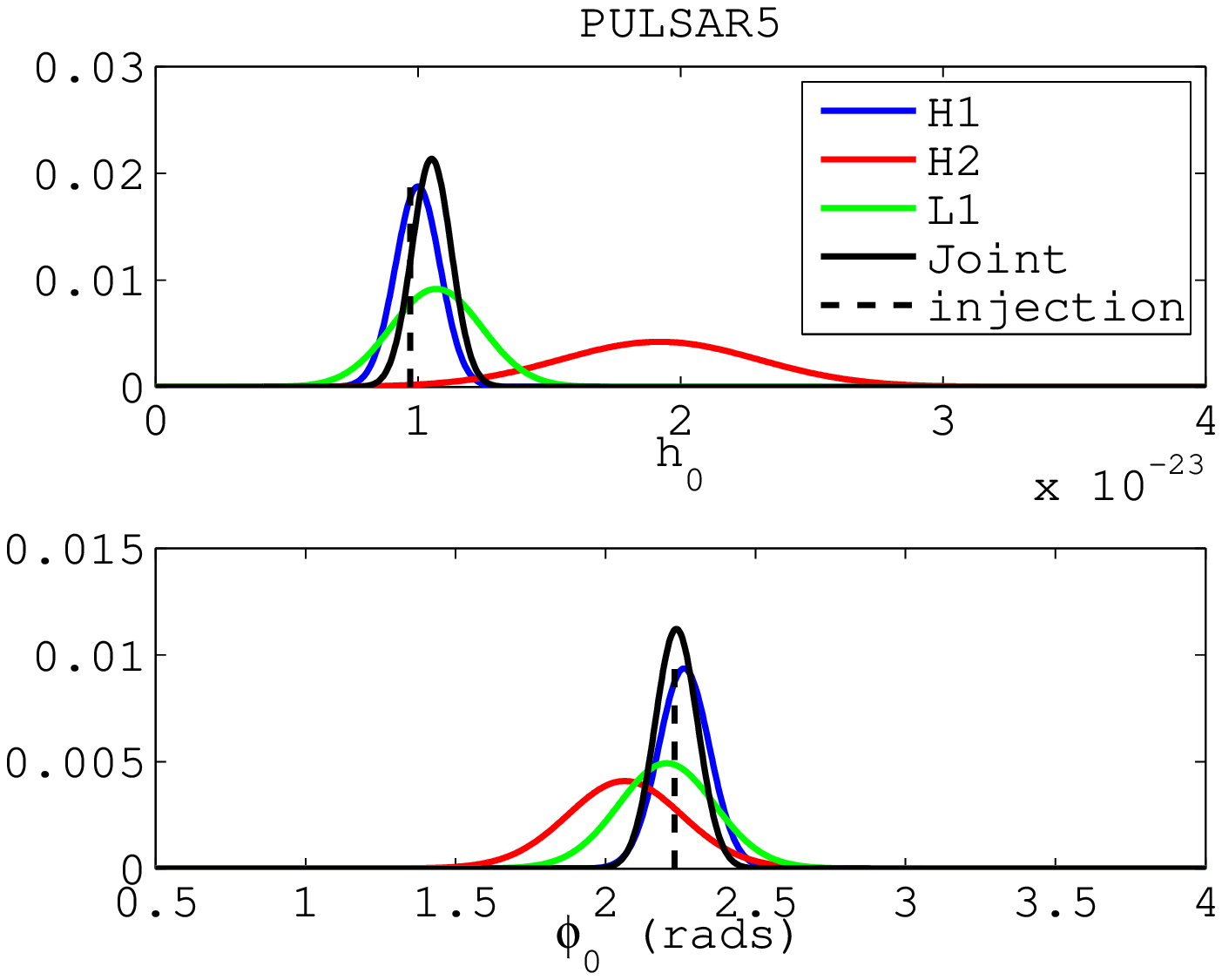} \\
\includegraphics[width=0.3\textwidth]{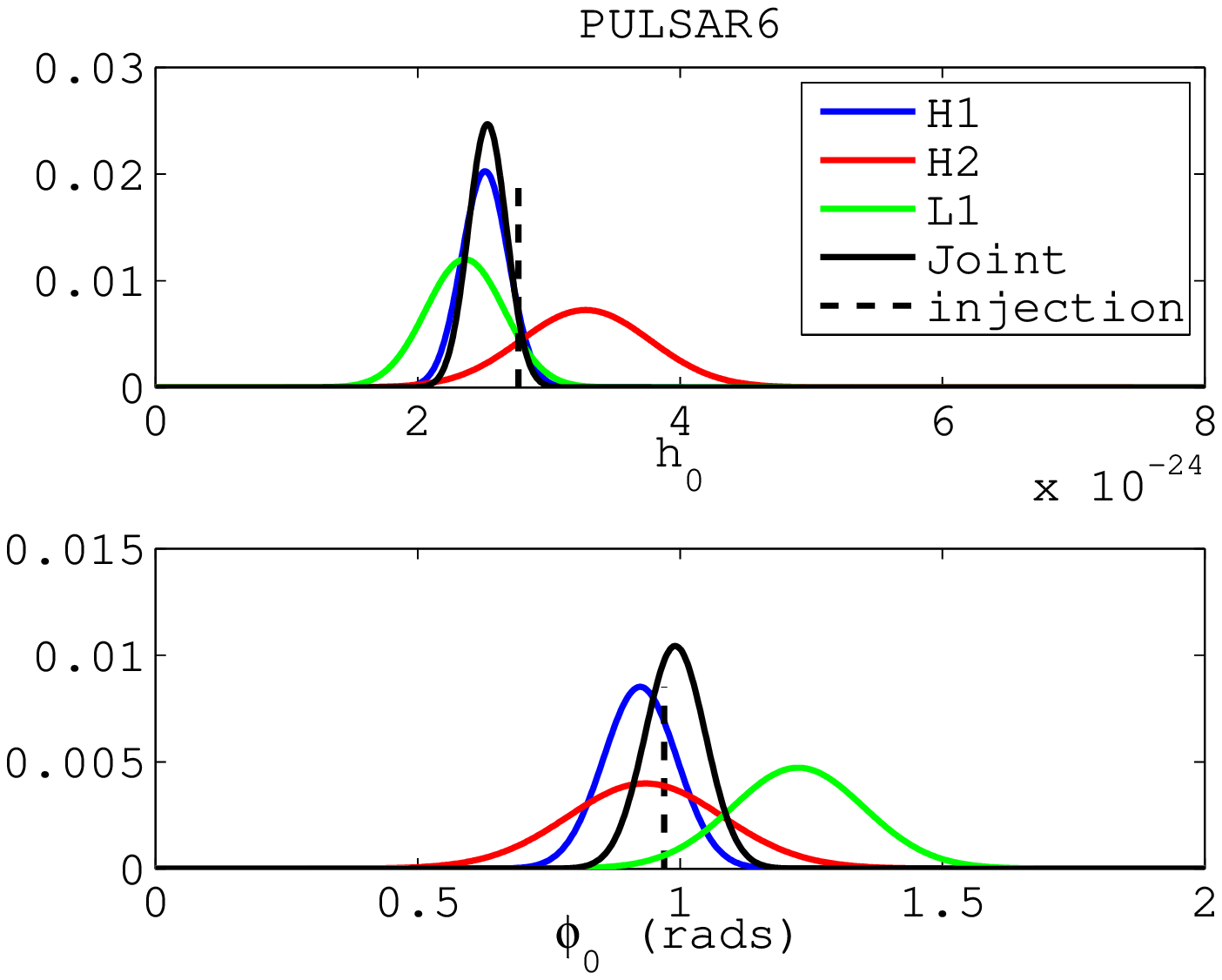} &
\includegraphics[width=0.3\textwidth]{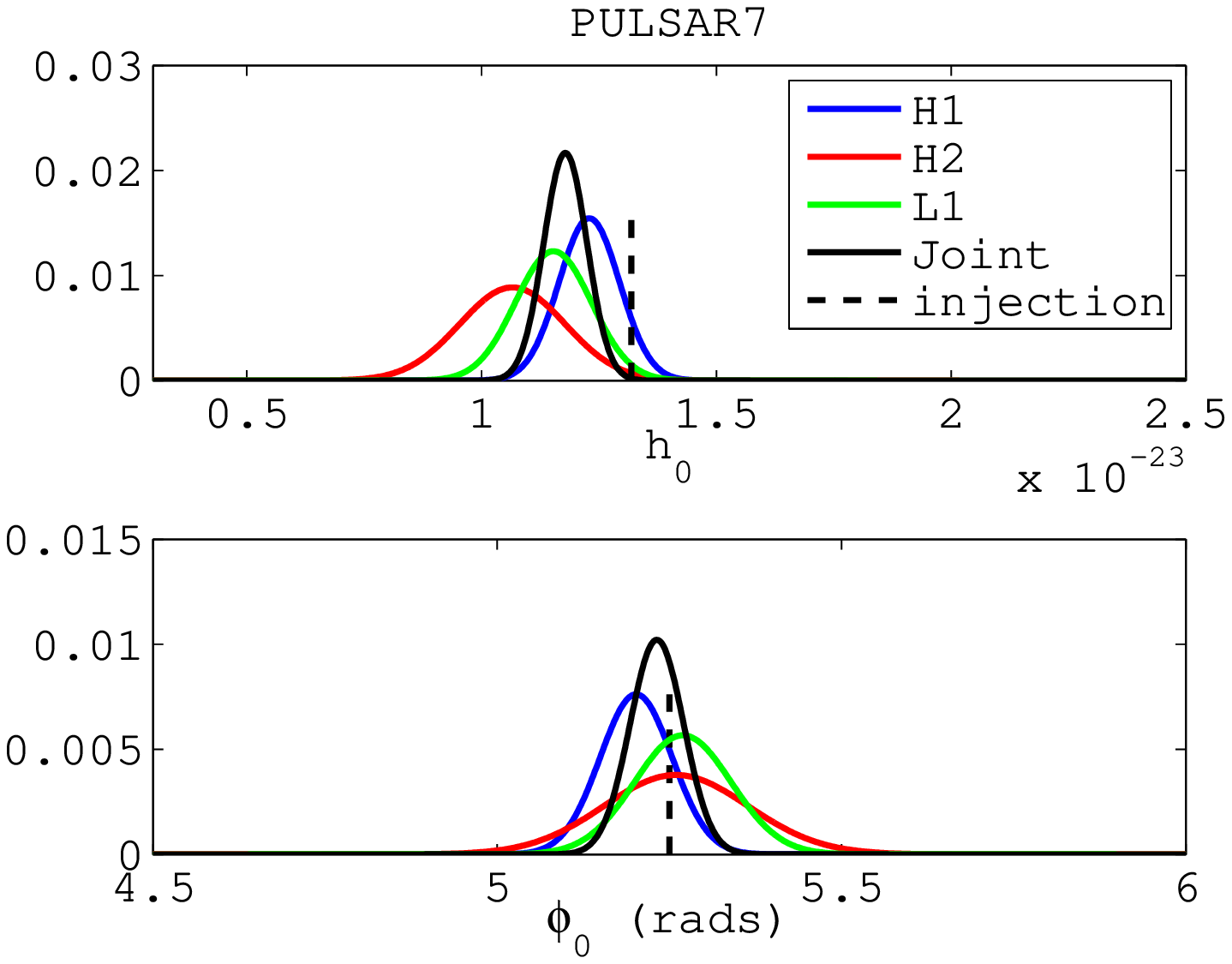} &
\includegraphics[width=0.3\textwidth]{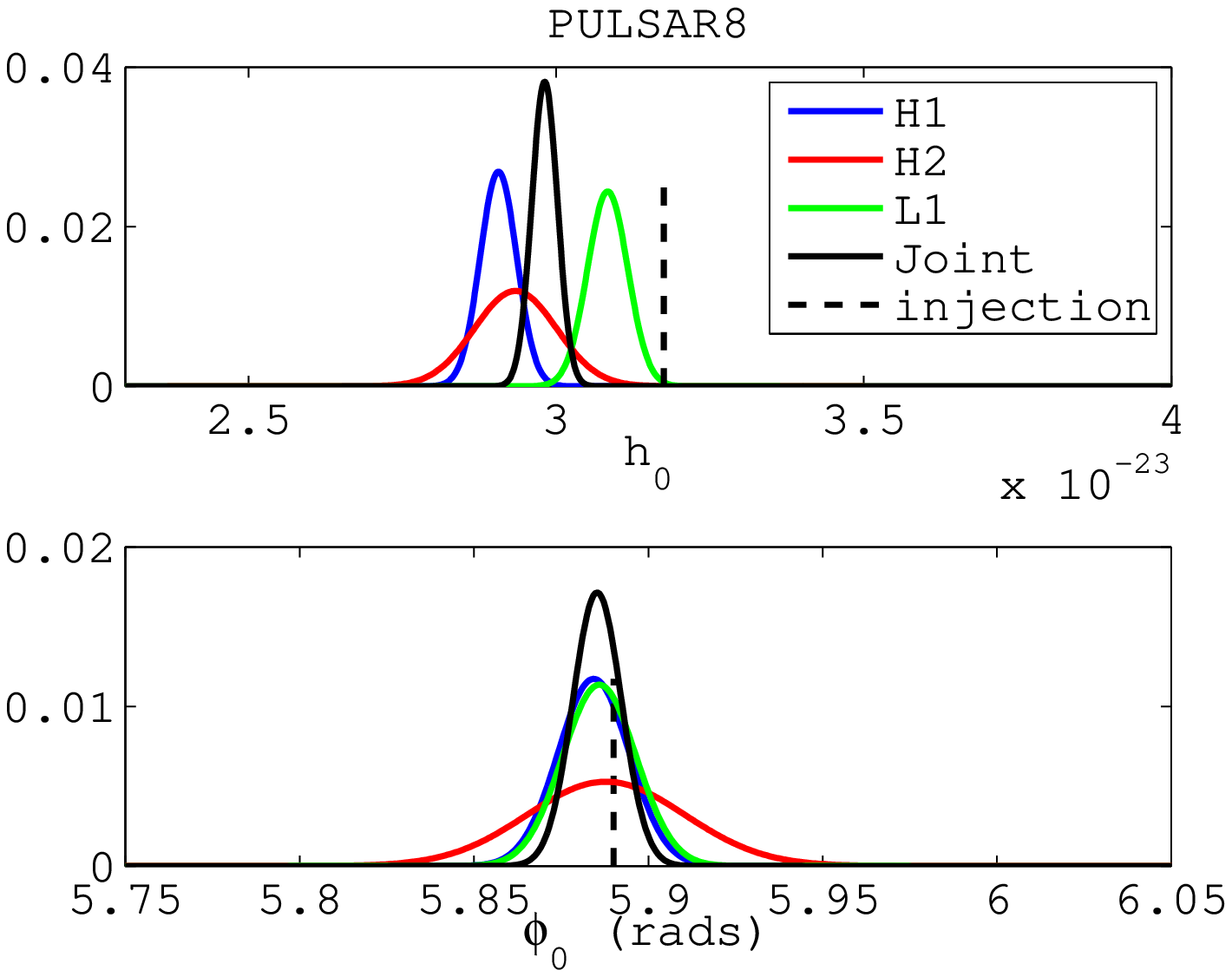} \\
\includegraphics[width=0.3\textwidth]{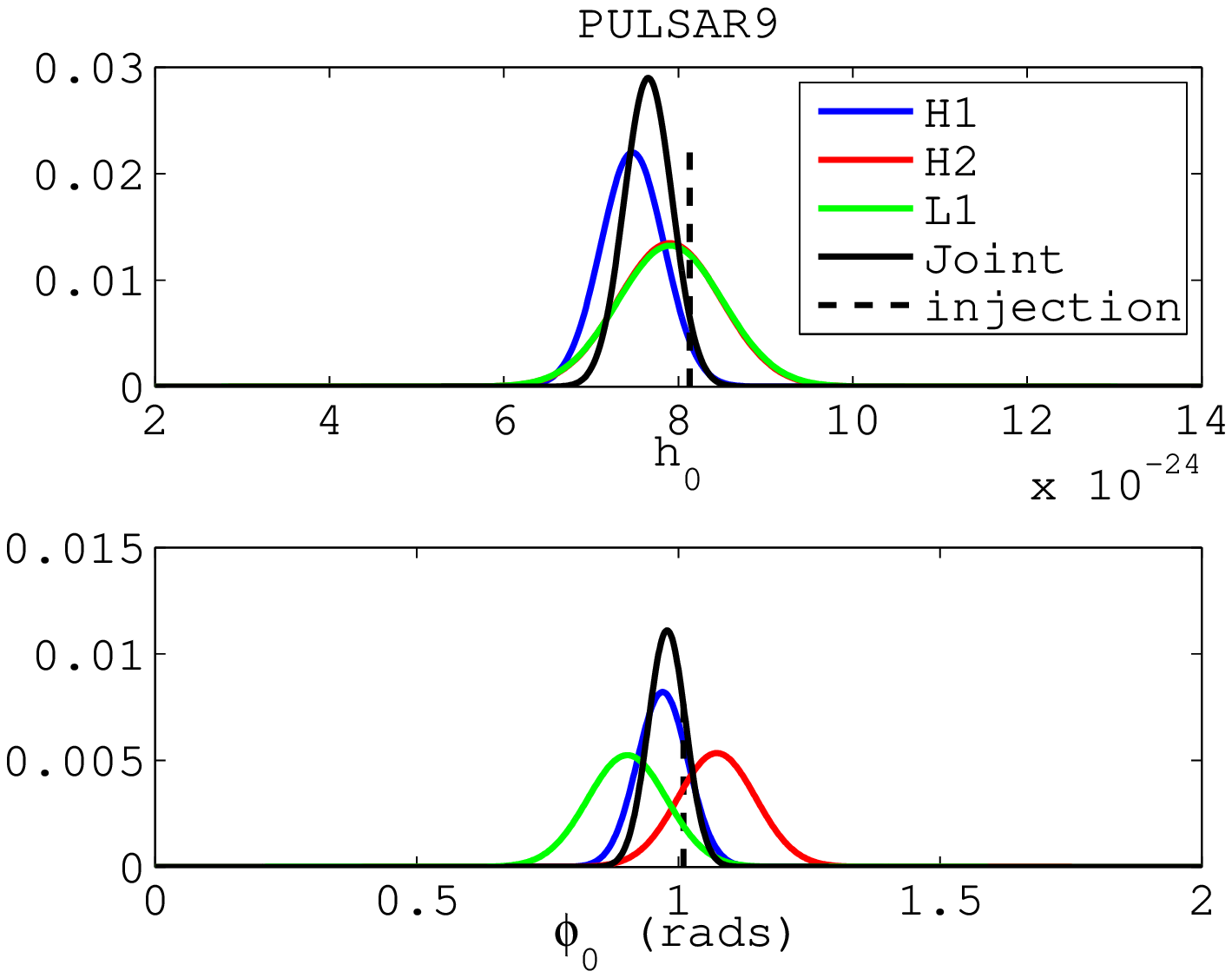} &
\includegraphics[width=0.3\textwidth]{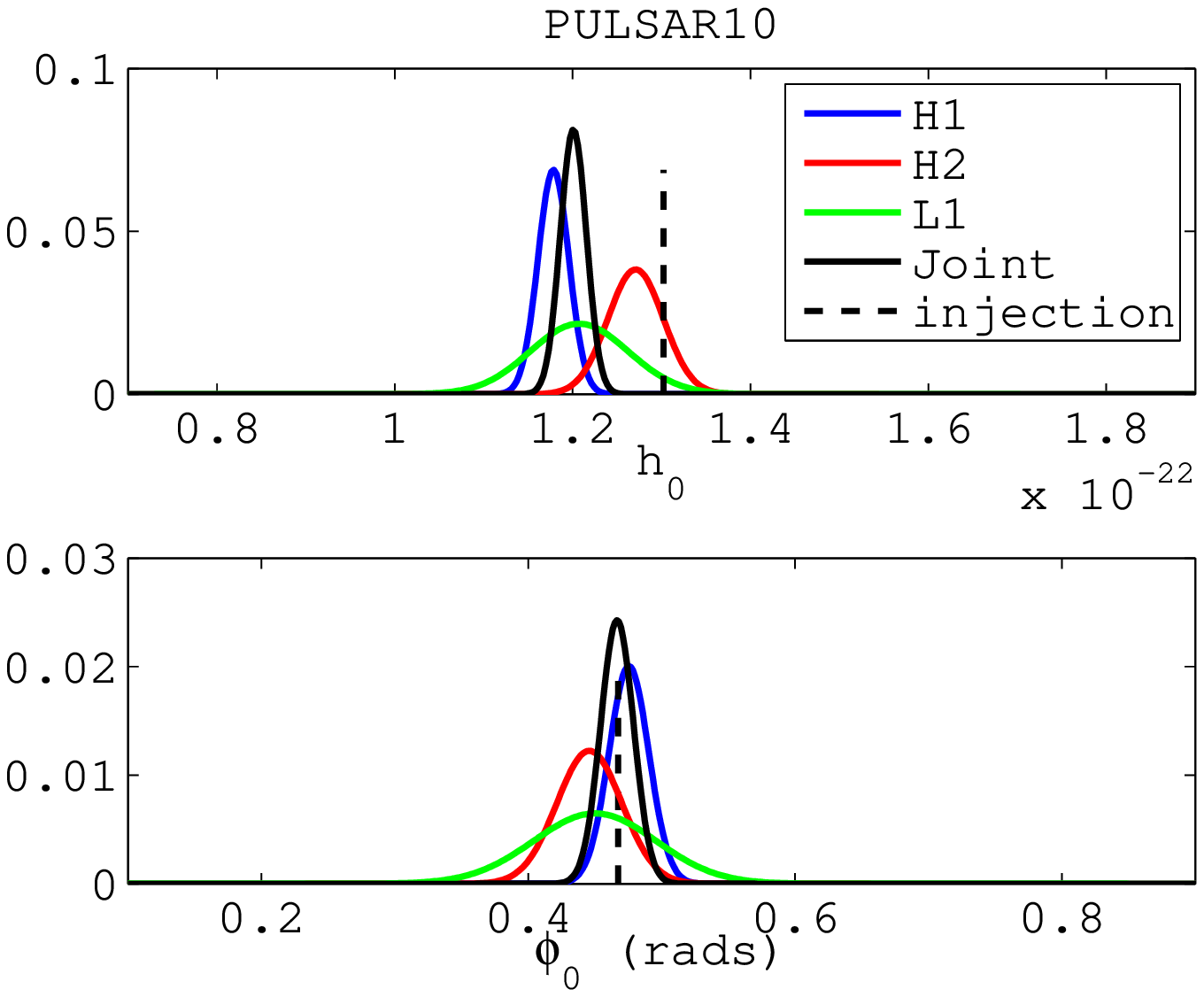} &
\includegraphics[width=0.3\textwidth]{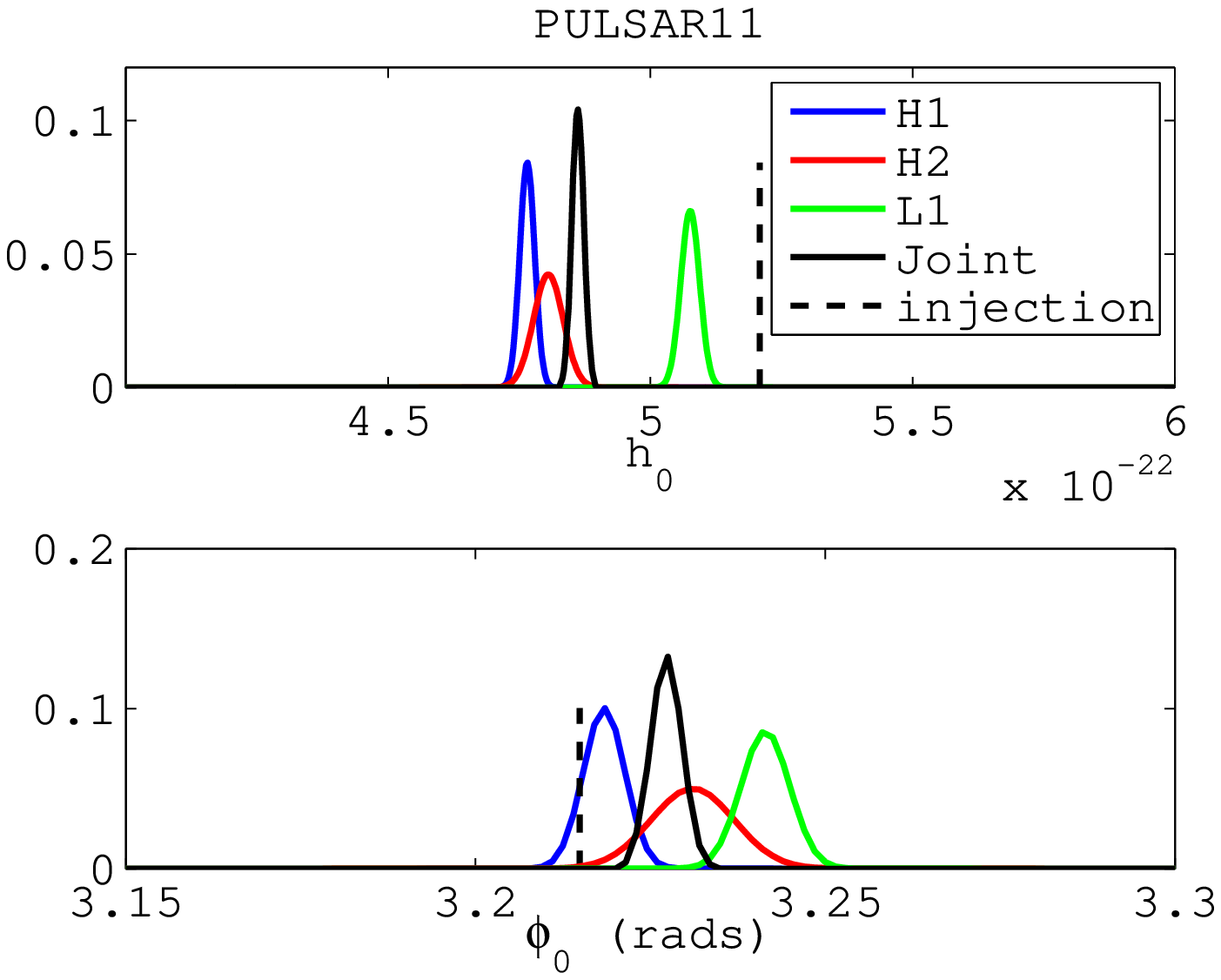} \\
\end{tabular}
\caption{The pdfs of $h_0$ and $\phi_0$ for 10 isolated and 2 binary pulsar injections into the
LIGO detectors during S4.}\label{S4PulsarInj}
\end{figure*}

The extracted amplitude and phase pdfs, after corrections described
below in \S\ref{sec:calissues}, are shown in Fig.~\ref{S4PulsarInj}
The observed phase consistency between the detectors, means that
joint likelihoods, using all three detectors, can be calculated. In
general the values of $h_0$ are well matched with the injection
values. It can again be seen that for the strongest signals the
narrow pdfs are offset from the injected value in $h_0$ reflecting
the calibration uncertainties of 5-10\%.

The binary pulsar injections show matches to their injected values.
This is a good confirmation that the binary timing code can track
the phase well and has no significant errors.

\subsection{Calibration Issues}\label{sec:calissues}
A brief note should be made of the effect of calibrations on the
above extracted pulsar hardware injections. The injections were for
the most part analysed using exactly the same pipeline as applied to
the general known pulsar analysis. However due to the nature of the
hardware injections some additional post-processing of the results
has had to be applied. To calculate the amplitude and phase of the
injections, when applying forces to the interferometer end test
masses, a reference calibration must be used. These reference
calibrations are different for each interferometer. For both S3 and
S4 these reference calibrations differed by small, but not
insignificant amounts, from the final calibration used when
extracting the signals, meaning that upon extracting the signals the
amplitude and phase appear offset from the injected values. As the
difference between the reference and final calibrations are
different for each interferometer there will also be slight offsets
between the extracted parameters between detectors. The extracted
signals from each interferometer therefore have had to be adjusted
to reflect these differences, determined independently of the 
hardware injections, and correct them so as to give the same
input signal. This allows the combined joint upper limits to be
produced.

\begin{turnpage}
\begingroup
\squeezetable
\begin{table*}
\section{Tables of upper limit results}\label{resultstables}
\caption{\label{table:allresults} Pulsar upper limits using
LIGO data from the S3 and S4 runs. The approximate pulsar spin
frequencies and spin-down rates are given. A * denotes globular
cluster pulsars for which no spin-down upper limit could be set. The
values marked with a $\dagger$ represent pulsars for which the
spin-down limit has been corrected for the Shklovskii effect. The
ratio column gives the ratio of our experimental upper limits to the
spin-down upper limits.}
\begin{tabular}{l | c | c | c | c | c | c | c | c | c | c | c | c | c | c | c | c | c | c | c | c}
\hline \hline
\multicolumn{3}{c|}{~} & \multicolumn{6}{c|}{S3} & \multicolumn{6}{c|}{S4} & \multicolumn{6}{c}{S3
and S4} \\ \hline
\multicolumn{3}{c|}{~}  & \multicolumn{4}{c|}{{$\log{h_0^{95\%}}$}} & {$\log{\varepsilon}$} & ratio
& \multicolumn{4}{|c|}{{$\log{h_0^{95\%}}$}} & {$\log{\varepsilon}$} & ratio &
\multicolumn{4}{|c|}{{$\log{h_0^{95\%}}$}} & {$\log{\varepsilon}$} & ratio \\ \hline
{P\textsc{ulsar}} & $\nu$ (Hz) & $\dot{\nu}$ (Hz\,s$^{-1}$) & {H1} & {H2} & {L1} &
\multicolumn{3}{|c|}{Joint} & {H1} & {H2} & {L1} & \multicolumn{3}{|c|}{Joint} & {H1} & {H2} & {L1}
& \multicolumn{3}{|c}{Joint} \\
\hline \hline
{\tt{J0024-7204C}} & {$\tt{173.71}$} & {$\tt{+1.50\ee{-15}}$} & {$\tt{-23.30}$} & {$\tt{-23.55}$} & {$\tt{-23.04}$} & {$\tt{-23.64}$} & {$\tt{-4.06}$} & {\tt{*}} &  {$\tt{-23.08}$} & {$\tt{-23.53}$} & {$\tt{-23.31}$} & {$\tt{-23.53}$} & {$\tt{-3.96}$} & {\tt{*}} &  {$\tt{-23.41}$} & {$\tt{-23.69}$} & {$\tt{-23.34}$} & {$\tt{-23.75}$} & {$\tt{-4.18}$} & {\tt{*}} \\ 
{\tt{J0024-7204D}} & {$\tt{186.65}$} & {$\tt{+1.20\ee{-16}}$} & {$\tt{-23.69}$} & {$\tt{-23.72}$} & {$\tt{-23.24}$} & {$\tt{-23.82}$} & {$\tt{-4.31}$} & {\tt{*}} &  {$\tt{-23.97}$} & {$\tt{-23.92}$} & {$\tt{-23.95}$} & {$\tt{-24.14}$} & {$\tt{-4.63}$} & {\tt{*}} &  {$\tt{-24.15}$} & {$\tt{-23.99}$} & {$\tt{-23.91}$} & {$\tt{-24.36}$} & {$\tt{-4.85}$} & {\tt{*}} \\ 
{\tt{J0024-7204E}} & {$\tt{282.78}$} & {$\tt{-7.88\ee{-15}}$} & {$\tt{-23.90}$} & {$\tt{-23.53}$} & {$\tt{-23.19}$} & {$\tt{-23.92}$} & {$\tt{-4.76}$} & {$\tt{1380^{\dagger}}$} & {$\tt{-24.02}$} & {$\tt{-23.93}$} & {$\tt{-23.83}$} & {$\tt{-24.15}$} & {$\tt{-4.99}$} & {$\tt{815^{\dagger}}$} & {$\tt{-24.06}$} & {$\tt{-24.03}$} & {$\tt{-23.84}$} &  {$\tt{-24.16}$} & {$\tt{-5.01}$} & {$\tt{786^{\dagger}}$} \\ 
{\tt{J0024-7204F}} & {$\tt{381.16}$} & {$\tt{-9.37\ee{-15}}$} & {$\tt{-23.64}$} & {$\tt{-23.47}$} & {$\tt{-22.98}$} & {$\tt{-23.70}$} & {$\tt{-4.81}$} & {$\tt{2403^{\dagger}}$} & {$\tt{-23.91}$} & {$\tt{-23.63}$} & {$\tt{-23.53}$} & {$\tt{-23.99}$} & {$\tt{-5.10}$} & {$\tt{1237^{\dagger}}$} & {$\tt{-24.05}$} & {$\tt{-23.63}$} & {$\tt{-23.51}$} &  {$\tt{-24.16}$} & {$\tt{-5.26}$} & {$\tt{845^{\dagger}}$} \\ 
{\tt{J0024-7204G}} & {$\tt{247.50}$} & {$\tt{+2.58\ee{-15}}$} & {$\tt{-24.05}$} & {$\tt{-23.66}$} & {$\tt{-23.14}$} & {$\tt{-24.12}$} & {$\tt{-4.85}$} & {\tt{*}} &  {$\tt{-24.09}$} & {$\tt{-23.99}$} & {$\tt{-23.87}$} & {$\tt{-24.29}$} & {$\tt{-5.02}$} & {\tt{*}} &  {$\tt{-24.16}$} & {$\tt{-24.04}$} & {$\tt{-23.88}$} & {$\tt{-24.37}$} & {$\tt{-5.10}$} & {\tt{*}} \\ 
&&&&&&&&&&&&&&&&&&&\\[-7pt] 
{\tt{J0024-7204I}} & {$\tt{286.94}$} & {$\tt{+3.78\ee{-15}}$} & {$\tt{-23.74}$} & {$\tt{-23.36}$} & {$\tt{-23.13}$} & {$\tt{-23.81}$} & {$\tt{-4.67}$} & {\tt{*}} &  {$\tt{-24.05}$} & {$\tt{-23.96}$} & {$\tt{-23.53}$} & {$\tt{-24.04}$} & {$\tt{-4.90}$} & {\tt{*}} &  {$\tt{-23.97}$} & {$\tt{-23.97}$} & {$\tt{-23.56}$} & {$\tt{-24.02}$} & {$\tt{-4.88}$} & {\tt{*}} \\ 
{\tt{J0024-7204J}} & {$\tt{476.05}$} & {$\tt{+2.22\ee{-15}}$} & {$\tt{-23.65}$} & {$\tt{-23.10}$} & {$\tt{-22.86}$} & {$\tt{-23.63}$} & {$\tt{-4.93}$} & {\tt{*}} &  {$\tt{-23.86}$} & {$\tt{-23.78}$} & {$\tt{-23.34}$} & {$\tt{-24.06}$} & {$\tt{-5.36}$} & {\tt{*}} &  {$\tt{-23.89}$} & {$\tt{-23.76}$} & {$\tt{-23.34}$} & {$\tt{-24.13}$} & {$\tt{-5.43}$} & {\tt{*}} \\ 
{\tt{J0024-7204L}} & {$\tt{230.09}$} & {$\tt{+6.46\ee{-15}}$} & {$\tt{-23.99}$} & {$\tt{-23.56}$} & {$\tt{-23.25}$} & {$\tt{-24.02}$} & {$\tt{-4.69}$} & {\tt{*}} &  {$\tt{-23.96}$} & {$\tt{-23.96}$} & {$\tt{-23.78}$} & {$\tt{-23.97}$} & {$\tt{-4.64}$} & {\tt{*}} &  {$\tt{-24.12}$} & {$\tt{-24.02}$} & {$\tt{-23.78}$} & {$\tt{-24.07}$} & {$\tt{-4.74}$} & {\tt{*}} \\ 
{\tt{J0024-7204M}} & {$\tt{271.99}$} & {$\tt{+2.84\ee{-15}}$} & {$\tt{-23.93}$} & {$\tt{-23.52}$} & {$\tt{-23.00}$} & {$\tt{-23.93}$} & {$\tt{-4.75}$} & {\tt{*}} &  {$\tt{-24.05}$} & {$\tt{-24.01}$} & {$\tt{-23.87}$} & {$\tt{-24.20}$} & {$\tt{-5.01}$} & {\tt{*}} &  {$\tt{-24.06}$} & {$\tt{-24.09}$} & {$\tt{-23.81}$} & {$\tt{-24.22}$} & {$\tt{-5.03}$} & {\tt{*}} \\ 
{\tt{J0024-7204N}} & {$\tt{327.44}$} & {$\tt{+2.34\ee{-15}}$} & {$\tt{-23.81}$} & {$\tt{-23.43}$} & {$\tt{-23.11}$} & {$\tt{-23.82}$} & {$\tt{-4.79}$} & {\tt{*}} &  {$\tt{-23.80}$} & {$\tt{-23.76}$} & {$\tt{-23.72}$} & {$\tt{-24.13}$} & {$\tt{-5.10}$} & {\tt{*}} &  {$\tt{-24.00}$} & {$\tt{-23.91}$} & {$\tt{-23.70}$} & {$\tt{-24.27}$} & {$\tt{-5.25}$} & {\tt{*}} \\ 
&&&&&&&&&&&&&&&&&&&\\[-7pt] 
{\tt{J0024-7204Q}} & {$\tt{247.94}$} & {$\tt{-2.09\ee{-15}}$} & {$\tt{-23.94}$} & {$\tt{-23.47}$} & {$\tt{-23.29}$} & {$\tt{-24.07}$} & {$\tt{-4.81}$} & {\tt{1734}} & {$\tt{-24.06}$} & {$\tt{-24.01}$} & {$\tt{-23.65}$} & {$\tt{-24.09}$} & {$\tt{-4.82}$} & {\tt{1669}} & {$\tt{-24.14}$} & {$\tt{-24.06}$} & {$\tt{-23.66}$} & {$\tt{-24.23}$} & {$\tt{-4.96}$} & {\tt{1215}} \\ 
{\tt{J0024-7204S}} & {$\tt{353.31}$} & {$\tt{+1.50\ee{-14}}$} & {$\tt{-23.58}$} & {$\tt{-23.35}$} & {$\tt{-23.10}$} & {$\tt{-23.60}$} & {$\tt{-4.64}$} & {\tt{*}} &  {$\tt{-23.97}$} & {$\tt{-23.89}$} & {$\tt{-23.62}$} & {$\tt{-24.17}$} & {$\tt{-5.21}$} & {\tt{*}} &  {$\tt{-23.97}$} & {$\tt{-23.97}$} & {$\tt{-23.62}$} & {$\tt{-24.11}$} & {$\tt{-5.15}$} & {\tt{*}} \\ 
{\tt{J0024-7204T}} & {$\tt{131.78}$} & {$\tt{-5.10\ee{-15}}$} & {$\tt{-24.17}$} & {$\tt{-23.74}$} & {$\tt{-23.47}$} & {$\tt{-24.21}$} & {$\tt{-4.39}$} & {\tt{591}} & {$\tt{-24.26}$} & {$\tt{-24.19}$} & {$\tt{-24.15}$} & {$\tt{-24.45}$} & {$\tt{-4.63}$} & {\tt{340}} & {$\tt{-24.29}$} & {$\tt{-24.22}$} & {$\tt{-24.17}$} & {$\tt{-24.48}$} & {$\tt{-4.66}$} & {\tt{319}} \\ 
{\tt{J0024-7204U}} & {$\tt{230.26}$} & {$\tt{-5.05\ee{-15}}$} & {$\tt{-23.73}$} & {$\tt{-23.64}$} & {$\tt{-23.29}$} & {$\tt{-23.83}$} & {$\tt{-4.50}$} & {$\tt{1886^{\dagger}}$} & {$\tt{-24.01}$} & {$\tt{-24.04}$} & {$\tt{-23.68}$} & {$\tt{-24.27}$} & {$\tt{-4.94}$} & {$\tt{693^{\dagger}}$} & {$\tt{-23.94}$} & {$\tt{-24.10}$} & {$\tt{-23.70}$} &  {$\tt{-24.16}$} & {$\tt{-4.83}$} & {$\tt{900^{\dagger}}$} \\ 
{\tt{J0034-0534}} & {$\tt{532.71}$} & {$\tt{-1.41\ee{-15}}$} & {$\tt{-23.20}$} & {$\tt{-23.20}$} & {$\tt{-22.69}$} & {$\tt{-23.45}$} & {$\tt{-5.54}$} & {\tt{2653}} & {$\tt{-23.38}$} & {$\tt{-23.36}$} & {$\tt{-23.45}$} & {$\tt{-23.81}$} & {$\tt{-5.89}$} & {\tt{1171}} & {$\tt{-23.51}$} & {$\tt{-23.43}$} & {$\tt{-23.43}$} & {$\tt{-23.87}$} & {$\tt{-5.96}$} & {\tt{999}} \\ 
&&&&&&&&&&&&&&&&&&&\\[-7pt] 
{\tt{J0218+4232}} & {$\tt{430.46}$} & {$\tt{-1.43\ee{-14}}$} & {$\tt{-23.56}$} & {$\tt{-23.41}$} & {$\tt{-22.93}$} & {$\tt{-23.66}$} & {$\tt{-4.79}$} & {\tt{2740}} & {$\tt{-23.85}$} & {$\tt{-23.49}$} & {$\tt{-23.54}$} & {$\tt{-23.84}$} & {$\tt{-4.97}$} & {\tt{1821}} & {$\tt{-23.95}$} & {$\tt{-23.59}$} & {$\tt{-23.58}$} & {$\tt{-23.94}$} & {$\tt{-5.07}$} & {\tt{1430}} \\ 
{\tt{J0534+2200}} & {$\tt{29.80}$} & {$\tt{-3.73\ee{-10}}$} & {$\tt{-23.18}$} & {$\tt{-22.04}$} & {$\tt{-22.40}$} & {$\tt{-23.22}$} & {$\tt{-2.49}$} & {$\tt{4.23^{\dagger}}$} & {$\tt{-23.42}$} & {$\tt{-23.19}$} & {$\tt{-22.96}$} & {$\tt{-23.46}$} & {$\tt{-2.73}$} & {$\tt{2.45^{\dagger}}$} & {$\tt{-23.49}$} & {$\tt{-23.19}$} & {$\tt{-22.96}$} & {$\tt{-23.51}$} & {$\tt{-2.78}$} & {$\tt{2.18^{\dagger}}$} \\ 
{\tt{J0613-0200}} & {$\tt{326.60}$} & {$\tt{-1.02\ee{-15}}$} & {$\tt{-23.77}$} & {$\tt{-23.33}$} & {$\tt{-23.07}$} & {$\tt{-23.80}$} & {$\tt{-5.11}$} & {$\tt{2571^{\dagger}}$} & {$\tt{-23.72}$} & {$\tt{-23.82}$} & {$\tt{-23.73}$} & {$\tt{-24.00}$} & {$\tt{-5.32}$} & {$\tt{1597^{\dagger}}$} & {$\tt{-23.96}$} & {$\tt{-23.86}$} & {$\tt{-23.73}$} &  {$\tt{-24.07}$} & {$\tt{-5.39}$} & {$\tt{1365^{\dagger}}$} \\ 
{\tt{J0621+1002}} & {$\tt{34.66}$} & {$\tt{-5.68\ee{-17}}$} & {$\tt{-23.55}$} & {$\tt{-22.56}$} & {$\tt{-23.04}$} & {$\tt{-23.60}$} & {$\tt{-3.03}$} & {$\tt{4675^{\dagger}}$} & {$\tt{-23.83}$} & {$\tt{-23.34}$} & {$\tt{-23.95}$} & {$\tt{-24.17}$} & {$\tt{-3.61}$} & {$\tt{1241^{\dagger}}$} & {$\tt{-23.89}$} & {$\tt{-23.34}$} & {$\tt{-23.96}$} &  {$\tt{-24.15}$} & {$\tt{-3.59}$} & {$\tt{1301^{\dagger}}$} \\ 
{\tt{J0711-6830}} & {$\tt{182.11}$} & {$\tt{-4.94\ee{-16}}$} & {$\tt{-23.95}$} & {$\tt{-23.21}$} & {$\tt{-23.23}$} & {$\tt{-24.01}$} & {$\tt{-5.14}$} & {$\tt{1018^{\dagger}}$} & {$\tt{-24.10}$} & {$\tt{-23.99}$} & {$\tt{-23.88}$} & {$\tt{-24.16}$} & {$\tt{-5.29}$} & {$\tt{733^{\dagger}}$} & {$\tt{-24.25}$} & {$\tt{-24.07}$} & {$\tt{-23.85}$} &  {$\tt{-24.31}$} & {$\tt{-5.44}$} & {$\tt{513^{\dagger}}$} \\ 
&&&&&&&&&&&&&&&&&&&\\[-7pt] 
{\tt{J0737-3039A}} & {$\tt{44.05}$} & {$\tt{-3.38\ee{-15}}$} & {$\tt{-24.05}$} & {$\tt{-23.20}$} & {$\tt{-23.28}$} & {$\tt{-24.03}$} & {$\tt{-4.19}$} & {\tt{75}} & {$\tt{-24.13}$} & {$\tt{-23.80}$} & {$\tt{-24.08}$} & {$\tt{-24.28}$} & {$\tt{-4.44}$} & {\tt{42}} & {$\tt{-24.27}$} & {$\tt{-23.83}$} & {$\tt{-24.08}$} & {$\tt{-24.34}$} & {$\tt{-4.50}$} & {\tt{37}} \\ 
{\tt{J0751+1807}} & {$\tt{287.46}$} & {$\tt{-6.43\ee{-16}}$} & {$\tt{-23.87}$} & {$\tt{-23.51}$} & {$\tt{-23.22}$} & {$\tt{-23.94}$} & {$\tt{-5.69}$} & {$\tt{604^{\dagger}}$} & {$\tt{-23.78}$} & {$\tt{-23.80}$} & {$\tt{-23.66}$} & {$\tt{-23.95}$} & {$\tt{-5.70}$} & {$\tt{590^{\dagger}}$} & {$\tt{-23.91}$} & {$\tt{-23.87}$} & {$\tt{-23.63}$} &  {$\tt{-24.02}$} & {$\tt{-5.77}$} & {$\tt{496^{\dagger}}$} \\ 
{\tt{J1012+5307}} & {$\tt{190.27}$} & {$\tt{-6.20\ee{-16}}$} & {$\tt{-24.08}$} & {$\tt{-23.75}$} & {$\tt{-23.23}$} & {$\tt{-24.06}$} & {$\tt{-5.53}$} & {$\tt{357^{\dagger}}$} & {$\tt{-24.22}$} & {$\tt{-24.03}$} & {$\tt{-23.89}$} & {$\tt{-24.42}$} & {$\tt{-5.89}$} & {$\tt{156^{\dagger}}$} & {$\tt{-24.32}$} & {$\tt{-24.11}$} & {$\tt{-23.93}$} &  {$\tt{-24.49}$} & {$\tt{-5.96}$} & {$\tt{135^{\dagger}}$} \\ 
{\tt{J1022+1001}} & {$\tt{60.78}$} & {$\tt{-1.60\ee{-16}}$} & {$\tt{-24.15}$} & {$\tt{-22.88}$} & {$\tt{-23.11}$} & {$\tt{-24.15}$} & {$\tt{-4.87}$} & {\tt{161}} & {$\tt{-23.99}$} & {$\tt{-23.80}$} & {$\tt{-24.24}$} & {$\tt{-24.31}$} & {$\tt{-5.03}$} & {\tt{113}} & {$\tt{-24.16}$} & {$\tt{-23.80}$} & {$\tt{-24.26}$} & {$\tt{-24.37}$} & {$\tt{-5.09}$} & {\tt{98}} \\ 
{\tt{J1024-0719}} & {$\tt{193.72}$} & {$\tt{-6.95\ee{-16}}$} & {$\tt{-23.79}$} & {$\tt{-23.46}$} & {$\tt{-23.46}$} & {$\tt{-23.97}$} & {$\tt{-5.63}$} & {\tt{243}} & {$\tt{-23.92}$} & {$\tt{-23.95}$} & {$\tt{-23.94}$} & {$\tt{-24.19}$} & {$\tt{-5.85}$} & {\tt{147}} & {$\tt{-24.05}$} & {$\tt{-23.99}$} & {$\tt{-23.95}$} & {$\tt{-24.33}$} & {$\tt{-5.98}$} & {\tt{109}} \\ 
&&&&&&&&&&&&&&&&&&&\\[-7pt] 
{\tt{J1045-4509}} & {$\tt{133.79}$} & {$\tt{-3.16\ee{-16}}$} & {$\tt{-23.92}$} & {$\tt{-23.61}$} & {$\tt{-23.46}$} & {$\tt{-23.96}$} & {$\tt{-4.32}$} & {$\tt{3243^{\dagger}}$} & {$\tt{-24.13}$} & {$\tt{-24.02}$} & {$\tt{-23.94}$} & {$\tt{-24.24}$} & {$\tt{-4.61}$} & {$\tt{1684^{\dagger}}$} & {$\tt{-24.23}$} & {$\tt{-24.01}$} & {$\tt{-23.99}$} &  {$\tt{-24.22}$} & {$\tt{-4.59}$} & {$\tt{1755^{\dagger}}$} \\ 
{\tt{J1300+1240}} & {$\tt{160.81}$} & {$\tt{-2.95\ee{-15}}$} & {$\tt{-23.97}$} & {$\tt{-23.52}$} & {$\tt{-23.44}$} & {$\tt{-24.00}$} & {$\tt{-5.15}$} & {$\tt{734^{\dagger}}$} & {$\tt{-23.84}$} & {$\tt{-23.79}$} & {$\tt{-24.13}$} & {$\tt{-24.08}$} & {$\tt{-5.23}$} & {$\tt{611^{\dagger}}$} & {$\tt{-23.95}$} & {$\tt{-23.79}$} & {$\tt{-24.10}$} &  {$\tt{-24.10}$} & {$\tt{-5.25}$} & {$\tt{577^{\dagger}}$} \\ 
{\tt{J1435-6100}} & {$\tt{106.98}$} & {$\tt{-2.80\ee{-16}}$} & {$\tt{-24.29}$} & {$\tt{-23.64}$} & {$\tt{-23.54}$} & {$\tt{-24.31}$} & {$\tt{-4.48}$} & {\tt{1217}} & {$\tt{-24.20}$} & {$\tt{-24.04}$} & {$\tt{-24.17}$} & {$\tt{-24.48}$} & {$\tt{-4.66}$} & {\tt{819}} & {$\tt{-24.40}$} & {$\tt{-24.18}$} & {$\tt{-24.22}$} & {$\tt{-24.57}$} & {$\tt{-4.74}$} & {\tt{668}} \\ 
{\tt{J1455-3330}} & {$\tt{125.20}$} & {$\tt{-3.81\ee{-16}}$} & {$\tt{-24.12}$} & {$\tt{-23.37}$} & {$\tt{-23.25}$} & {$\tt{-24.14}$} & {$\tt{-5.09}$} & {$\tt{475^{\dagger}}$} & {$\tt{-24.29}$} & {$\tt{-24.13}$} & {$\tt{-23.89}$} & {$\tt{-24.38}$} & {$\tt{-5.33}$} & {$\tt{275^{\dagger}}$} & {$\tt{-24.38}$} & {$\tt{-24.07}$} & {$\tt{-23.88}$} &  {$\tt{-24.41}$} & {$\tt{-5.37}$} & {$\tt{253^{\dagger}}$} \\ 
{\tt{J1518+0205A}} & {$\tt{180.06}$} & {$\tt{-1.34\ee{-15}}$} & {$\tt{-23.68}$} & {$\tt{-22.61}$} & {$\tt{-22.86}$} & {$\tt{-23.73}$} & {$\tt{-3.97}$} & {\tt{6640}} & {$\tt{-23.53}$} & {$\tt{-23.74}$} & {$\tt{-23.79}$} & {$\tt{-23.85}$} & {$\tt{-4.09}$} & {\tt{5045}} & {$\tt{-23.85}$} & {$\tt{-23.74}$} & {$\tt{-23.79}$} & {$\tt{-23.96}$} & {$\tt{-4.20}$} & {\tt{3897}} \\ 
\end{tabular}
\end{table*}
\endgroup
\end{turnpage}

\begin{turnpage}
\begingroup 
\squeezetable 
\begin{table*}
\begin{center}{TABLE~\ref{table:allresults} cont.}\end{center}
\begin{tabular}{l | c | c | c | c | c | c | c | c | c | c | c | c | c | c | c | c | c | c | c | c}
\hline \hline
\multicolumn{3}{c|}{~} & \multicolumn{6}{c|}{S3} & \multicolumn{6}{c|}{S4} & \multicolumn{6}{c}{S3
and S4} \\ \hline
\multicolumn{3}{c|}{~} & \multicolumn{4}{c|}{{$\log{h_0^{95\%}}$}} & {$\log{\varepsilon}$} & ratio &
\multicolumn{4}{|c|}{{$\log{h_0^{95\%}}$}} & {$\log{\varepsilon}$} & ratio &
\multicolumn{4}{|c|}{{$\log{h_0^{95\%}}$}} & {$\log{\varepsilon}$} & ratio \\ \hline
{P\textsc{ulsar}} & $\nu$ (Hz) & $\dot{\nu}$ (Hz\,s$^{-1}$) & {H1} & {H2} & {L1} &
\multicolumn{3}{|c|}{Joint} &  {H1} & {H2} & {L1} & \multicolumn{3}{|c|}{Joint} & {H1} & {H2} & {L1}
& \multicolumn{3}{|c}{Joint} \\ 
\hline \hline
{\tt{J1537+1155}} & {$\tt{26.38}$} & {$\tt{-1.69\ee{-15}}$} & {$\tt{-22.87}$} & {$\tt{-22.09}$} & {$\tt{-22.25}$} & {$\tt{-22.85}$} & {$\tt{-2.36}$} & {$\tt{1998^{\dagger}}$} & {$\tt{-23.39}$} & {$\tt{-22.95}$} & {$\tt{-23.38}$} & {$\tt{-23.68}$} & {$\tt{-3.19}$} & {$\tt{297^{\dagger}}$} & {$\tt{-23.40}$} & {$\tt{-22.95}$} & {$\tt{-23.39}$} &  {$\tt{-23.70}$} & {$\tt{-3.21}$} & {$\tt{282^{\dagger}}$} \\ 
{\tt{J1603-7202}} & {$\tt{67.38}$} & {$\tt{-7.10\ee{-17}}$} & {$\tt{-24.33}$} & {$\tt{-23.60}$} & {$\tt{-23.52}$} & {$\tt{-24.35}$} & {$\tt{-4.42}$} & {$\tt{1040^{\dagger}}$} & {$\tt{-24.39}$} & {$\tt{-23.90}$} & {$\tt{-24.07}$} & {$\tt{-24.43}$} & {$\tt{-4.49}$} & {$\tt{876^{\dagger}}$} & {$\tt{-24.50}$} & {$\tt{-23.95}$} & {$\tt{-24.07}$} &  {$\tt{-24.58}$} & {$\tt{-4.65}$} & {$\tt{613^{\dagger}}$} \\ 
{\tt{J1623-2631}} & {$\tt{90.29}$} & {$\tt{-5.47\ee{-15}}$} & {$\tt{-23.95}$} & {$\tt{-23.45}$} & {$\tt{-23.52}$} & {$\tt{-24.00}$} & {$\tt{-4.20}$} & {$\tt{364^{\dagger}}$} & {$\tt{-24.16}$} & {$\tt{-23.76}$} & {$\tt{-24.16}$} & {$\tt{-24.16}$} & {$\tt{-4.36}$} & {$\tt{250^{\dagger}}$} & {$\tt{-24.24}$} & {$\tt{-23.79}$} & {$\tt{-24.15}$} &  {$\tt{-24.49}$} & {$\tt{-4.69}$} & {$\tt{117^{\dagger}}$} \\ 
{\tt{J1629-6902}} & {$\tt{166.65}$} & {$\tt{-2.78\ee{-16}}$} & {$\tt{-24.13}$} & {$\tt{-23.70}$} & {$\tt{-23.29}$} & {$\tt{-24.23}$} & {$\tt{-5.17}$} & {\tt{771}} & {$\tt{-24.14}$} & {$\tt{-24.12}$} & {$\tt{-24.11}$} & {$\tt{-24.37}$} & {$\tt{-5.31}$} & {\tt{559}} & {$\tt{-24.36}$} & {$\tt{-24.18}$} & {$\tt{-24.11}$} & {$\tt{-24.49}$} & {$\tt{-5.43}$} & {\tt{423}} \\ 
{\tt{J1640+2224}} & {$\tt{316.12}$} & {$\tt{-2.83\ee{-16}}$} & {$\tt{-23.58}$} & {$\tt{-23.52}$} & {$\tt{-22.98}$} & {$\tt{-23.56}$} & {$\tt{-5.11}$} & {$\tt{5659^{\dagger}}$} & {$\tt{-23.79}$} & {$\tt{-23.85}$} & {$\tt{-23.73}$} & {$\tt{-24.02}$} & {$\tt{-5.57}$} & {$\tt{1954^{\dagger}}$} & {$\tt{-23.84}$} & {$\tt{-23.86}$} & {$\tt{-23.73}$} &  {$\tt{-24.05}$} & {$\tt{-5.60}$} & {$\tt{1819^{\dagger}}$} \\ 
&&&&&&&&&&&&&&&&&&&\\[-7pt] 
{\tt{J1643-1224}} & {$\tt{216.37}$} & {$\tt{-8.66\ee{-16}}$} & {$\tt{-23.69}$} & {$\tt{-23.36}$} & {$\tt{-23.19}$} & {$\tt{-23.80}$} & {$\tt{-4.41}$} & {$\tt{5447^{\dagger}}$} & {$\tt{-23.95}$} & {$\tt{-23.80}$} & {$\tt{-23.86}$} & {$\tt{-23.97}$} & {$\tt{-4.58}$} & {$\tt{3658^{\dagger}}$} & {$\tt{-23.97}$} & {$\tt{-23.78}$} & {$\tt{-23.96}$} &  {$\tt{-23.97}$} & {$\tt{-4.58}$} & {$\tt{3627^{\dagger}}$} \\ 
{\tt{J1701-3006A}} & {$\tt{190.78}$} & {$\tt{+4.80\ee{-15}}$} & {$\tt{-24.10}$} & {$\tt{-23.37}$} & {$\tt{-23.42}$} & {$\tt{-24.01}$} & {$\tt{-4.36}$} & {\tt{*}} &  {$\tt{-24.11}$} & {$\tt{-23.88}$} & {$\tt{-23.97}$} & {$\tt{-24.23}$} & {$\tt{-4.58}$} & {\tt{*}} &  {$\tt{-24.27}$} & {$\tt{-23.85}$} & {$\tt{-24.00}$} & {$\tt{-24.41}$} & {$\tt{-4.76}$} & {\tt{*}} \\ 
{\tt{J1701-3006B}} & {$\tt{278.25}$} & {$\tt{+2.71\ee{-14}}$} & {$\tt{-23.96}$} & {$\tt{-23.48}$} & {$\tt{-23.07}$} & {$\tt{-23.94}$} & {$\tt{-4.62}$} & {\tt{*}} &  {$\tt{-24.06}$} & {$\tt{-23.89}$} & {$\tt{-23.54}$} & {$\tt{-24.05}$} & {$\tt{-4.73}$} & {\tt{*}} &  {$\tt{-24.16}$} & {$\tt{-23.92}$} & {$\tt{-23.56}$} & {$\tt{-24.16}$} & {$\tt{-4.84}$} & {\tt{*}} \\ 
{\tt{J1701-3006C}} & {$\tt{262.71}$} & {$\tt{+2.20\ee{-15}}$} & {$\tt{-23.91}$} & {$\tt{-23.45}$} & {$\tt{-23.18}$} & {$\tt{-23.84}$} & {$\tt{-4.47}$} & {\tt{*}} &  {$\tt{-23.94}$} & {$\tt{-23.94}$} & {$\tt{-23.78}$} & {$\tt{-24.11}$} & {$\tt{-4.74}$} & {\tt{*}} &  {$\tt{-24.13}$} & {$\tt{-23.93}$} & {$\tt{-23.76}$} & {$\tt{-24.20}$} & {$\tt{-4.82}$} & {\tt{*}} \\ 
{\tt{J1713+0747}} & {$\tt{218.81}$} & {$\tt{-4.08\ee{-16}}$} & {$\tt{-23.65}$} & {$\tt{-23.16}$} & {$\tt{-23.13}$} & {$\tt{-23.64}$} & {$\tt{-4.90}$} & {$\tt{2401^{\dagger}}$} & {$\tt{-23.98}$} & {$\tt{-24.02}$} & {$\tt{-23.81}$} & {$\tt{-24.08}$} & {$\tt{-5.34}$} & {$\tt{865^{\dagger}}$} & {$\tt{-24.01}$} & {$\tt{-23.94}$} & {$\tt{-23.78}$} &  {$\tt{-24.15}$} & {$\tt{-5.40}$} & {$\tt{748^{\dagger}}$} \\ 
&&&&&&&&&&&&&&&&&&&\\[-7pt] 
{\tt{J1744-1134}} & {$\tt{245.43}$} & {$\tt{-5.39\ee{-16}}$} & {$\tt{-23.88}$} & {$\tt{-23.47}$} & {$\tt{-23.39}$} & {$\tt{-23.93}$} & {$\tt{-5.78}$} & {$\tt{386^{\dagger}}$} & {$\tt{-23.67}$} & {$\tt{-23.91}$} & {$\tt{-23.95}$} & {$\tt{-23.98}$} & {$\tt{-5.83}$} & {$\tt{345^{\dagger}}$} & {$\tt{-23.83}$} & {$\tt{-23.87}$} & {$\tt{-23.97}$} &  {$\tt{-24.05}$} & {$\tt{-5.90}$} & {$\tt{296^{\dagger}}$} \\ 
{\tt{J1745-0952}} & {$\tt{51.61}$} & {$\tt{-2.53\ee{-16}}$} & {$\tt{-23.99}$} & {$\tt{-23.42}$} & {$\tt{-23.40}$} & {$\tt{-24.00}$} & {$\tt{-3.68}$} & {\tt{1328}} & {$\tt{-24.33}$} & {$\tt{-23.83}$} & {$\tt{-24.15}$} & {$\tt{-24.38}$} & {$\tt{-4.06}$} & {\tt{552}} & {$\tt{-24.28}$} & {$\tt{-23.83}$} & {$\tt{-24.16}$} & {$\tt{-24.38}$} & {$\tt{-4.06}$} & {\tt{551}} \\ 
{\tt{J1748-2446A}} & {$\tt{172.96}$} & {$\tt{+2.54\ee{-16}}$} & {$\tt{-24.19}$} & {$\tt{-23.66}$} & {$\tt{-23.61}$} & {$\tt{-24.19}$} & {$\tt{-3.75}$} & {\tt{*}} &  {\tt{-}} & {\tt{-}} & {\tt{-}} & {\tt{-}} & {\tt{-}} & {\tt{-}} &  {\tt{-}} & {\tt{-}} & {\tt{-}} & {\tt{-}} & {\tt{-}} & {\tt{-}} \\ 
{\tt{J1748-2446C}} & {$\tt{118.54}$} & {$\tt{+8.52\ee{-15}}$} & {$\tt{-24.18}$} & {$\tt{-23.12}$} & {$\tt{-23.58}$} & {$\tt{-24.18}$} & {$\tt{-4.01}$} & {\tt{*}} &  {$\tt{-24.27}$} & {$\tt{-23.95}$} & {$\tt{-24.23}$} & {$\tt{-24.50}$} & {$\tt{-4.34}$} & {\tt{*}} &  {$\tt{-24.41}$} & {$\tt{-23.96}$} & {$\tt{-24.24}$} & {$\tt{-24.54}$} & {$\tt{-4.37}$} & {\tt{*}} \\ 
{\tt{J1756-2251}} & {$\tt{35.14}$} & {$\tt{-1.26\ee{-15}}$} & {$\tt{-23.71}$} & {$\tt{-22.84}$} & {$\tt{-23.04}$} & {$\tt{-23.70}$} & {$\tt{-2.95}$} & {\tt{1209}} & {$\tt{-23.83}$} & {$\tt{-23.25}$} & {$\tt{-23.59}$} & {$\tt{-23.78}$} & {$\tt{-3.04}$} & {\tt{999}} & {$\tt{-23.84}$} & {$\tt{-23.26}$} & {$\tt{-23.59}$} & {$\tt{-23.80}$} & {$\tt{-3.05}$} & {\tt{971}} \\ 
&&&&&&&&&&&&&&&&&&&\\[-7pt]
{\tt{J1757-5322}} & {$\tt{112.74}$} & {$\tt{-3.34\ee{-16}}$} & {$\tt{-24.16}$} & {$\tt{-23.50}$} & {$\tt{-23.45}$} & {$\tt{-24.21}$} & {$\tt{-4.81}$} & {\tt{605}} & {$\tt{-24.16}$} & {$\tt{-24.14}$} & {$\tt{-24.16}$} & {$\tt{-24.44}$} & {$\tt{-5.04}$} & {\tt{353}} & {$\tt{-24.24}$} & {$\tt{-24.13}$} & {$\tt{-24.13}$} & {$\tt{-24.43}$} & {$\tt{-5.03}$} & {\tt{360}} \\ 
{\tt{J1801-1417}} & {$\tt{275.85}$} & {$\tt{-4.02\ee{-16}}$} & {$\tt{-23.70}$} & {$\tt{-23.40}$} & {$\tt{-23.15}$} & {$\tt{-23.82}$} & {$\tt{-5.07}$} & {\tt{2783}} & {$\tt{-23.92}$} & {$\tt{-23.82}$} & {$\tt{-23.89}$} & {$\tt{-24.14}$} & {$\tt{-5.39}$} & {\tt{1347}} & {$\tt{-23.99}$} & {$\tt{-23.87}$} & {$\tt{-23.86}$} & {$\tt{-24.20}$} & {$\tt{-5.46}$} & {\tt{1156}} \\ 
{\tt{J1802-2124}} & {$\tt{79.07}$} & {$\tt{-4.50\ee{-16}}$} & {$\tt{-24.05}$} & {$\tt{-23.43}$} & {$\tt{-23.53}$} & {$\tt{-24.01}$} & {$\tt{-3.91}$} & {\tt{1688}} & {$\tt{-24.11}$} & {$\tt{-24.13}$} & {$\tt{-24.14}$} & {$\tt{-24.34}$} & {$\tt{-4.24}$} & {\tt{791}} & {$\tt{-24.17}$} & {$\tt{-24.09}$} & {$\tt{-24.15}$} & {$\tt{-24.23}$} & {$\tt{-4.13}$} & {\tt{1022}} \\ 
{\tt{J1804-0735}} & {$\tt{43.29}$} & {$\tt{-8.75\ee{-16}}$} & {$\tt{-23.86}$} & {$\tt{-23.24}$} & {$\tt{-23.13}$} & {$\tt{-23.83}$} & {$\tt{-2.81}$} & {\tt{3409}} & {$\tt{-24.10}$} & {$\tt{-23.70}$} & {$\tt{-24.07}$} & {$\tt{-24.26}$} & {$\tt{-3.24}$} & {\tt{1266}} & {$\tt{-24.24}$} & {$\tt{-23.72}$} & {$\tt{-24.07}$} & {$\tt{-24.31}$} & {$\tt{-3.28}$} & {\tt{1146}} \\ 
{\tt{J1804-2717}} & {$\tt{107.03}$} & {$\tt{-4.68\ee{-16}}$} & {$\tt{-24.08}$} & {$\tt{-23.47}$} & {$\tt{-23.53}$} & {$\tt{-24.02}$} & {$\tt{-4.64}$} & {\tt{663}} & {$\tt{-24.15}$} & {$\tt{-24.15}$} & {$\tt{-24.23}$} & {$\tt{-24.36}$} & {$\tt{-4.97}$} & {\tt{305}} & {$\tt{-24.28}$} & {$\tt{-24.16}$} & {$\tt{-24.22}$} & {$\tt{-24.32}$} & {$\tt{-4.94}$} & {\tt{329}} \\ 
&&&&&&&&&&&&&&&&&&&\\[-7pt]
{\tt{J1807-2459A}} & {$\tt{326.86}$} & {$\tt{+4.87\ee{-16}}$} & {$\tt{-23.73}$} & {$\tt{-23.30}$} & {$\tt{-23.02}$} & {$\tt{-23.79}$} & {$\tt{-5.01}$} & {\tt{*}} &  {$\tt{-23.98}$} & {$\tt{-23.79}$} & {$\tt{-23.85}$} & {$\tt{-24.12}$} & {$\tt{-5.35}$} & {\tt{*}} &  {$\tt{-24.07}$} & {$\tt{-23.84}$} & {$\tt{-23.86}$} & {$\tt{-24.20}$} & {$\tt{-5.42}$} & {\tt{*}} \\  
{\tt{J1810-2005}} & {$\tt{30.47}$} & {$\tt{-1.40\ee{-16}}$} & {$\tt{-23.22}$} & {$\tt{-22.24}$} & {$\tt{-22.70}$} & {$\tt{-23.24}$} & {$\tt{-2.23}$} & {\tt{13340}} & {$\tt{-23.40}$} & {$\tt{-23.13}$} & {$\tt{-23.36}$} & {$\tt{-23.68}$} & {$\tt{-2.66}$} & {\tt{4920}} & {$\tt{-23.57}$} & {$\tt{-23.13}$} & {$\tt{-23.37}$} & {$\tt{-23.69}$} & {$\tt{-2.68}$} & {\tt{4768}} \\ 
{\tt{J1823-3021A}} & {$\tt{367.65}$} & {$\tt{-1.14\ee{-13}}$} & {$\tt{-23.74}$} & {$\tt{-23.46}$} & {$\tt{-23.36}$} & {$\tt{-23.77}$} & {$\tt{-4.02}$} & {$\tt{671}$} & {\tt{-}} & {\tt{-}} & {\tt{-}} & {\tt{-}} & {\tt{-}} & {\tt{-}} & {\tt{-}} & {\tt{-}} & {\tt{-}} &  {\tt{-}} & {\tt{-}} & {\tt{-}} \\ 
{\tt{J1824-2452}} & {$\tt{327.41}$} & {$\tt{-1.74\ee{-13}}$} & {$\tt{-23.93}$} & {$\tt{-23.41}$} & {$\tt{-23.05}$} & {$\tt{-23.92}$} & {$\tt{-4.88}$} & {$\tt{321^{\dagger}}$} & {$\tt{-23.87}$} & {$\tt{-23.77}$} & {$\tt{-23.68}$} & {$\tt{-24.09}$} & {$\tt{-5.06}$} & {$\tt{214^{\dagger}}$} & {$\tt{-24.03}$} & {$\tt{-23.87}$} & {$\tt{-23.70}$} &  {$\tt{-24.19}$} & {$\tt{-5.16}$} & {$\tt{171^{\dagger}}$} \\ 
{\tt{J1843-1113}} & {$\tt{541.81}$} & {$\tt{-2.82\ee{-15}}$} & {$\tt{-23.24}$} & {$\tt{-23.14}$} & {$\tt{-22.77}$} & {$\tt{-23.30}$} & {$\tt{-5.09}$} & {\tt{5429}} & {$\tt{-23.59}$} & {$\tt{-23.57}$} & {$\tt{-23.43}$} & {$\tt{-23.85}$} & {$\tt{-5.65}$} & {\tt{1508}} & {$\tt{-23.61}$} & {$\tt{-23.61}$} & {$\tt{-23.42}$} & {$\tt{-23.89}$} & {$\tt{-5.69}$} & {\tt{1370}} \\ 
&&&&&&&&&&&&&&&&&&&\\[-7pt]
{\tt{J1857+0943}} & {$\tt{186.49}$} & {$\tt{-6.20\ee{-16}}$} & {$\tt{-23.72}$} & {$\tt{-23.50}$} & {$\tt{-23.41}$} & {$\tt{-23.71}$} & {$\tt{-4.92}$} & {$\tt{1223^{\dagger}}$} & {$\tt{-23.97}$} & {$\tt{-23.80}$} & {$\tt{-24.04}$} & {$\tt{-24.31}$} & {$\tt{-5.52}$} & {$\tt{309^{\dagger}}$} & {$\tt{-24.00}$} & {$\tt{-23.82}$} & {$\tt{-24.07}$} &  {$\tt{-24.30}$} & {$\tt{-5.51}$} & {$\tt{313^{\dagger}}$} \\ 
{\tt{J1905+0400}} & {$\tt{264.24}$} & {$\tt{-3.39\ee{-16}}$} & {$\tt{-23.89}$} & {$\tt{-23.22}$} & {$\tt{-23.27}$} & {$\tt{-23.83}$} & {$\tt{-5.17}$} & {\tt{2185}} & {$\tt{-23.88}$} & {$\tt{-23.72}$} & {$\tt{-23.81}$} & {$\tt{-23.88}$} & {$\tt{-5.22}$} & {\tt{1937}} & {$\tt{-24.05}$} & {$\tt{-23.84}$} & {$\tt{-23.84}$} & {$\tt{-24.02}$} & {$\tt{-5.36}$} & {\tt{1410}} \\ 
{\tt{J1909-3744}} & {$\tt{339.32}$} & {$\tt{-1.61\ee{-15}}$} & {$\tt{-23.82}$} & {$\tt{-23.41}$} & {$\tt{-22.97}$} & {$\tt{-23.92}$} & {$\tt{-5.55}$} & {$\tt{1725^{\dagger}}$} & {$\tt{-23.84}$} & {$\tt{-23.89}$} & {$\tt{-23.67}$} & {$\tt{-24.00}$} & {$\tt{-5.63}$} & {$\tt{1437^{\dagger}}$} & {$\tt{-23.88}$} & {$\tt{-23.97}$} & {$\tt{-23.72}$} &  {$\tt{-24.13}$} & {$\tt{-5.76}$} & {$\tt{1060^{\dagger}}$} \\ 
{\tt{J1910-5959A}} & {$\tt{306.17}$} & {$\tt{-2.88\ee{-16}}$} & {$\tt{-23.93}$} & {$\tt{-23.35}$} & {$\tt{-23.10}$} & {$\tt{-24.01}$} & {$\tt{-5.01}$} & {\tt{4950}} & {$\tt{-23.79}$} & {$\tt{-23.59}$} & {$\tt{-23.76}$} & {$\tt{-24.07}$} & {$\tt{-5.07}$} & {\tt{4330}} & {$\tt{-24.01}$} & {$\tt{-23.80}$} & {$\tt{-23.78}$} & {$\tt{-24.21}$} & {$\tt{-5.20}$} & {\tt{3178}} \\ 
{\tt{J1910-5959B}} & {$\tt{119.65}$} & {$\tt{+1.14\ee{-14}}$} & {$\tt{-24.28}$} & {$\tt{-23.04}$} & {$\tt{-23.36}$} & {$\tt{-24.29}$} & {$\tt{-4.47}$} & {\tt{*}} &  {$\tt{-23.61}$} & {$\tt{-24.02}$} & {$\tt{-24.06}$} & {$\tt{-24.20}$} & {$\tt{-4.38}$} & {\tt{*}} &  {$\tt{-24.30}$} & {$\tt{-24.02}$} & {$\tt{-24.07}$} & {$\tt{-24.39}$} & {$\tt{-4.57}$} & {\tt{*}} \\ 
\end{tabular}
\end{table*}
\endgroup
\end{turnpage}

\begin{turnpage}
\begingroup 
\squeezetable 
\begin{table*}
\begin{center}{TABLE~\ref{table:allresults} cont.}\end{center}
\begin{tabular}{l | c | c | c | c | c | c | c | c | c | c | c | c | c | c | c | c | c | c | c | c}
\hline \hline 
\multicolumn{3}{c|}{~} & \multicolumn{6}{c|}{S3} & \multicolumn{6}{c|}{S4} & \multicolumn{6}{c}{S3
and S4} \\ \hline
\multicolumn{3}{c|}{~} & \multicolumn{4}{c|}{{$\log{h_0^{95\%}}$}} & {$\log{\varepsilon}$} & ratio &
\multicolumn{4}{|c|}{{$\log{h_0^{95\%}}$}} & {$\log{\varepsilon}$} & ratio &
\multicolumn{4}{|c|}{{$\log{h_0^{95\%}}$}} & {$\log{\varepsilon}$} & ratio \\ \hline
{P\textsc{ulsar}} & $\nu$ (Hz) & $\dot{\nu}$ (Hz\,s$^{-1}$) & {H1} & {H2} & {L1} &
\multicolumn{3}{|c|}{Joint} &  {H1} & {H2} & {L1} & \multicolumn{3}{|c|}{Joint} & {H1} & {H2} & {L1}
& \multicolumn{3}{|c}{Joint} \\ 
\hline \hline
{\tt{J1910-5959C}} & {$\tt{189.49}$} & {$\tt{-7.90\ee{-17}}$} & {$\tt{-24.03}$} & {$\tt{-23.60}$} & {$\tt{-23.37}$} & {$\tt{-24.04}$} & {$\tt{-4.62}$} & {\tt{7018}} & {$\tt{-24.18}$} & {$\tt{-24.07}$} & {$\tt{-23.97}$} & {$\tt{-24.23}$} & {$\tt{-4.81}$} & {\tt{4470}} & {$\tt{-24.32}$} & {$\tt{-24.10}$} & {$\tt{-24.00}$} & {$\tt{-24.39}$} & {$\tt{-4.97}$} & {\tt{3107}} \\ 
{\tt{J1910-5959D}} & {$\tt{110.68}$} & {$\tt{-1.18\ee{-14}}$} & {$\tt{-24.20}$} & {$\tt{-23.35}$} & {$\tt{-23.26}$} & {$\tt{-24.12}$} & {$\tt{-4.23}$} & {\tt{369}} & {$\tt{-24.40}$} & {$\tt{-24.05}$} & {$\tt{-24.17}$} & {$\tt{-24.48}$} & {$\tt{-4.60}$} & {\tt{158}} & {$\tt{-24.41}$} & {$\tt{-24.07}$} & {$\tt{-24.12}$} & {$\tt{-24.42}$} & {$\tt{-4.53}$} & {\tt{182}} \\ 
{\tt{J1910-5959E}} & {$\tt{218.73}$} & {$\tt{+2.09\ee{-14}}$} & {$\tt{-23.98}$} & {$\tt{-23.71}$} & {$\tt{-23.30}$} & {$\tt{-24.05}$} & {$\tt{-4.75}$} & {\tt{*}} &  {$\tt{-24.09}$} & {$\tt{-23.96}$} & {$\tt{-23.99}$} & {$\tt{-24.35}$} & {$\tt{-5.05}$} & {\tt{*}} &  {$\tt{-24.15}$} & {$\tt{-24.01}$} & {$\tt{-23.99}$} & {$\tt{-24.34}$} & {$\tt{-5.04}$} & {\tt{*}} \\ 
{\tt{J1911+0101A}} & {$\tt{276.36}$} & {$\tt{+5.03\ee{-16}}$} & {$\tt{-23.56}$} & {$\tt{-23.38}$} & {$\tt{-23.05}$} & {$\tt{-23.79}$} & {$\tt{-4.43}$} & {\tt{*}} &  {$\tt{-23.89}$} & {$\tt{-23.83}$} & {$\tt{-23.83}$} & {$\tt{-24.12}$} & {$\tt{-4.76}$} & {\tt{*}} &  {$\tt{-24.07}$} & {$\tt{-23.84}$} & {$\tt{-23.81}$} & {$\tt{-24.22}$} & {$\tt{-4.86}$} & {\tt{*}} \\ 
{\tt{J1911+0101B}} & {$\tt{185.72}$} & {$\tt{+6.90\ee{-17}}$} & {$\tt{-23.70}$} & {$\tt{-23.43}$} & {$\tt{-23.30}$} & {$\tt{-23.78}$} & {$\tt{-4.07}$} & {\tt{*}} &  {$\tt{-24.05}$} & {$\tt{-24.01}$} & {$\tt{-24.05}$} & {$\tt{-24.24}$} & {$\tt{-4.53}$} & {\tt{*}} &  {$\tt{-23.96}$} & {$\tt{-24.08}$} & {$\tt{-23.99}$} & {$\tt{-24.13}$} & {$\tt{-4.43}$} & {\tt{*}} \\ 
&&&&&&&&&&&&&&&&&&&\\[-7pt]
{\tt{J1911-1114}} & {$\tt{275.81}$} & {$\tt{-1.08\ee{-15}}$} & {$\tt{-23.85}$} & {$\tt{-23.51}$} & {$\tt{-23.00}$} & {$\tt{-23.84}$} & {$\tt{-5.14}$} & {$\tt{2188^{\dagger}}$} & {$\tt{-23.99}$} & {$\tt{-23.80}$} & {$\tt{-23.71}$} & {$\tt{-24.07}$} & {$\tt{-5.37}$} & {$\tt{1288^{\dagger}}$} & {$\tt{-24.04}$} & {$\tt{-23.86}$} & {$\tt{-23.70}$} &  {$\tt{-24.12}$} & {$\tt{-5.43}$} & {$\tt{1139^{\dagger}}$} \\ 
{\tt{J1939+2134}} & {$\tt{641.93}$} & {$\tt{-4.33\ee{-14}}$} & {$\tt{-23.17}$} & {$\tt{-23.22}$} & {$\tt{-22.51}$} & {$\tt{-23.37}$} & {$\tt{-5.05}$} & {$\tt{2323^{\dagger}}$} & {$\tt{-23.41}$} & {$\tt{-23.44}$} & {$\tt{-23.33}$} & {$\tt{-23.68}$} & {$\tt{-5.36}$} & {$\tt{1146^{\dagger}}$} & {$\tt{-23.46}$} & {$\tt{-23.49}$} & {$\tt{-23.31}$} &  {$\tt{-23.78}$} & {$\tt{-5.47}$} & {$\tt{899^{\dagger}}$} \\ 
{\tt{J1955+2908}} & {$\tt{163.05}$} & {$\tt{-7.91\ee{-16}}$} & {$\tt{-23.82}$} & {$\tt{-23.61}$} & {$\tt{-23.35}$} & {$\tt{-23.88}$} & {$\tt{-4.20}$} & {$\tt{4088^{\dagger}}$} & {$\tt{-24.22}$} & {$\tt{-23.94}$} & {$\tt{-24.02}$} & {$\tt{-24.32}$} & {$\tt{-4.64}$} & {$\tt{1487^{\dagger}}$} & {$\tt{-24.26}$} & {$\tt{-23.96}$} & {$\tt{-23.99}$} &  {$\tt{-24.40}$} & {$\tt{-4.72}$} & {$\tt{1223^{\dagger}}$} \\ 
{\tt{J1959+2048}} & {$\tt{622.12}$} & {$\tt{-6.52\ee{-15}}$} & {$\tt{-23.30}$} & {$\tt{-23.04}$} & {$\tt{-22.65}$} & {$\tt{-23.35}$} & {$\tt{-5.38}$} & {$\tt{3220^{\dagger}}$} & {$\tt{-23.66}$} & {$\tt{-23.37}$} & {$\tt{-23.54}$} & {$\tt{-23.79}$} & {$\tt{-5.82}$} & {$\tt{1163^{\dagger}}$} & {$\tt{-23.68}$} & {$\tt{-23.36}$} & {$\tt{-23.54}$} &  {$\tt{-23.82}$} & {$\tt{-5.85}$} & {$\tt{1080^{\dagger}}$} \\ 
{\tt{J2019+2425}} & {$\tt{254.16}$} & {$\tt{-4.54\ee{-16}}$} & {$\tt{-23.79}$} & {$\tt{-23.49}$} & {$\tt{-23.23}$} & {$\tt{-23.93}$} & {$\tt{-5.41}$} & {$\tt{1628^{\dagger}}$} & {$\tt{-23.93}$} & {$\tt{-23.89}$} & {$\tt{-23.79}$} & {$\tt{-24.16}$} & {$\tt{-5.64}$} & {$\tt{970^{\dagger}}$} & {$\tt{-23.93}$} & {$\tt{-23.91}$} & {$\tt{-23.75}$} &  {$\tt{-24.17}$} & {$\tt{-5.65}$} & {$\tt{950^{\dagger}}$} \\ 
&&&&&&&&&&&&&&&&&&&\\[-7pt]
{\tt{J2051-0827}} & {$\tt{221.80}$} & {$\tt{-6.27\ee{-16}}$} & {$\tt{-23.97}$} & {$\tt{-23.49}$} & {$\tt{-23.16}$} & {$\tt{-24.05}$} & {$\tt{-5.26}$} & {$\tt{852^{\dagger}}$} & {$\tt{-23.83}$} & {$\tt{-23.79}$} & {$\tt{-23.77}$} & {$\tt{-24.02}$} & {$\tt{-5.23}$} & {$\tt{917^{\dagger}}$} & {$\tt{-24.00}$} & {$\tt{-23.83}$} & {$\tt{-23.77}$} &  {$\tt{-24.11}$} & {$\tt{-5.32}$} & {$\tt{740^{\dagger}}$} \\ 
{\tt{J2124-3358}} & {$\tt{202.79}$} & {$\tt{-8.45\ee{-16}}$} & {$\tt{-24.08}$} & {$\tt{-23.56}$} & {$\tt{-23.35}$} & {$\tt{-24.06}$} & {$\tt{-5.90}$} & {$\tt{165^{\dagger}}$} & {$\tt{-23.91}$} & {$\tt{-23.91}$} & {$\tt{-24.00}$} & {$\tt{-24.13}$} & {$\tt{-5.98}$} & {$\tt{139^{\dagger}}$} & {$\tt{-24.20}$} & {$\tt{-24.03}$} & {$\tt{-23.99}$} &  {$\tt{-24.31}$} & {$\tt{-6.15}$} & {$\tt{93^{\dagger}}$} \\ 
{\tt{J2129-5721}} & {$\tt{268.36}$} & {$\tt{-1.49\ee{-15}}$} & {$\tt{-23.82}$} & {$\tt{-23.45}$} & {$\tt{-23.22}$} & {$\tt{-23.89}$} & {$\tt{-4.96}$} & {$\tt{1809^{\dagger}}$} & {$\tt{-24.02}$} & {$\tt{-23.95}$} & {$\tt{-23.90}$} & {$\tt{-24.19}$} & {$\tt{-5.27}$} & {$\tt{892^{\dagger}}$} & {$\tt{-24.01}$} & {$\tt{-23.91}$} & {$\tt{-23.89}$} &  {$\tt{-24.18}$} & {$\tt{-5.25}$} & {$\tt{926^{\dagger}}$} \\ 
{\tt{J2140-2310A}} & {$\tt{90.75}$} & {$\tt{+4.27\ee{-16}}$} & {$\tt{-24.25}$} & {$\tt{-23.48}$} & {$\tt{-23.52}$} & {$\tt{-24.23}$} & {$\tt{-3.82}$} & {\tt{*}} &  {$\tt{-24.31}$} & {$\tt{-24.00}$} & {$\tt{-24.06}$} & {$\tt{-24.45}$} & {$\tt{-4.04}$} & {\tt{*}} &  {$\tt{-24.45}$} & {$\tt{-23.99}$} & {$\tt{-24.10}$} & {$\tt{-24.55}$} & {$\tt{-4.14}$} & {\tt{*}} \\ 
{\tt{J2145-0750}} & {$\tt{62.30}$} & {$\tt{-1.15\ee{-16}}$} & {$\tt{-24.14}$} & {$\tt{-23.18}$} & {$\tt{-23.53}$} & {$\tt{-24.18}$} & {$\tt{-4.69}$} & {$\tt{326^{\dagger}}$} & {$\tt{-24.23}$} & {$\tt{-23.76}$} & {$\tt{-24.24}$} & {$\tt{-24.45}$} & {$\tt{-4.97}$} & {$\tt{173^{\dagger}}$} & {$\tt{-24.31}$} & {$\tt{-23.78}$} & {$\tt{-24.26}$} &  {$\tt{-24.47}$} & {$\tt{-4.98}$} & {$\tt{167^{\dagger}}$} \\ 
&&&&&&&&&&&&&&&&&&&\\[-7pt]
{\tt{J2229+2643}} & {$\tt{335.82}$} & {$\tt{-1.65\ee{-16}}$} & {$\tt{-23.58}$} & {$\tt{-23.31}$} & {$\tt{-23.08}$} & {$\tt{-23.64}$} & {$\tt{-5.17}$} & {\tt{5747}} & {$\tt{-23.77}$} & {$\tt{-23.51}$} & {$\tt{-23.72}$} & {$\tt{-23.89}$} & {$\tt{-5.41}$} & {\tt{3253}} & {$\tt{-23.89}$} & {$\tt{-23.56}$} & {$\tt{-23.71}$} & {$\tt{-24.13}$} & {$\tt{-5.65}$} & {\tt{1885}} \\ 
{\tt{J2317+1439}} & {$\tt{290.25}$} & {$\tt{-2.04\ee{-16}}$} & {$\tt{-23.41}$} & {$\tt{-23.40}$} & {$\tt{-23.17}$} & {$\tt{-23.55}$} & {$\tt{-4.82}$} & {$\tt{9996^{\dagger}}$} & {$\tt{-23.89}$} & {$\tt{-23.91}$} & {$\tt{-23.79}$} & {$\tt{-24.17}$} & {$\tt{-5.44}$} & {$\tt{2406^{\dagger}}$} & {$\tt{-23.86}$} & {$\tt{-23.90}$} & {$\tt{-23.80}$} &  {$\tt{-24.15}$} & {$\tt{-5.43}$} & {$\tt{2500^{\dagger}}$} \\ 
{\tt{J2322+2057}} & {$\tt{207.97}$} & {$\tt{-4.20\ee{-16}}$} & {$\tt{-23.86}$} & {$\tt{-23.48}$} & {$\tt{-23.17}$} & {$\tt{-24.07}$} & {$\tt{-5.44}$} & {$\tt{900^{\dagger}}$} & {$\tt{-24.01}$} & {$\tt{-23.91}$} & {$\tt{-23.85}$} & {$\tt{-24.19}$} & {$\tt{-5.56}$} & {$\tt{673^{\dagger}}$} & {$\tt{-24.07}$} & {$\tt{-23.96}$} & {$\tt{-23.96}$} &  {$\tt{-24.26}$} & {$\tt{-5.63}$} & {$\tt{578^{\dagger}}$} \\ 
\hline \hline
\end{tabular}

~ \\
~ \\
 
\caption{\label{table:allresultsGEO} Upper limits including GEO\,600 for the two fastest pulsars in
the analysis.}
\begin{tabular}{l | c | c | c | c | c | c | c | c | c | c | c | c | c | c | c | c | c | c | c | c |
c | c |
c}
\hline \hline
\multicolumn{3}{c|}{~} & \multicolumn{7}{c|}{S3} & \multicolumn{7}{c|}{S4} & \multicolumn{7}{c}{S3
and S4} \\ \hline
\multicolumn{3}{c|}{~} & \multicolumn{5}{c|}{{$\log{h_0^{95\%}}$}} & {$\log{\varepsilon}$} & ratio &
 \multicolumn{5}{|c|}{{$\log{h_0^{95\%}}$}} & {$\log{\varepsilon}$} & ratio & \multicolumn{5}{|c|}{{$\log{h_0^{95\%}}$}} & {$\log{\varepsilon}$} & ratio \\ \hline
{P\textsc{ulsar}} & $\nu$ (Hz) & $\dot{\nu}$ (Hz\,s$^{-1}$) & {H1} & {H2} & {L1} & {GEO} &
\multicolumn{3}{|c|}{Joint} &  {H1} & {H2} & {L1} & {GEO} &
\multicolumn{3}{|c|}{Joint} & {H1} & {H2} & {L1} & {GEO} & \multicolumn{3}{|c}{Joint} \\ 
\hline \hline
{\tt{J1843-1113}} & {$\tt{541.81}$} & {$\tt{-2.82\ee{-15}}$} & {$\tt{-23.24}$} & {$\tt{-23.14}$}
& {$\tt{-22.77}$} & {\tt{*}} & {$\tt{-23.30}$} & {$\tt{-5.09}$} & {$\tt{5429}$} & {$\tt{-23.59}$} &
{$\tt{-23.57}$} & {$\tt{-23.43}$} & {$\tt{-22.45}$} & {$\tt{-23.85}$} & {$\tt{-5.65}$} & {$\tt{1508}$} &
{$\tt{-23.61}$} & {$\tt{-23.61}$} & {$\tt{-23.42}$} & {$\tt{-22.45}$} & {$\tt{-23.90}$} &
{$\tt{-5.67}$} & {$\tt{1348}$} \\  
{\tt{J1939+2134}} & {$\tt{641.93}$} & {$\tt{-4.33\ee{-14}}$} & {$\tt{-23.17}$} & {$\tt{-23.22}$}
& {$\tt{-22.51}$} & {$\tt{-22.23}$} & {$\tt{-23.37}$} & {$\tt{-5.05}$} & {$\tt{2323^{\dagger}}$} &
{$\tt{-23.41}$} & {$\tt{-23.44}$} & {$\tt{-23.33}$} & {$\tt{-22.68}$} &
{$\tt{-23.66}$} & {$\tt{-5.34}$} & {$\tt{1189^{\dagger}}$} & {$\tt{-23.46}$} & {$\tt{-23.49}$} & {$\tt{-23.31}$} & {$\tt{-22.71}$} &
{$\tt{-23.77}$} & {$\tt{-5.46}$} & {$\tt{914^{\dagger}}$} \\ 

\hline \hline
\end{tabular}
\end{table*} 
\endgroup
\end{turnpage}

\end{document}